%% file: PaperFeDHCAL_20190121.tex
\documentclass[review]{elsarticle}

\usepackage{amssymb}
\usepackage[LGRgreek]{mathastext}

\usepackage{subfig}
\usepackage{url}
\usepackage[applemac]{inputenc}
\usepackage{microtype}

\journal{Nuclear Physics A}

\begin{document}

\begin{frontmatter}

\input{authors.tex}

\clearpage
\vfill

\begin{abstract}
We present a study of the response of the highly granular Digital Hadronic Calorimeter with steel absorbers, the Fe-DHCAL, to positrons, muons, and pions with momenta ranging from 2 to 60\,GeV/c. Developed in the context of the CALICE collaboration, this hadron calorimeter utilises Resistive Plate Chambers as active media, interspersed with steel absorber plates. With a transverse granularity of 1$\,\times\,$1\,cm$^{2}$ and a longitudinal segmentation of 38 layers, the calorimeter counted 350,208 readout channels, each read out with single-bit resolution (digital readout). The data were recorded in the Fermilab test beam in 2010-11. The analysis includes measurements of the calorimeter response and the energy resolution to positrons and muons, as well as detailed studies of various shower shape quantities. The results are compared to simulations based on \textsc{Geant4}, which utilise different electromagnetic and hadronic physics lists.
\end{abstract}


\end{frontmatter}


\section{Introduction}
\label{sec:intro}
For experiments at a future $e^+e^-$ linear collider such as the International Linear Collider (ILC)~\cite{ILC} or the Compact Linear Collider (CLIC)~\cite{CLIC}, new calorimeter systems are being developed with the goal to achieve jet energy resolutions of 30\,\%/$\sqrt{E}$ to perform precision measurements like the determination of the various Higgs couplings. This ambitious goal can be achieved using Particle Flow Algorithms~\cite{Thomson200925} for event and particle reconstruction. These reconstruction algorithms require calorimeter systems with high transversal and longitudinal granularity, to distinguish between close by particles and to match the signals between the tracking and calorimetric detector systems. The CALICE collaboration~\cite{CALICE} developed and tested different technological choices to address the challenge of calorimeters with  multi-million channel readouts. 

This paper presents the performance study of a highly granular Digital Hadron Calorimeter prototype (DHCAL) that was designed to fulfil the ILC and CLIC requirement of a 3-4\,\% jet energy resolution. The construction and subsequent tests of the prototype served to validate both the technological approach and the detailed simulation of hadron shower models. 

This paper focusses on the analysis of single particle events obtained with the Fe-DHCAL in beam tests at Fermilab. This study is complemented by the validation of the simulation of the RPC response tuned to muon and positron data and the comparison to several electromagnetic and hadronic physics lists of \textsc{Geant4}. The expected performance of the Fe-DHCAL within a full-size experiment is also discussed. The validation using a full jet reconstruction chain lies beyond the reach of the presented analysis.

\section{The Digital Hadron Calorimeter with steel absorbers}
The Digital Hadron Calorimeter (DHCAL)~\cite{Adams} is a sampling calorimeter with Resistive Plate Chambers (RPCs)~\cite{RPCs} as active medium. Each DHCAL layer consists of three RPCs with dimensions of $32\,\times\,96\,cm^2$ that are stacked on top of each other within a cassette consisting of a 2\,mm Copper front plate and a 2\,mm Steel back plate. Thus a layer covers an area of approximately $1\times1$\,m$^2$. These layers are inserted in 1.4\,cm wide gaps of a Steel absorber structure consisting of 39 absorber plates of 17.4\,mm thickness. The RPCs consist of two glass plates enclosing a 1.15\,mm gap filled with the standard RPC gas mixture~\cite{Adams} for operation in avalanche mode. 

Charged particles traversing the RPC gap ionise the molecules of the gas. The ionisation is amplified through avalanche processes induced by the high bias voltage of 6.3\,kV applied through a resistive coating on the outside of the glass plates. The avalanche is quenched by the high bulk resistivity of the glass of around $4.7\cdot10^{13}\,\Omega cm$ and the Isobutane and SF$_{6}$ components of the gas mixture. The avalanche induces a charge on the array of 1$\,\times\,$1\,cm$^{2}$ readout pads. If the charge exceeds a threshold of 110\,fC, a hit is time-stamped and registered. The electronic readout system is pulsed at 10\,MHz, thus providing time bins with a width of 100\,ns. The spatial dispersion of the charge avalanche within the gas gap results in an average hit multiplicity larger than 1 for Minimum Ionising Particles (MIPs).

\section{Experimental setup}
\label{sec:expSetup}
The data samples of the Fe-DHCAL were recorded in 2010-11 at the Fermilab Test Beam Facility (FTBF)~\cite{Fermilab}, using a positively charged secondary beam composed of muons, pions, protons, kaons and positrons.

The testbeam setup consisted of a main stack with 38 DHCAL layers and up to 14 DHCAL layers inserted in a so-called Tail Catcher Muon Tracker (TCMT)~\cite{TCMT} located downstream of the main stack. The Fe-DHCAL thickness corresponded to $5.3$ nuclear interaction lengths $\lambda_{n}$ and $57.6$ radiation lengths $X_{0}$. The TCMT added another $5.8$ interaction lengths, which ensured full shower containment with a total thickness of $11.1\,\lambda_{n}$. In addition, the signals of Cherenkov threshold counters, tuned to be responsive to electrons but not to heavier particles, were included into the data stream.

The applied threshold on the pads was kept constant during the operation to about 110\,fC. 
A set of two scintillator paddles of 19\,$\times$\,19\,cm$^2$ was placed directly behind each other, one meter upstream of the Fe-DHCAL. The coincidence of their signals was used to trigger the data acquisition and thus collect the beam data. Additional scintillator panels of 1\,$\times$\,1\,m$^2$ were placed 4 meters upstream of the Fe-DHCAL and downstream of the TCMT structure, which enabled the identification of muons using the coincidence of their signals~\cite{DHCALmuons_pro}. 

The present analysis focusses on the Fe-DHCAL. Since for part of the data sets the TCMT was not fully equipped, the TCMT data have been excluded from the analysis. 

\section{Equalisation of the response}
\label{sec:Calib}
The testbeam data were recorded in 101 separate data taking runs spanning the beam energies from 2 to 60\,GeV. During the data taking period, the operational conditions of the RPCs i.e. the temperature and ambient air pressure, changed, which impacted both the single particle detection efficiency and the average pad multiplicity for single particles~\cite{EnvDepRPCs}. To ensure a homogenous response over all RPCs, an offline calibration procedure is applied to the data set. This procedure applies a time dependent correction factor $c_{i,j}$ to all hits in RPC $j$ of layer $i$ 
\begin{equation}
c_{i,j}=\frac{\epsilon_{0}\cdot\mu_{0}}{\epsilon_{i,j}\cdot\mu_{i,j}},
\end{equation}
where $\epsilon_{0}=0.97$ and $\mu_{0}=1.69$ are the average detection efficiency and pad multiplicity for single particles of all chambers and all runs. The detection efficiency $\epsilon_{i,j}$ of RPC j in layer i is defined as the probability to measure at least one hit per traversing minimum-ionising particle. The pad multiplicity $\mu_{i,j}$ of a RPC is defined as the average number of hits measured per traversing minimum-ionising particle. The efficiency and multiplicity of a RPC can be determined using muons or track segments originating from MIPs within the hadronic showers~\cite{CalibrationDhcalPaper, tracksAHCAL}. The conditions are assumed to be constant during a given data taking run. This analysis uses track segments, since these reflect the conditions of the chambers during the exact same time as the data taking run. However, the disadvantage is the limited statistics especially for the top and bottom RPCs due to the location of the beam at the centre of the front face of the calorimeter. To ensure a meaningful extraction of calibration constants, the minimum number of track measurements per RPC is set to 500. In case one RPC does not reach the necessary number of measurements, the calibration constant of the center RPC in the same layer, which always contains the minimum number of tracks, is assigned. This is a reasonable choice since the gas flow is the same and the temperature variation within one layer is negligible. 

\begin{figure}
\centering
	\includegraphics[width=.75\textwidth]{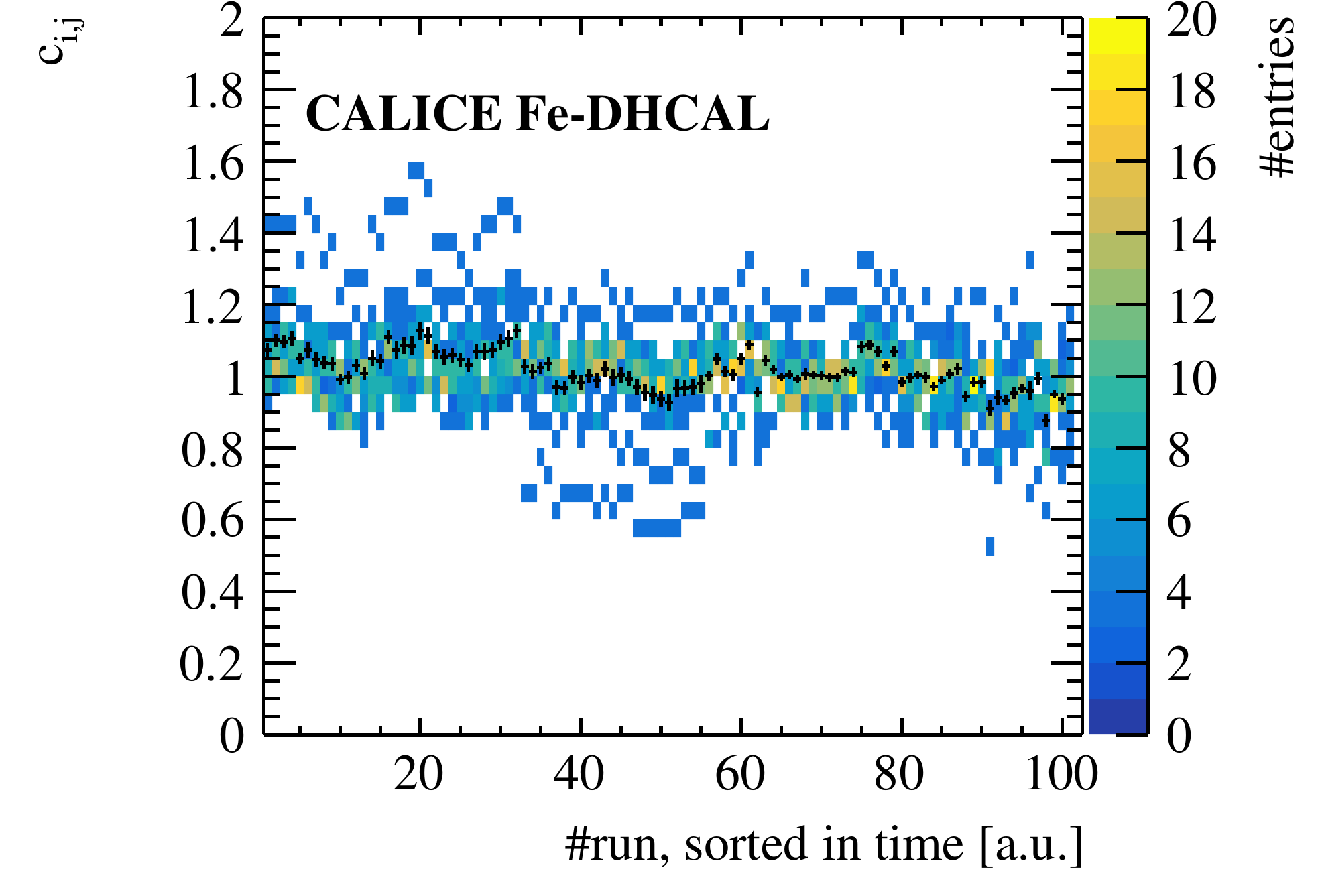}
\caption{The calibration coefficients $c_{i,j}$ for all RPCs of the recorded data runs. The mean values per run are shown in black.}
\label{fig:Calibrations}
\end{figure}

Figure~\ref{fig:Calibrations} shows the calibration coefficients $c_{i,j}$ for all runs and RPCs per run. The fluctuations around 1 display the corrections to the determined average hit multiplies and efficiencies.  
Further information about the calibration procedure can be found in~\cite{CalibrationDhcalPaper, PhDthesis}.

\section{Event selection}
\label{sec:events}
The FTBF provides momentum selected secondary beams with a mixture of $\mu^{+}$, $e^{+}$, $\pi^{+}$, protons and kaons, where the fraction of each particle type depends on the beam energy. While for beam energies below 10\,GeV the positron content is dominant, the beam is composed of more than 50\,\% pions for beam energies between 10 and 40\,GeV~\cite{Fermilab}. For beam energies above 40\,GeV the proton and kaon content becomes dominant. In general, positrons and pions were identified with a Cherenkov threshold counter for beam momenta below and above 32\,GeV, respectively. For part of the runs at 2, 4, 25 and 32\,GeV, the Cherenkov information was however not available, and particles are identified by selection rules based on event topologies. 

\begin{figure}
\centering
\subfloat[]{
	\includegraphics[width=1.\textwidth]{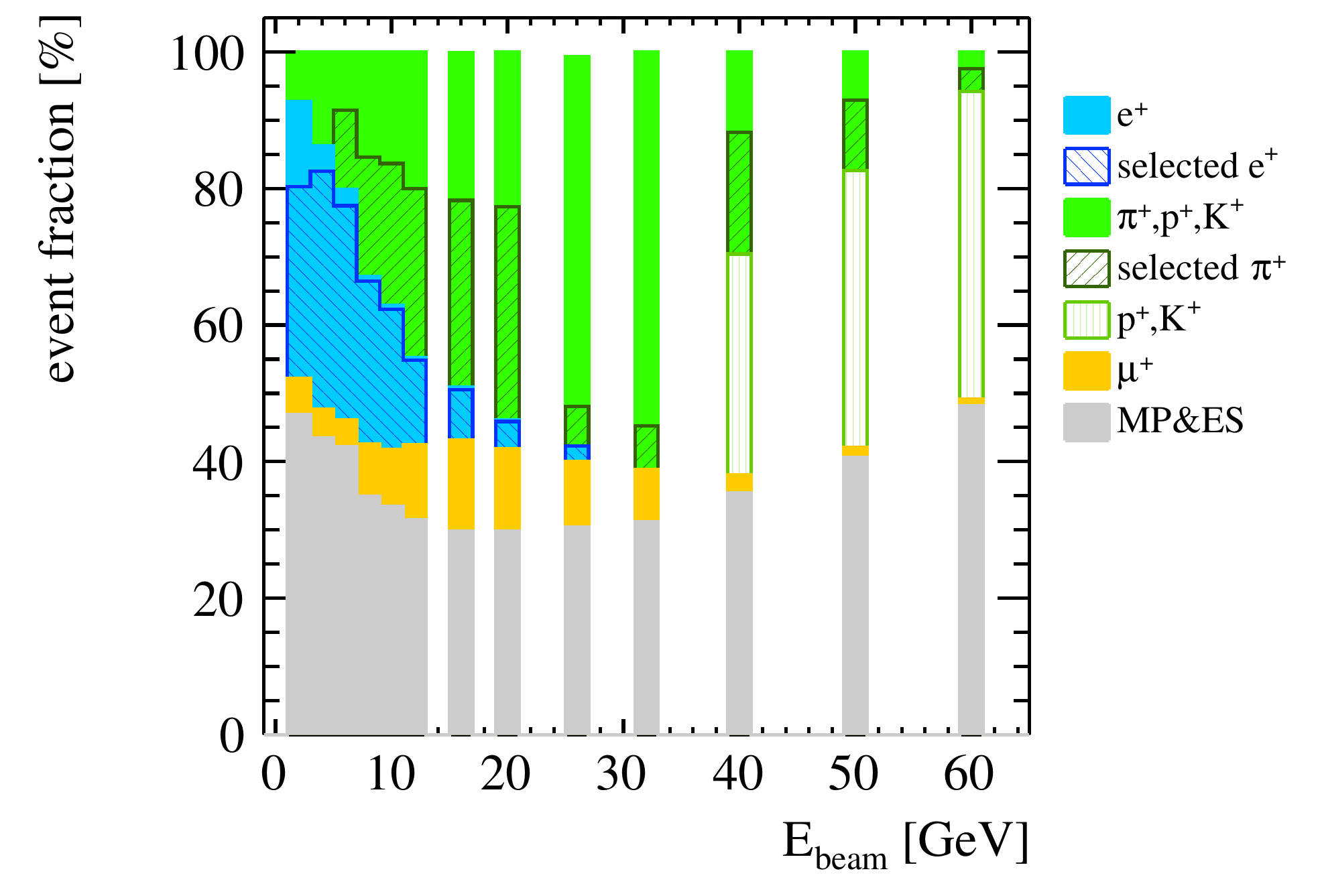}
	}\\
\subfloat[]{
	\includegraphics[width=.9\textwidth]{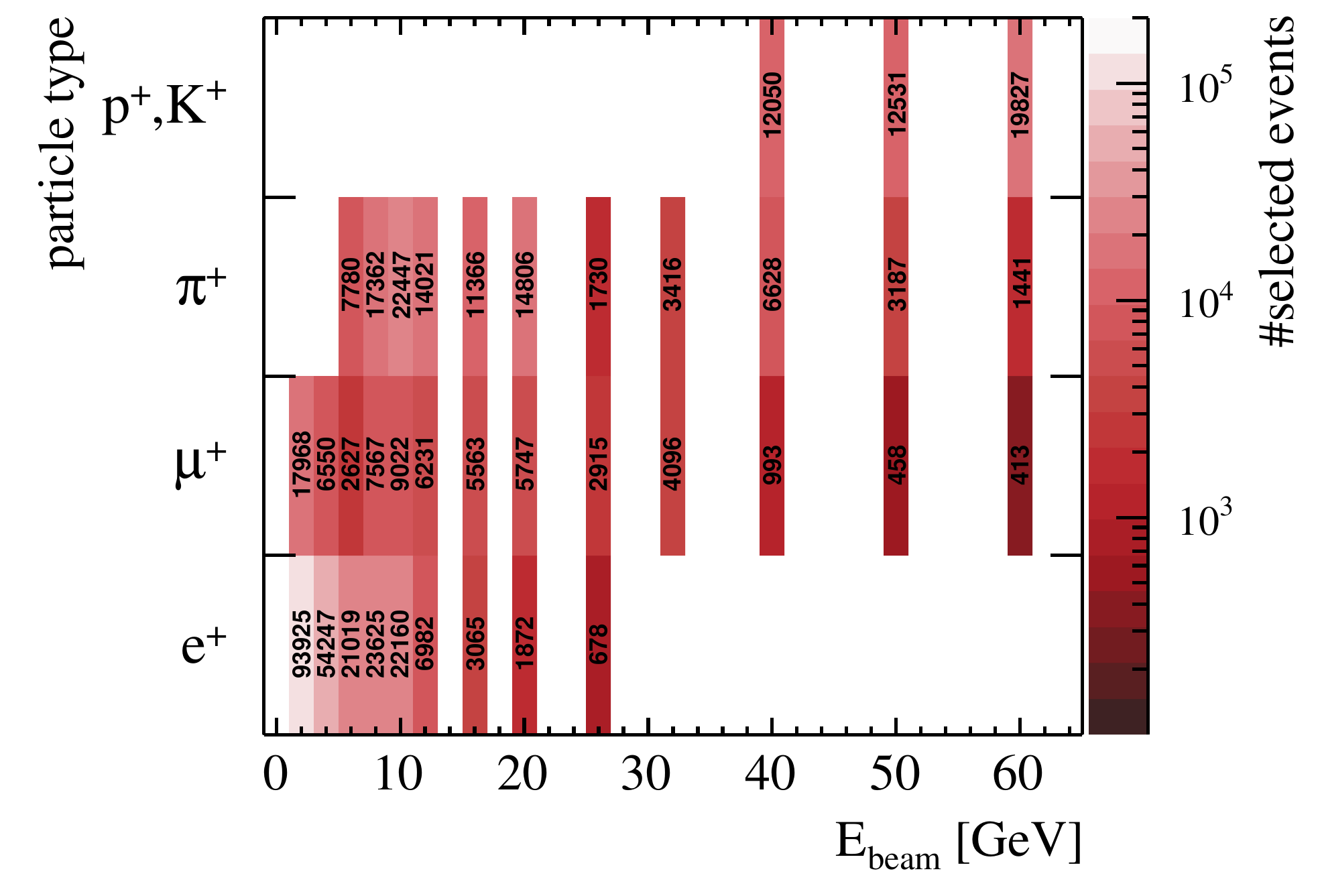}
	}
\caption{(a) Event fractions of identified and selected particle types, with \textit{MP\&ES} standing for multiple particle and early showers. (b) Number of events after the final selection of muons, positrons and pions.}
\label{fig:EveSel}
\end{figure}

A significant fraction of events contained more than one particle per trigger. In addition, some events featured particles which had initiated showers upstream of the calorimeter. The contamination from these events was effectively eliminated by requiring exactly one cluster with at most four hits in the first layer of the Fe-DHCAL. A cluster is defined as either one isolated hit or a combination of hits that are connected through a common pad border. On average this requirement removed 36\,\% of the events, see Table~\ref{tab:DHCALeventSelection} and Fig.~\ref{fig:EveSel}(a). 

Through-going muons are identified by the $1\times1$\,m$^2$ large scintillator planes located upstream and downstream of the Fe-DHCAL. This technique works well for beam energies up to 32\,GeV. For higher energies, late-showering or punch-through pions can trigger the second plane, leading to a mis-identification as through-going muons. Therefore, above 32\,GeV, muons are identified instead by requiring the centre of gravity $cog_{z}$ in the beam direction to be larger than layer 15 and the average number of hits per layer to be $>0.5$ and $<2.5$. The former is defined as the weighted z position of all hits 
\begin{equation}
cog_{z}=\frac{1}{N_{hits}} \sum_{i=1}^{38}z_{i}\sum_{j=1}^{3}\sum_{k=1}^{3,072}h_{i,j,k} \cdot c_{i,j},
\end{equation} 
with $z_{i}$ being the longitudinal position of layer i. The number of hits per event is defined as the sum over all layers $\mathrm{i}$, RPCs $\mathrm{j}$, and pads $\mathrm{k}$ reading out that RPC, 
\begin{equation}
\mathrm{N_{hits} = \sum_{i=1}^{38}\sum_{j=1}^{3}{\sum_{k=1}^{3,072}h_{i,j,k}\cdot c_{i,j}}},
\end{equation}
where $\mathrm{h_{i,j,k}}=1$ if the pad charge is above threshold and $\mathrm{h_{i,j,k}}=0$ otherwise.

For the identification of electromagnetic showers the centre of gravity $\mathrm{cog_{z}}$ and the average shower widths $rms_{x}$ and $rms_{y}$ of the events are used. The latter are defined as the standard deviation of the x and y positions of all hits in an event. Typically, positrons initiate an electromagnetic shower within the first layers of the calorimeter and deposit their energy within a cylinder of 5\,cm radius (the Moli\`ere radius of the Fe-DHCAL is about $1.8$\,cm). Positrons are therefore selected requiring $rms_{x,y}<5\,cm$. Finally, the shower is required to start within the first 5 layers, which is equivalent to 8 radiation lengths. This ensures the full EM shower containment and an additional separation from pions ($1\,\lambda_{n}$ correspond to $\thicksim7.2$ DHCAL layer).

Proton and kaon events are identified for beam energies of 40, 50 and 60\,GeV, see the light green shaded area in Fig.~\ref{fig:EveSel}~(a). Pions are distinguished from protons and kaons using the Cherenkov counter signals.

The remaining events after the muon, positron, and proton/kaon selections described above are identified as pions. To minimise longitudinal leakage, pion events are required to initiate showering in the first 10 layers of the calorimeter. The so-called interaction layer is determined using an algorithm based on the average number of hits in three consecutive layers. The interaction layer is defined as the middle of such consecutive layers for which the average increased by at least a factor of two. If several triplets of consecutive layers show such an increase in the number of hits, the one closest to the front of the calorimeter is chosen as the interaction layer~\cite{DHCALinteractionLayer}.

The total fractions of identified muon, positron and pion events are summarised in Table~\ref{tab:DHCALeventSelection}. The final selection includes the requirement of a first hard interaction.

The final numbers of selected events are pictured in Fig.~\ref{fig:EveSel}. The efficiency of the topological cuts has been studied with MC simulations in the energy range where the Cherenkov signals have not always been available for the identification and separation of positron and pion events. Within the energy range of 20 to 40\,GeV, the purity of the pion selection has been determined to be better than 99\,\% at an electron identification efficiency better than 70\,\%. The mis-identification probability of pions as muons has been evaluated for the energy range from 6-60\,GeV and is on average $1.5\pm0.2$\,\%. The distinction between protons and pions in data is fully dependent on the Cherenkov signals and thus dominated by the Cherenkov counter efficiency for beam energies between 40 and 60\,GeV. A dedicated study of simulation sets with protons, pions, and a mix of both has shown a good agreement of the selected pions with the pure pion simulation set, thus validating the strategy of the pion selection.
  
\renewcommand*{\thefootnote}{\fnsymbol{footnote}}

\begin{table}
\begin{center}
\caption[]{Total event fractions of the multi particle and early shower events, the identified muon, positron and pion events as well as the selected positron and pion events (last 2 columns) in the data set. For beam energies of 2, 4, 25 and 32\,GeV, the Cherenkov information was not available. }
\label{tab:DHCALeventSelection}
\begin{tabular}{|p{.9cm}||p{1.3cm}|p{.5cm}|p{.5cm}|p{1.7cm}||p{.6cm}|p{.6cm}|}
energy [GeV] & MP\,\&\,ES [\%] & $\mu^+$ [\%] & $e^{+}$ [\%] & $\pi^{+}$/\,$p^{+}$/\,$K^{+}$ [\%] & $e^{+}_{final}$ [\%] & $\pi^{+}_{final}$ [\%] \\ 
\hline
2\footnotemark[1] & 46.9 & 5.4  & 40.5 &     -  & 28.0 & - \\
4\footnotemark[1] & 43.4 & 4.2  & 38.7 &     -  & 34.9 & - \\
6   & 42.2 & 3.9  & 33.8 & 20.1 		& 31.3 & 11.6 \\
8   & 34.9 & 7.6  & 24.5 & 32.9		& 23.9 & 17.5 \\
10 & 33.5 & 8.4  & 20.9 & 37.1 		& 20.5 & 20.8 \\
12 & 31.5 & 11.0 & 12.5 & 44.8 		& 12.4 & 24.8 \\
16 & 29.8 & 13.4 & 7.5  & 49.0 		& 7.4  & 27.3 \\
20 & 29.8 & 12.1 & 4.0  & 53.8 		& 4.0  & 31.2 \\
25\footnotemark[1]  & 30.4 & 9.7 & 56.9	& 56.9 & 2.3 & 5.7 \\
32\footnotemark[1]  & 31.2 & 7.6 & - 		& 61.2 & -     & 6.4 \\
40 & 35.4 & 2.7  & - & 61.9 & -      	& 17.8 \\
50 & 40.6 & 1.5  & - & 57.9 & -       	& 10.3 \\
60 & 48.2 & 0.9  & - & 50.1 & -      	& 3.3 \\
  \hline
  \multicolumn{7}{l}{\footnotesize{* Cherenkov not always available.}}
\end{tabular}
\end{center}
\end{table}

\section{Monte Carlo simulation}
\label{sec:Sim}
The Fe-DHCAL testbeam setup is simulated using the software package \textsc{Geant4}~\cite{GEANT4} version 10.01. The \textsc{Geant4} software toolkit describes the interaction of particles with matter using a variety of models. The development of electromagnetic showers involves electrons, positrons and photons, originating from Bremsstrahlung and $e^+e^-$ pair production. These processes and the ionising energy loss are well understood and modelled in great detail. 
This analysis focuses on the impact of different electromagnetic model options. These models differ in the accuracy of the description of multiple scattering and in the step limits used for the calculation of the ionising energy loss ranging from 0.1 to 1.0\,mm~\cite{Geant4_emLists}. 

Hadronic showers are exceedingly more complex, involving a large number of physical processes, which renders the simulation significantly more challenging. The most accurate description of hadronic showers is achieved by string models that are coupled to cascade models~\cite{Geant4_PartonString, Geant4_reccomendation}. Thus, the present study concentrates on the validation of the FTFP\_BERT and QGSP\_BERT physics lists, which have been the most successful in the description of other highly granular calorimeters~\cite{ValidationGeant4}. 

All 101 testbeam runs have been simulated individually. The dead channels identified in the data have been switched off as well in simulation to decouple the simulation accuracy from hardware effects. The differences in the hit multiplicity and efficiency per RPC are modelled on average over the whole prototype by the digitiser of the RPC response, which is described in more detail in the following section.
 
\subsection{Digitisation of the RPC response}
\label{sec:digi}
\begin{figure}
\begin{center}
	\hspace*{-1cm}
	\begin{minipage}{.5\textwidth}
  		\subfloat [$d_{cut}$] {\label{fig:Dcut} \includegraphics [width=0.9 \textwidth] {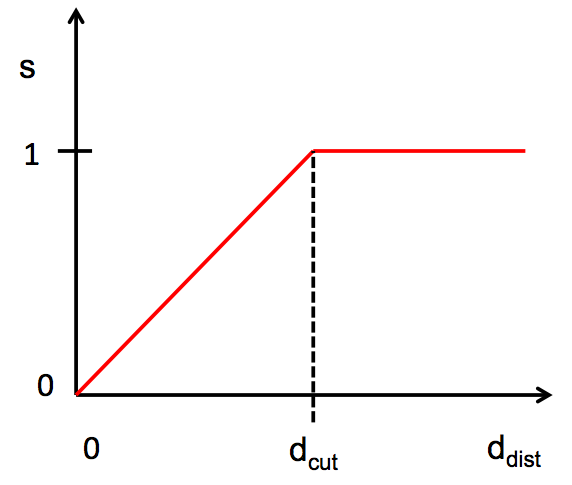}}
   	\end{minipage}
	\hspace*{-.75cm}
	\begin{minipage}{.45\textwidth}
  		\subfloat [$q_{ind}$] {\label{fig:ChargeDistr} \includegraphics [width=1.2 \textwidth] {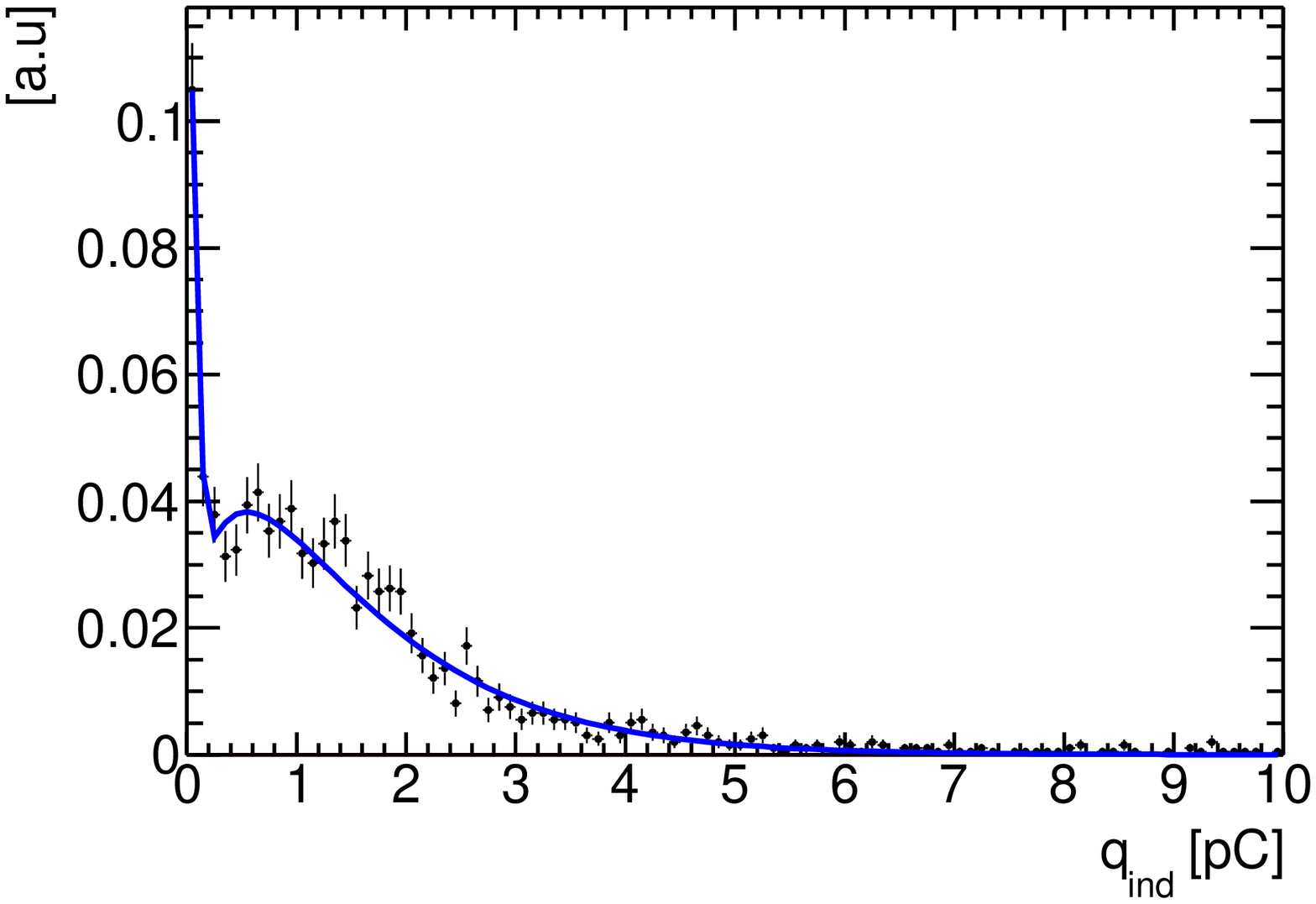}}
	\end{minipage}
\caption{a) Schematic of the dependence of the scaling factor \textit{s} on the distance between \textsc{Geant4} energy depositions $d_{dist}$. $d_{cut}$ marks the transition of depositions that get assigned a lowered charge by \textit{s}. b) The charge distribution of muons measured in testbeam~\cite{Burak}, and the corresponding fit (blue line). }
\end{center}
\end{figure}

The digitiser simulates the response of the RPCs to ionising radiation. The RPC response is emulated considering all energy depositions in the gas gap as seeds for avalanches. Since the size of the avalanche depends strongly on the location of the first ionisation in the gas gap, but only weakly on the energy deposited, the latter is not considered when generating a signal charge. 
Within the gas gap, the probability of an electron to gain enough energy to generate a Townsend avalanche decreases in the presence of an avalanche already developing close by due to the drop in the electric field strength. This limitation in spatial response of the RPCs is simulated by introducing a scaling factor \textit{s} that is assigned to one energy deposition if it is too close to another deposition and later in time. The timing information of the energy depositions is given by \textsc{Geant4}. To identify the affected energy depositions, the first step is to calculate the distances $d_{dist}$ between all energy depositions in the same layer. If two depositions are closer than a distance $d_{cut}$, the charge of the later energy deposition is scaled by \textit{s} between 0 and 1, increasing linearly with $d_{dist}$. A schematic of the scaling factor \textit{s} as a function of the distance $d_{dist}$ is shown in Fig.~\ref{fig:Dcut}. 

In the next step, the digitiser assigns a charge to each deposition according to the fit of the measured RPC charge spectrum shown in Fig.~\ref{fig:ChargeDistr}. Instead of using the theoretical description of the charge, following the approach of the CALICE Semi-Digial HCAL~\cite{SDHCALdigitizer}, this spectrum was recorded in a muon beam at Fermilab by one RPC that was also used for collecting the present data set. This RPC was read out with an analogue readout system~\cite{RPCs} and was operated in similar conditions as in the 2010 testbeam period. 

The measured charge distribution is shown in Fig.~\ref{fig:ChargeDistr}. The shape of the charge distribution strongly depends on the distance of the primary ionisation from the readout anode, which defines the induced signal height~\cite{RPCSimulation}. The closer a deposition is to the anode the smaller is the probability to generate a Townsend avalanche; the shorter the path length of an induced avalanche; the smaller the induced signal on the pad plane. This effect is seen in the large number of charges $<0.2\,pC$. 

Due to possible differences in operating conditions, an additional free, but universal, scaling factor $q_{0}$ is introduced multiplying the generated avalanche charge. 

In a next step, the generated avalanche charge is spread on the anode plane as a function of the lateral distance $r$ from the ionisation location:
\begin{equation}
f\left(r\right)=\left(1-R\right)\cdot\exp\left(-\frac{r^2}{2\sigma_{1}^2}\right)+R\cdot\exp\left(-\frac{r^2}{2\sigma_{2}^2}\right), 
\end{equation}
with three parameters: the ratio R weighting the contributions from the two Gaussians and the widths of the Gaussians $\sigma_{1}$ and $\sigma_{2}$. After all charges from all avalanches are distributed over the readout pads, the charges on each pad are summed up and a threshold $T$ is applied. 

\begin{figure}
\begin{center}
	\includegraphics[width=.6\textwidth]{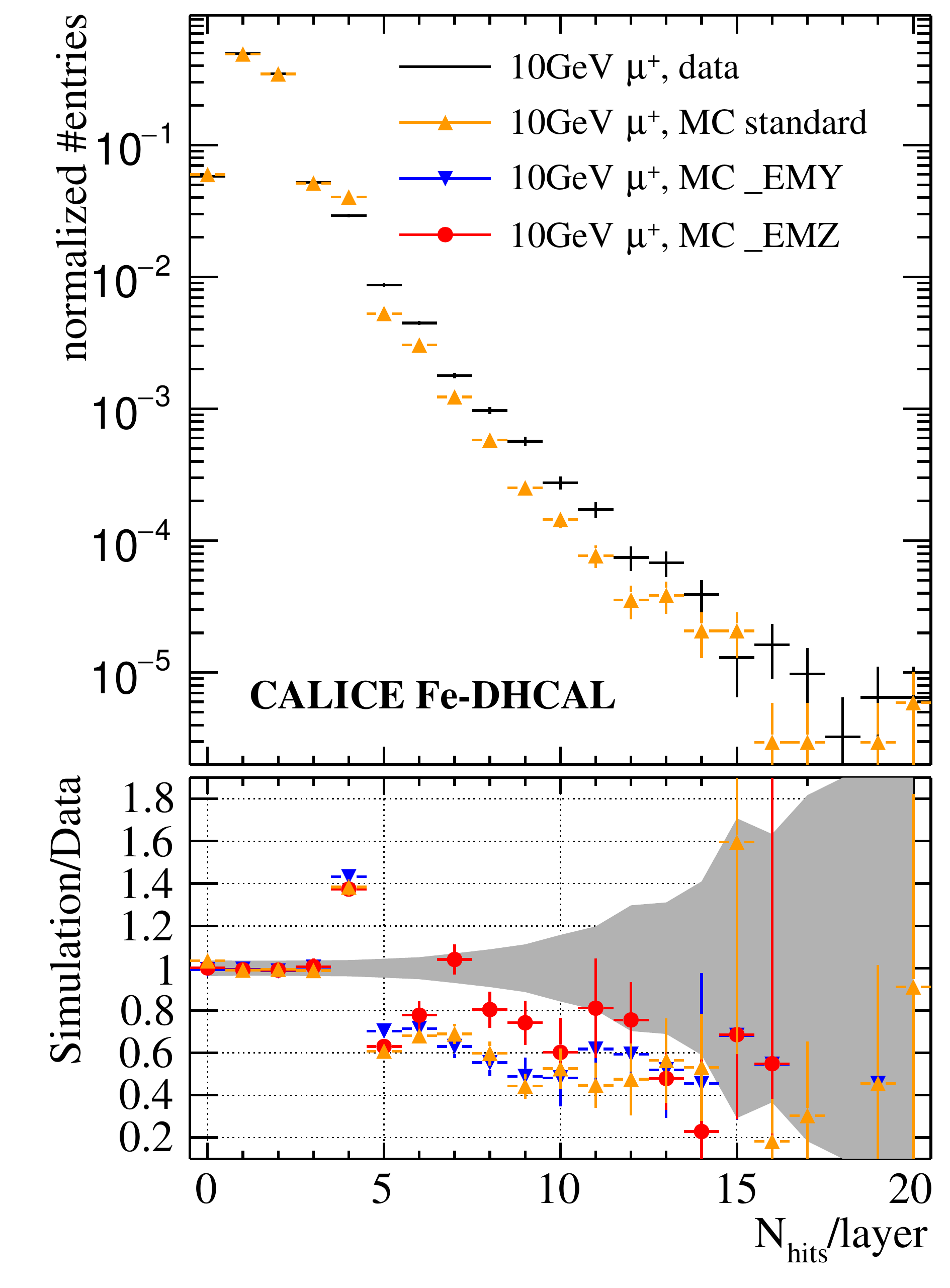}
\caption{The top plot shows the distribution of the number of hits per layer $N_{hits}/layer$ for 10\,GeV muons for data and the standard EM physics list. The bottom plot shows the ratio between all the simulations and data. The grey band indicates the statistical and systematic uncertainties of the data added in quadrature. }
\label{fig:tuning_FTFP_BERT}
\end{center}
\end{figure}

\begin{figure}
\begin{center}
	\includegraphics[width=.6\textwidth]{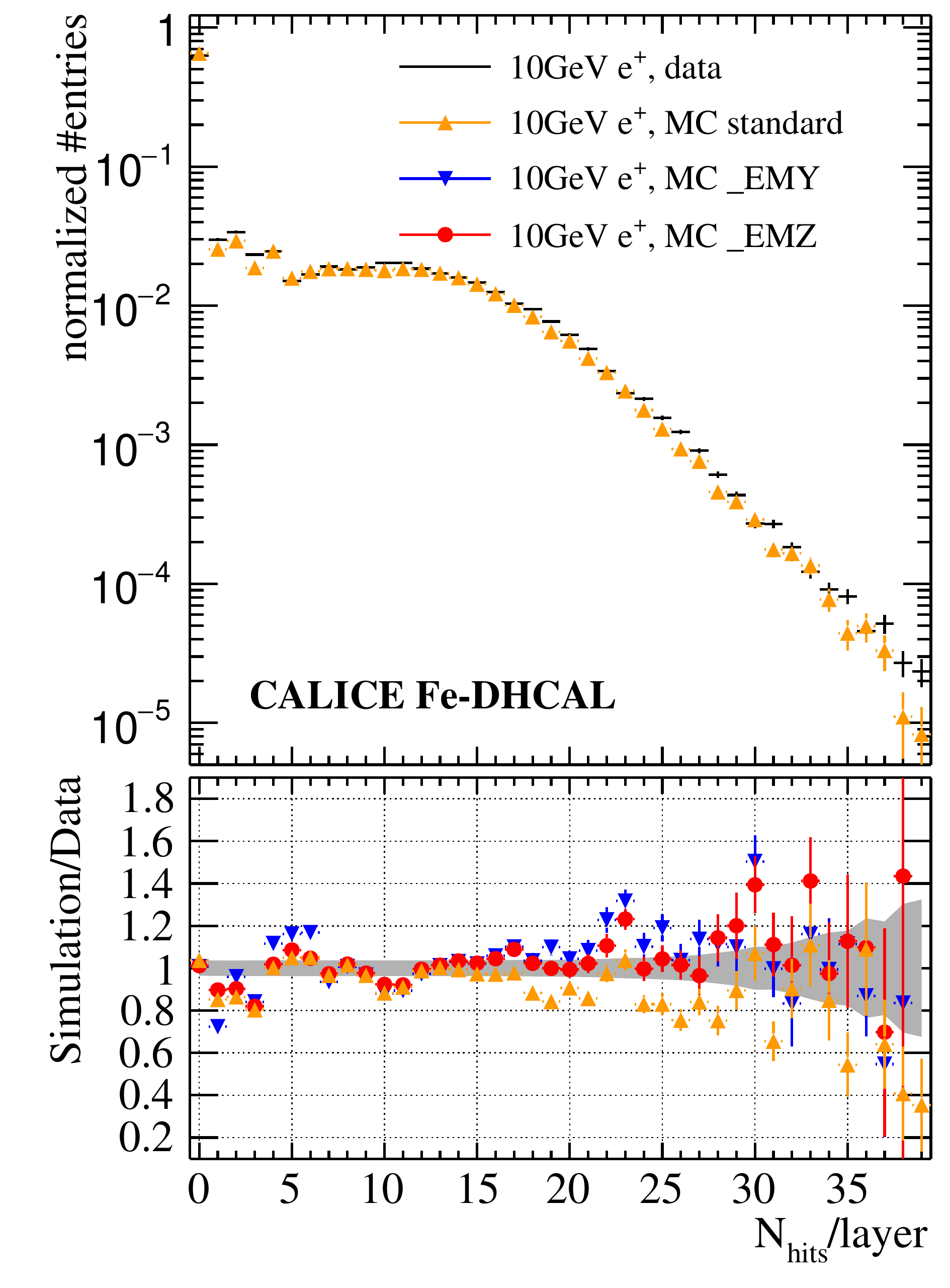}
\caption{The top plot shows the distribution of the number of hits per layer for 10\,GeV positrons for data and the standard EM physics list. The bottom plot shows the ratio between all the simulations and data. The grey bands indicate the statistical uncertainty of the data added in quadrature. }
\label{fig:tuning_FTFP_BERT_EMY}
\end{center}
\end{figure}

\begin{figure}
\begin{center}
	\includegraphics[width=.6\textwidth]{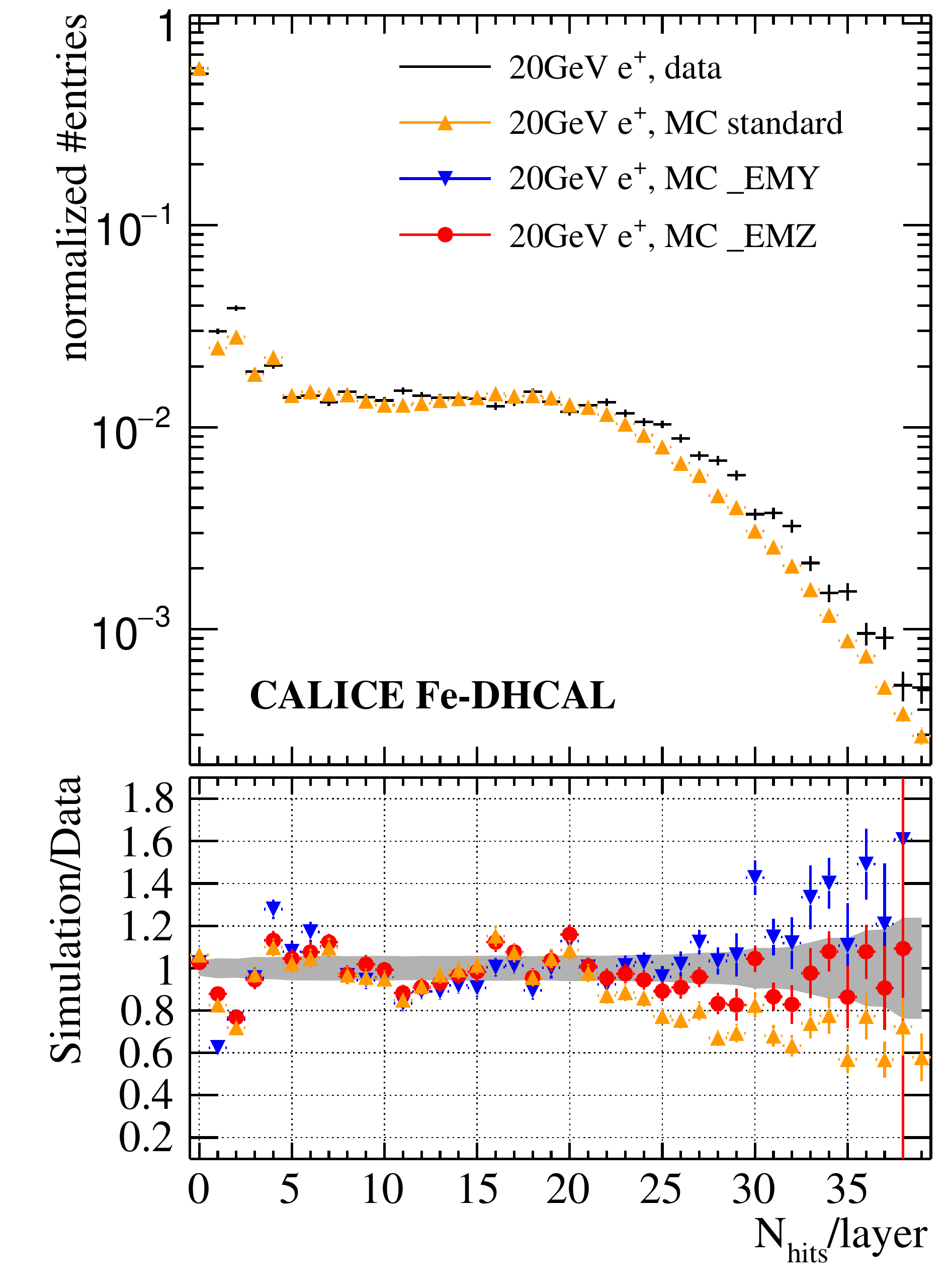}	
\caption{The top plot shows the distribution of the number of hits per layer for 20\,GeV positrons for data and the standard EM physics list. The bottom plot shows the ratio between all the simulations and data. The grey bands indicate the statistical and systematic uncertainties of the data added in quadrature. }
\label{fig:tuning_FTFP_BERT_EMZ}
\end{center}
\end{figure}

The 6 digitisation parameters ($d_{cut}$, $q_{0}$, $R$, $\sigma_{1}$, $\sigma_{2}$ and T) are highly correlated and have to be determined from data. The tuning of these parameters is done, matching the simulated number of hits per layer “$N_{hits}/layer$" of 10\,GeV muons and 10 and 20\,GeV positrons to the measured distributions, see Figs.~\ref{fig:tuning_FTFP_BERT} to~\ref{fig:tuning_FTFP_BERT_EMZ}. The parameter space has been explored by assigning to each parameter a value within a reasonable range and testing all possible combinations. The agreement between the data and the simulation is determined for each parameter combination using the $\chi^2$ values between the histograms as a measure of agreement. 

This procedure is repeated for three different versions of electromagnetic (EM) physics lists of \textsc{Geant4}~\cite{Geant4_emLists}; the “standard", the “option 3" or \_EMY, and “option 4" or \_EMZ physics lists. These options vary in accuracy, and most important for this analysis, in the step length for which the next ionising energy deposition is calculated for~\cite{Geant4_emLists}. Since the deposited energies themselves are not taken into account in the digitisation, but for each deposition point a charge is assigned and an avalanche is generated, the number of original depositions has a great effect on the generated total number of hits. From ~\cite{Geant4_private} the recommended EM physics list for gaseous detectors is \_EMY; with a reduced step length of 0.1\,mm for electrons and positrons compared to the standard EM list that calculates the ionising energy loss every 1\,mm. The EMZ physics list additionally describes the gamma conversion with higher accuracy~\cite{Geant4_emLists}.

The $N_{hits}/layer$ distributions for 10\,GeV muons are shown in Fig.~\ref{fig:tuning_FTFP_BERT} for the simulations with the standard, \_EMY and \_EMZ EM physics lists. The tuning parameter for all investigated EM physics lists are summarised in Table~\ref{tab:tuningParameters}. In addition to the digitisation parameters, the sum of $\chi^2/ndf$ values from the comparison to the data are given in the table. The best agreement with the data is found for the \_EMZ physics list. However, the $\chi^2/ndf$ values are still quite large, which can be explained by the necessary simplifications of the signal modelling and possible \textsc{Geant4} inaccuracies. 

The $N_{hits}/layer$ distributions for 10 and 20\,GeV positrons using the standard, \_EMY, and \_EMZ physics lists are shown in Fig.~\ref{fig:tuning_FTFP_BERT_EMY} and~\ref{fig:tuning_FTFP_BERT_EMZ}. The \_EMZ physics list reveals a better description of the data especially for high $N_{hits}/layer$ compared to the \_EMY option and the standard EM physics list.
\begin{table}
\caption{The digitisation parameters for the three EM physics lists of \textsc{Geant4} determined from the tuning process. The total $\chi^2/ndf$ describes the difference between the data and simulation in the distributions shown in Figs.~\ref{fig:tuning_FTFP_BERT}, ~\ref{fig:tuning_FTFP_BERT_EMY} and~\ref{fig:tuning_FTFP_BERT_EMZ}.}
\begin{center}
\begin{footnotesize}
\begin{tabular}{p{2cm}|c|c|c}
	 & standard & \_EMY & \_EMZ \\
	\hline
	\hline
	$\sigma_1$\,[mm] & 0.7 & 0.7 & 0.7 \\
	$\sigma_2$\,[mm] & 5.0 & 4.0 & 5.0 \\
	$R$ & 0.08 & 0.05 & 0.08 \\
	$T$\,[pC] & 0.07 & 0.08 & 0.07 \\
	$q_{0}$ & 1.0 & 1.2 & 1.0 \\
	$d_{cut}$\,[mm] & 0.05 & 0.01 & 0.05 \\
	\hline
	\hline
	$\left(\chi^{2}/ndf\right)_{tot}$ & 23.45 & 23.35 & 16.89 \\
\end{tabular}
\end{footnotesize}
\end{center}
\label{tab:tuningParameters}
\end{table}

\section{Systematic uncertainties}
\label{sec:Unc}
The systematic uncertainty on the data is dominated by the response equalisation procedure which is mostly affected by the limited statistics in the determination of the RPC efficiency $\epsilon_{i,j,}$ and hit multiplicity $\mu_{i,j}$. By propagating the statistical uncertainties on $\sigma_{\epsilon_{i,j,}}$ and $\sigma_{\mu_{i,j}}$ onto the equalisation coefficients $c_{i,j}$, the measurement is affected by at most $+2.6$ and $-2.4\,\%$. Additional systematic uncertainties originating from particle contamination, noise (0.1 hits per event~\cite{CAN-31}) and inefficiencies of the algorithm to find the first hard interaction have been found to be negligible~\cite{Min-DHCAL}.

The systematic uncertainty on the simulation originates from the tuning process of the digitisation parameters. The data samples of muons and positrons are used in the tuning process, thus preventing an assessment of systematic errors for the results based on these samples. 

For pions, the uncertainty on the shower observable $x$, $\sigma_{x}$, is estimated by the remaining deviations between the data and the simulation of positron showers $\Delta x$, following 
\begin{equation} 
	\sigma_{x}=\frac{1}{N_{E}N_{bins}}\cdot{\sum_{i=1}^{N_{E}}{\sum_{j=1}^{N_{bins}}\Delta x_{i,j}}}
\end{equation}
with $N_{E}$ the number of beam energies, and $N_{bins}$ the number of bins included having sufficient statistics.
This is a conservative approach and results in relatively large systematic uncertainties on the pion simulations. The values obtained through this procedure are summarised in Table~\ref{tab:e+Errors}. The shower observables will be described later in the text.

\begin{table}
\begin{center}
\begin{footnotesize}
\caption{Average uncertainty for simulated pion showers in percent on the 2D ($\sigma_{density}$) and 3D hit densities ($\sigma_{3D density}$), the longitudinal ($\sigma_{longProfile}$) and radial profile ($\sigma_{radProfile}$), as well as on the shower maximum ($\sigma_{t_{max}}$), the mean number of hits ($\sigma_{\left<N_{hits}\right>}$) and the resolution ($\sigma_{\sigma_{rec}/\left<E_{rec}\right>}$) for the standard, \_EMY and \_EMZ EM physics lists.}
\label{tab:e+Errors}
\begin{tabular}{p{2cm}|c|c|c}
 & standard & \_EMY & \_EMZ \\
 \hline
  \hline
$\sigma_{density}$ 					& 10.1 & 11.9 & 13.0 \\
$\sigma_{3D density}$ 				& 24.7 & 15.3 & 26.4 \\
$\sigma_{longProfile}$ 				& 29.9 & 22.8 & 20.9 \\
$\sigma_{radProfile}$ 				& 9 & 13 & 8 \\
$\sigma_{\left<N_{hits}\right>}$ 		& 6.6 & 6.3 & 0.2 \\
$\sigma_{\sigma_{rec}/\left<E_{rec}\right>}$ & 4.0 & 4.4 & 4.0 \\
  \hline
  \hline
\end{tabular}
\end{footnotesize}
\end{center}
\end{table}

\section{Positron shower analysis}
\label{sec:e}
In the following, the positron showers are studied for energies in the range of 2 to 25\,GeV and the data are compared to the simulation with different EM physics lists. 

\subsection{Response and energy reconstruction}\label{sec:ErecPositrons}
\begin{figure}
\begin{center}
\begin{minipage}{.65\textwidth}
  		\subfloat [$N_{hits}$] {\label{fig:DHCALeNhits} \includegraphics [width=1 \textwidth] {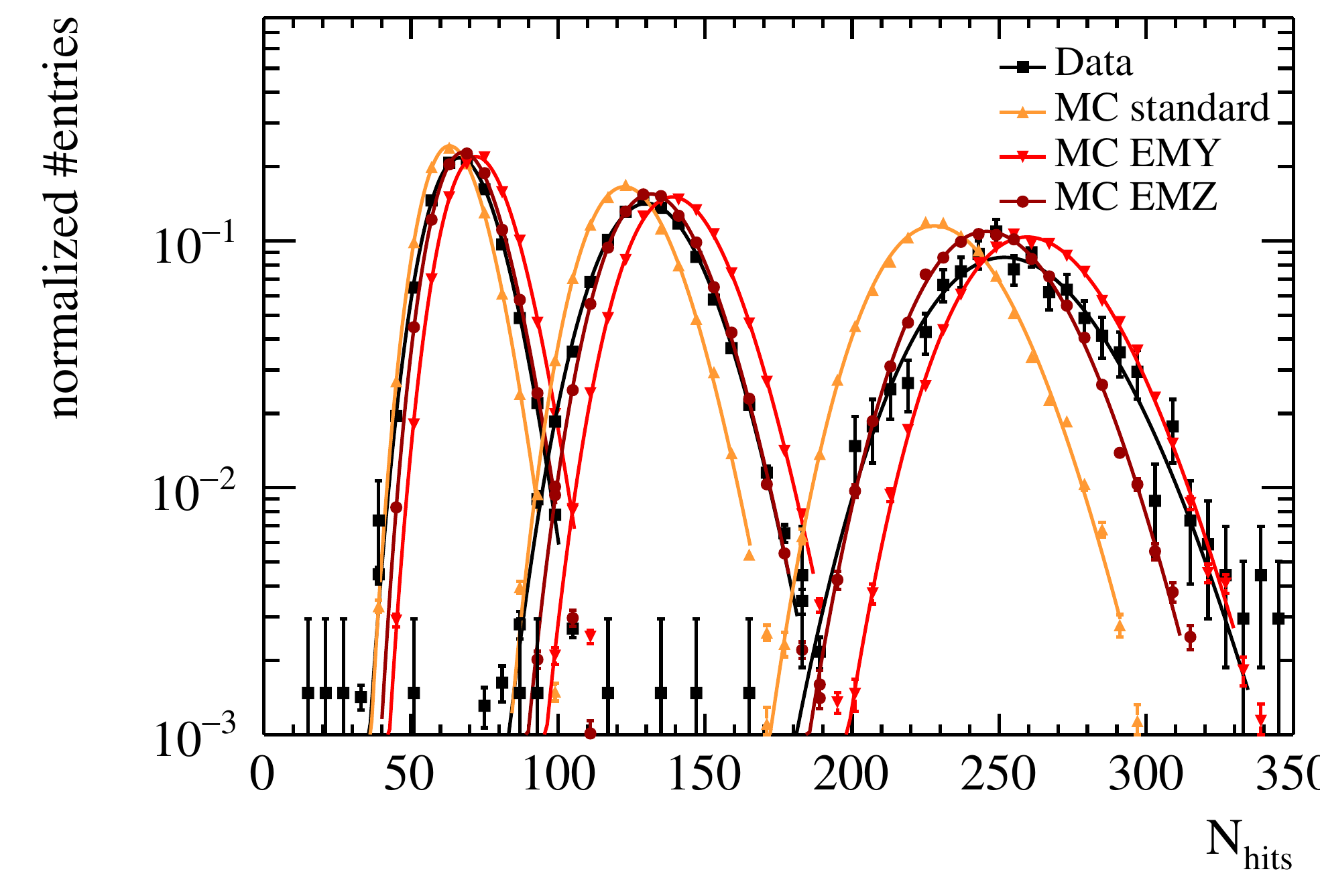}}
	\hfill
  		\subfloat [$\left<N_{hits}\right>$] {\label{fig:DHCALeResponse} \includegraphics [width=1 \textwidth] {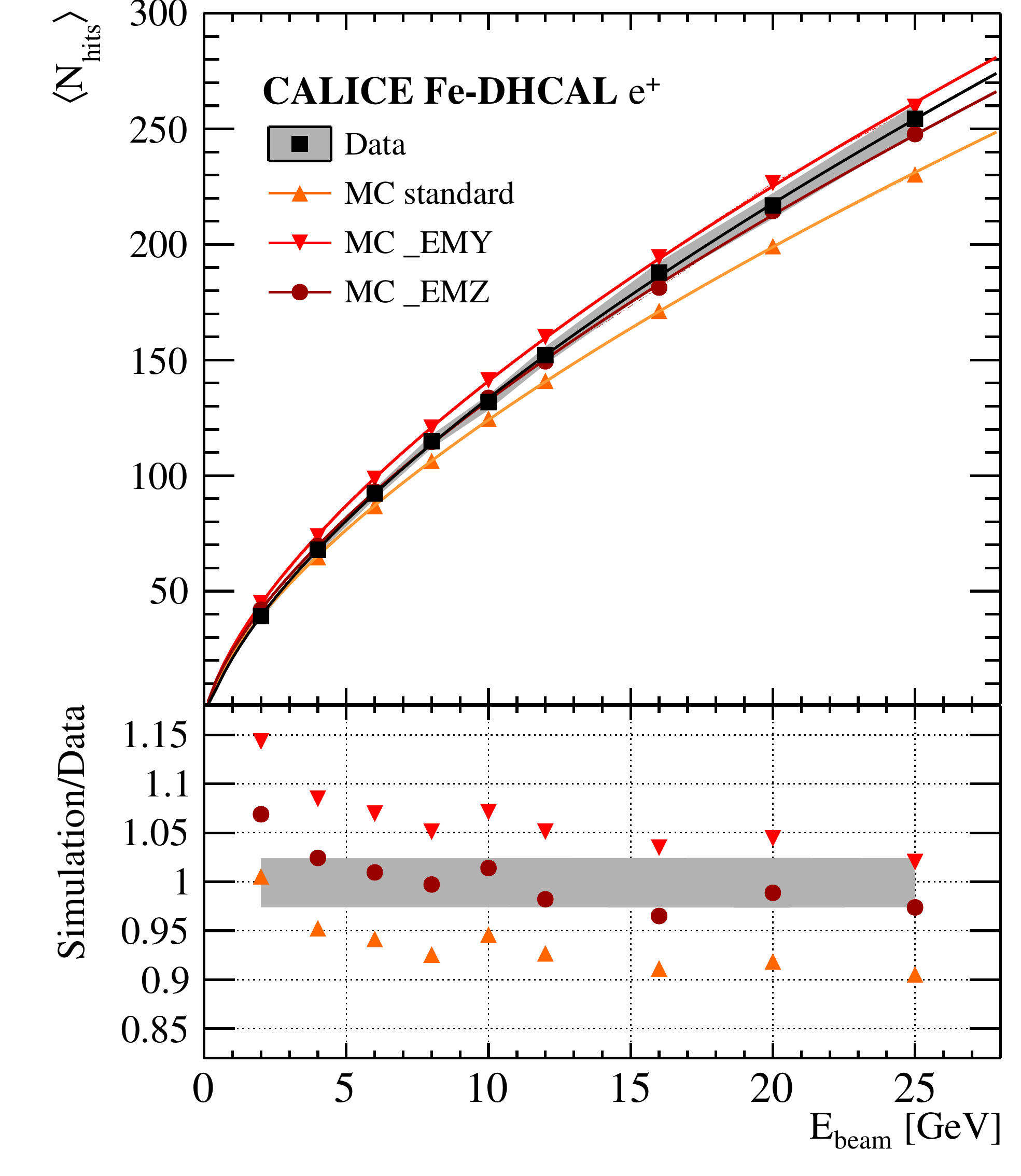}}
	\end{minipage}
\caption{a) The response distributions in number of hits for 2 to 25\,GeV positrons. The lines represent the Novosibirsk fits used for the determination of the mean response. \\ b) The mean response $\left<N_{hits}\right>$ before the correction for non-linearity to positron showers. The curves show the power law fit function. The plot on the bottom shows the ratio between data and simulation.
The grey bands indicate the statistical and systematic uncertainty of the data. The statistical errors of the simulations are smaller than the size of the markers. }
\label{fig:DHCALe}
\end{center}
\end{figure}

The positron response is measured in terms of the mean number of hits per event $\left<N_{hits}\right>$. To extract the mean number of hits for every $N_{hits}$ distribution as shown in Fig.~\ref{fig:DHCALeNhits}, the distribution is fitted to a Novosibirsk function~\cite{NovoFit} within a range of $\pm3\,\sigma$ around the peak position determined from a previous fit with a Gaussian function. The Novosibirsk function is used to describe the tails originating from e.g. leakage or saturation effects and to reduce the impact of outliers. A histogram is filled based on the results of the fit and the mean and RMS of that histogram are used as an estimate of the mean response and its standard deviation. A detailed description of the procedure can be found in~\cite{CAN-49a,PhDthesis}. 

Figure~\ref{fig:DHCALeResponse} shows the mean number of hits as a function of the beam energy. The comparison of the data with the three EM physics lists of \textsc{Geant4} reveals the best agreement for the \_EMZ simulation. The standard EM physics list and the \_EMY simulation show deviations of up to 15\,\%, where the simulation with the standard EM list shows systematically too few hits and the simulation with \_EMY too many. These deviations of the total number of hits are consistent with the observations made in the detailed shower analysis, see Section~\ref{sec:positronShowers}. 

Due to the high density and small lateral width of EM showers and the comparatively large pad size of the DHCAL readout, the data as well as the simulation show a saturation in the mean number of hits $\left<N_{hits}\right>$ versus beam energy. The mean response versus beam energy is fitted by a power law function
\begin{equation}
\left<N_{hits}\right>=a\cdot E_{beam}^{b}-c,
\end{equation}
where $a$ relates to the number of hits that correspond to a deposited energy of 1\,GeV, $b$ correlates with the saturation, and $c$ is related to the noise and the energy losses in front of the DHCAL. However, due to strong correlations between all three fit parameters, they are not an exact measure of these effects. 
 
\begin{figure}
\begin{center}
\begin{minipage}{.65\textwidth}
  		\subfloat [$E_{rec}$] {\label{fig:DHCALeErec} \includegraphics [width=1 \textwidth] {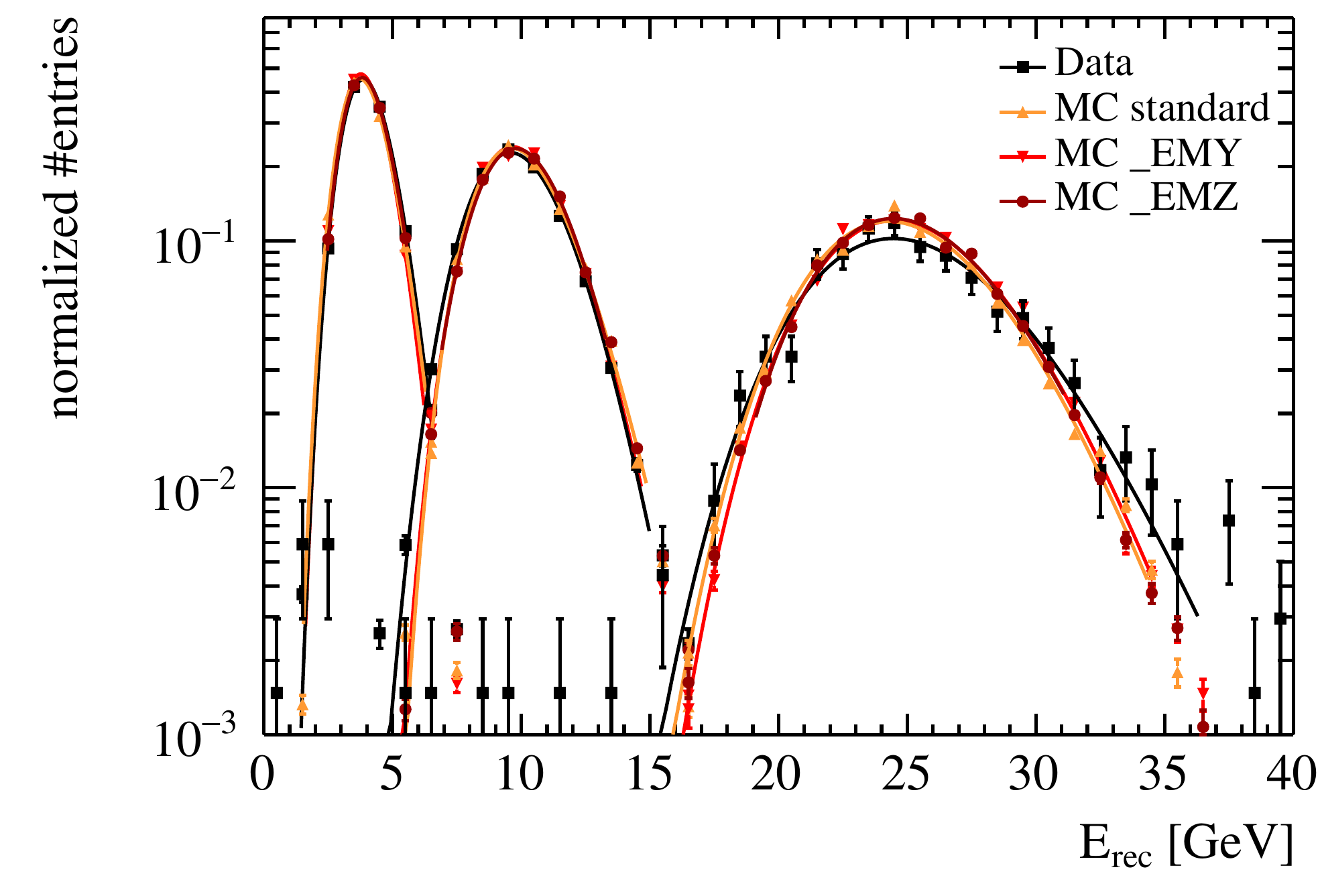}}
	\hfill
  		\subfloat [$\left<E_{rec}\right>$] {\label{fig:DHCALeLin} \includegraphics [width=1 \textwidth] {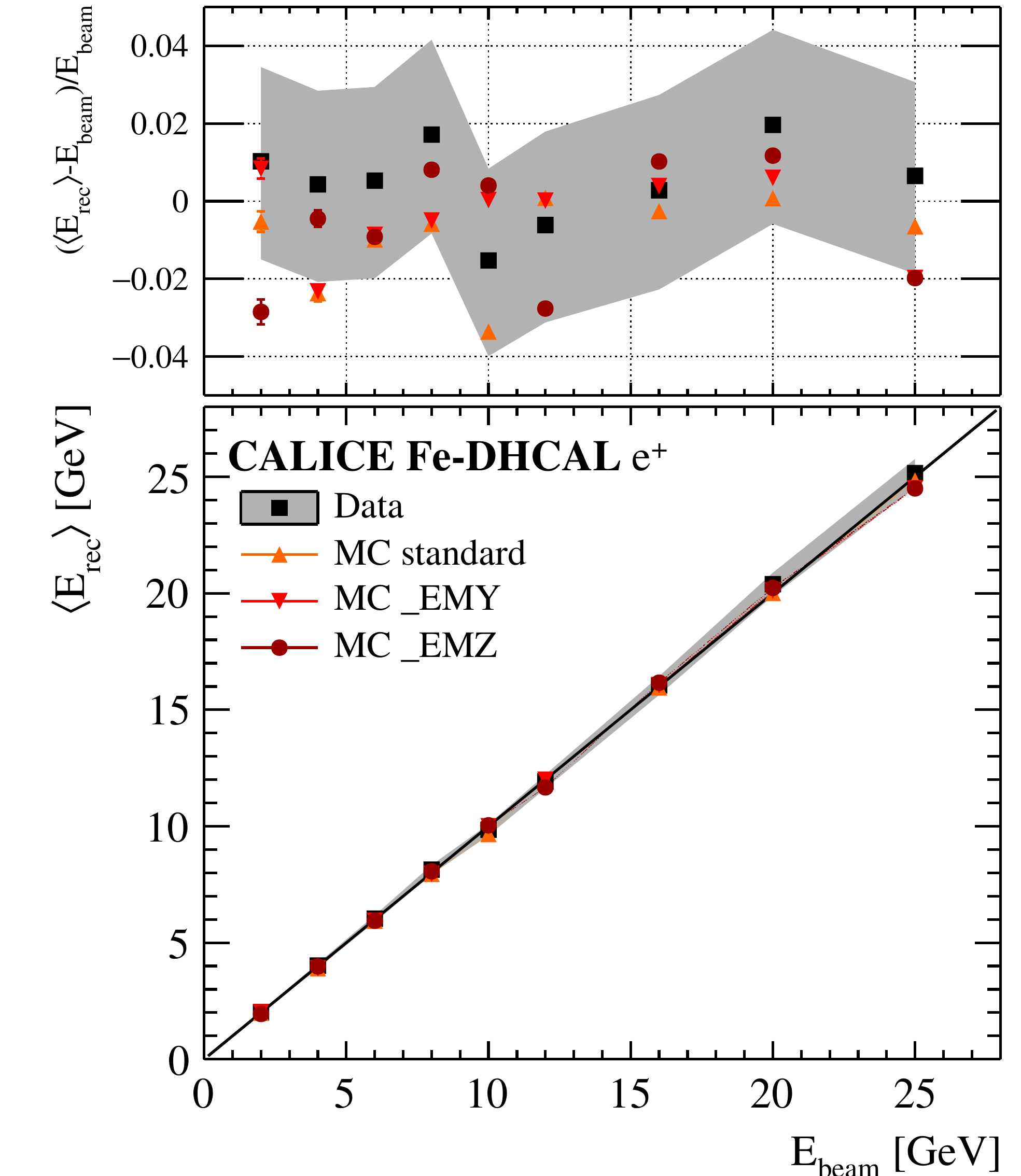}}
	\end{minipage}
\caption{a) The reconstructed energy distributions for 2 to 25\,GeV positrons. The lines represent the Novosibirsk fits used for the determination of the linearity. \\ b) The linearity after the correction for non-linearity to positron showers. The plot on the top shows the residuals to the beam energy.
The grey bands indicate the statistical and systematic uncertainty of the data. The statistical errors of the simulations are smaller than the size of the markers. }
\label{fig:DHCALe}
\end{center}
\end{figure}

For every event, the energy is reconstructed by inverting the power law function, replacing $E_{beam}$ with $E_{rec}$
\begin{equation}
	E_{rec}= \sqrt[b]{\frac{N_{\mathrm{hits}}+c}{a}}.
\end{equation}
The obtained parameters are listed in Table~\ref{tab:digiRecoParametersE} and the resulting energy distributions are shown in Fig.~\ref{fig:DHCALeErec}. A satisfactory linearity is achieved for all samples. The remaining non-linearities of the mean reconstructed energies are smaller than $\pm3.5$\,\%, see Fig.~\ref{fig:DHCALeLin}.


\begin{table}
\caption{The reconstruction parameters for $e^{+}$ events, extracted from the power law fit to the mean response in Fig.~\ref{fig:DHCALeResponse}. A value of $b=1$ would correspond to a perfectly linear response.}
\begin{center}
\begin{footnotesize}
\begin{tabular}{p{2cm}|c|c|c}
	 & $a$ [GeV$^{-b}$] & $b$ & $c$ [\#] \\
	\hline
	\hline
	Data 		& $30.6\pm4.1$  & $0.67\pm0.04$ & $9.5\pm5.7$ \\
	\hline
	MC standard 	& $33.4\pm0.2$  & $0.648\pm0.002$ & $7.8\pm0.3$ \\
	MC \_EMY 	& $28.9\pm0.2$ & $0.654\pm0.002$ & $6.2\pm0.3$ \\
	MC \_EMZ 	& $30.5\pm0.2$ & $0.658\pm0.002$ & $6.1\pm0.3$  \\
	\hline
	\hline
\end{tabular}
\end{footnotesize}
\end{center}
\label{tab:digiRecoParametersE}
\end{table}%

\subsection{Energy resolution}\label{sec:DHCALeResolution}
\begin{figure}
\begin{center}
\includegraphics[width=.65\textwidth]{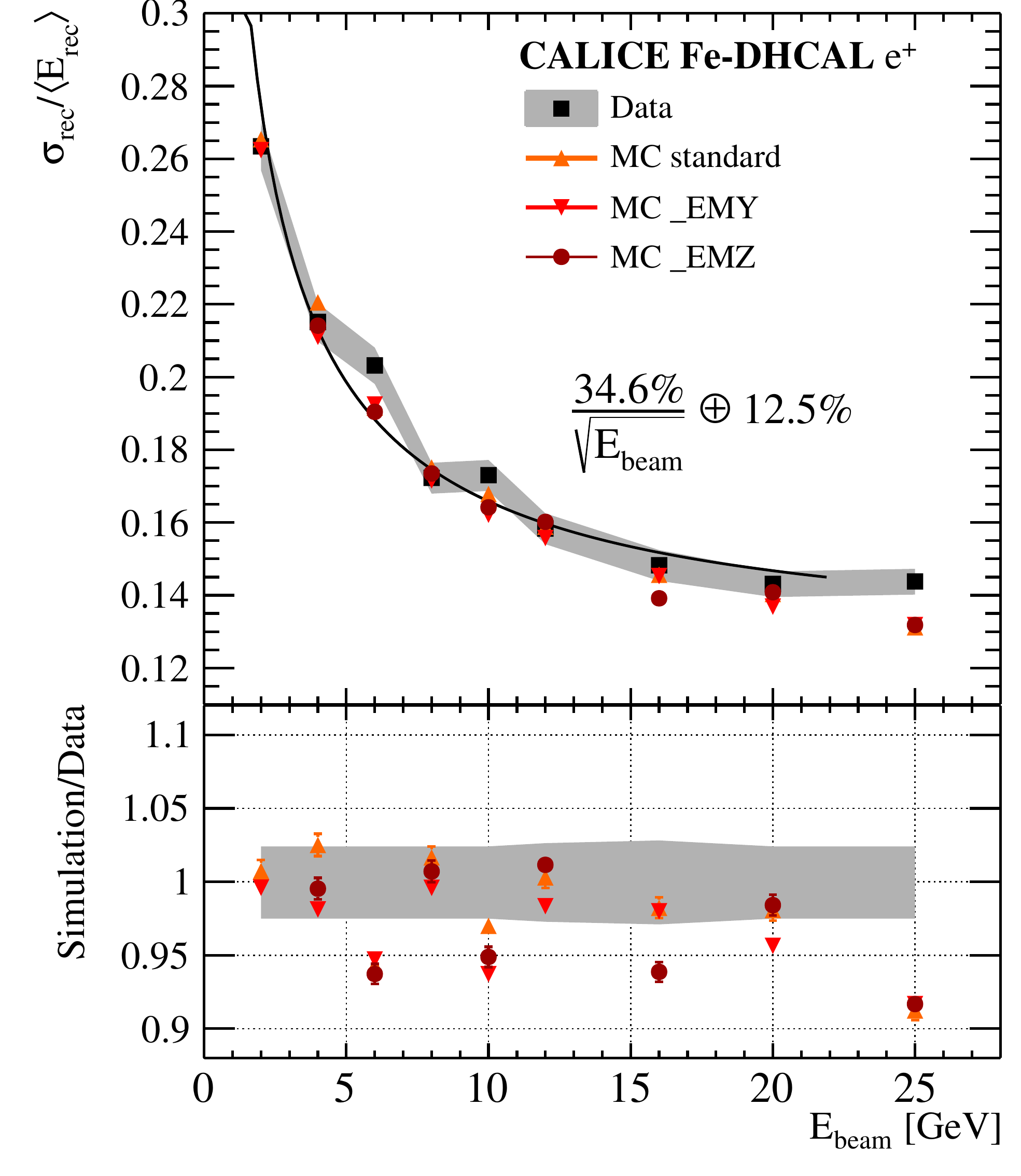}
\caption{The energy resolution for positrons with energies of 2 to 25\,GeV. The bottom plot shows the ratio of the simulations and data. The error bands show the systematic and statical uncertainty of the data added in quadrature. The statistical errors of the simulations are smaller than the size of the markers.}
\label{fig:Eresolution}
\end{center}
\end{figure}

The energy resolution for positron showers is obtained from the energy distributions, shown in Fig.~\ref{fig:DHCALeErec}, using the Novosibirsk fit function to reproduce a histogram from which the RMS is taken as $\sigma_{rec}$. The results are shown in Fig.~\ref{fig:Eresolution}, where the data points (black squares) are fitted to the convolution of a stochastic and a constant term
\begin{equation}\label{eq:EMresolution}
\frac{\sigma_{\mathrm{rec}}}{\langle E_{\mathrm{rec}}\rangle}=\frac{\alpha}{\sqrt{E_{\mathrm{beam}}[GeV]}}\oplus \beta.
\end{equation}
The ratio between the simulation and the data (bottom plot in Fig.~\ref{fig:Eresolution}) shows an agreement within $5\,\%$ for the energies of 2 to 20\,GeV. The simulated 25\,GeV positrons show a better resolution by around 10\,\%. 

This modest resolution of $\left(34.6\pm0.9\right)\,\%/\sqrt{E}$ and a constant term of $\left(12.5\pm0.3\right)\,\%$ is mostly due to the saturation caused by the dense EM showers and the digital readout of the $1\,\times\,1$\,cm$^2$ pads. 

However, by applying a weighting scheme based on the hit densities, following the method described in \cite{SCPaper,PhDthesis}, the saturation effect can be mitigated leading to an improvement of the energy resolution. This was achieved in the analysis of the data recorded with the DHCAL without absorbers~\cite{Min-DHCAL} but is beyond the scope of this paper.

\subsection{Positron shower shapes}\label{sec:positronShowers}
The longitudinal and lateral shower shapes as well as the hit densities of the EM showers are studied over the full energy range. In the following these observables are shown for 12\,GeV positrons. The differences seen for other energies are discussed in the text.

\begin{figure}
\begin{center}
	\begin{minipage}{.5\textwidth}
	\subfloat [2D hit density]  {\label{fig:digi12GeVeD}\includegraphics [width=1\textwidth]{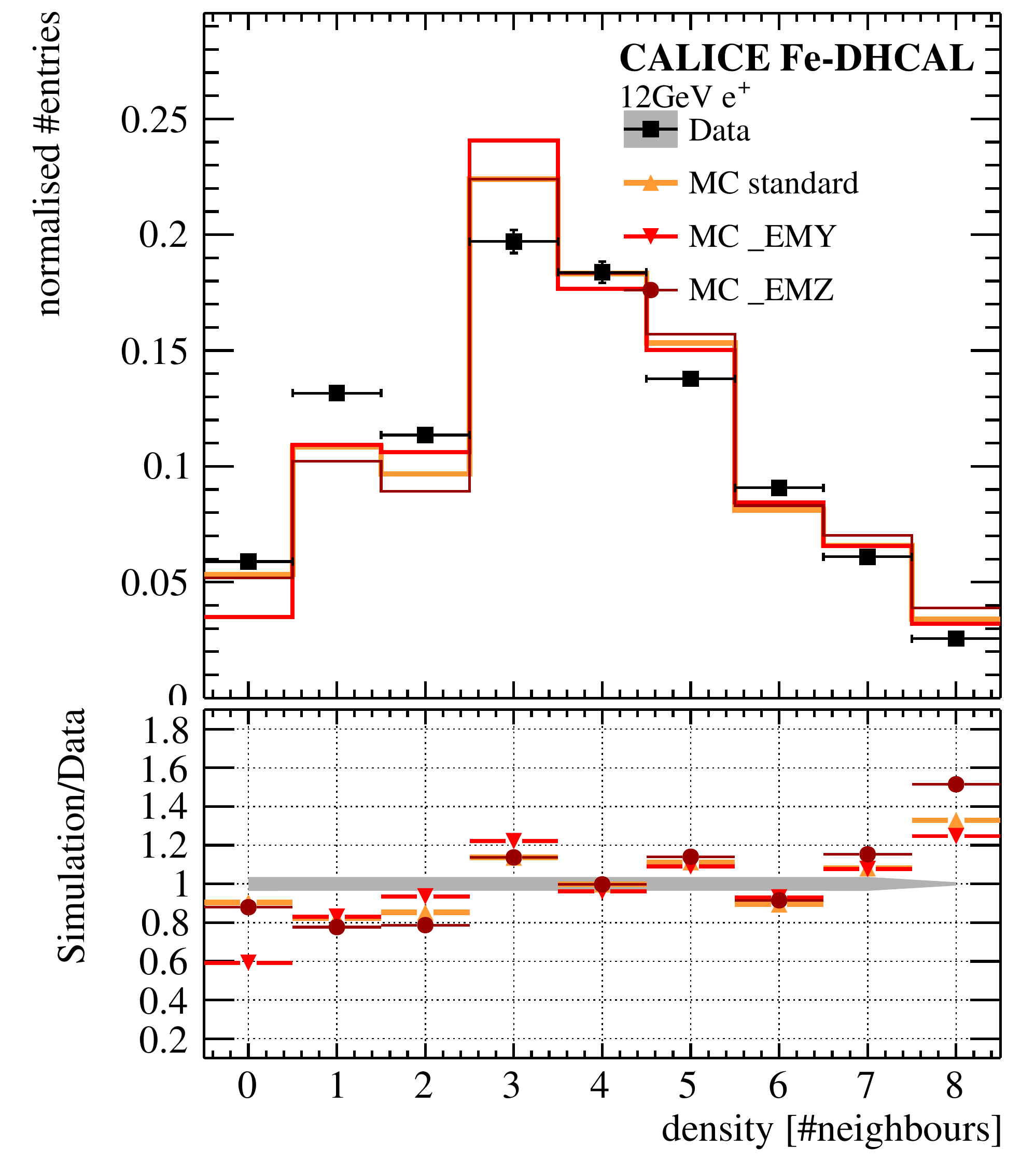}} 
	\\
	\subfloat [3D hit density]  {\label{fig:digi12GeVe3DD}\includegraphics [width=1\textwidth]{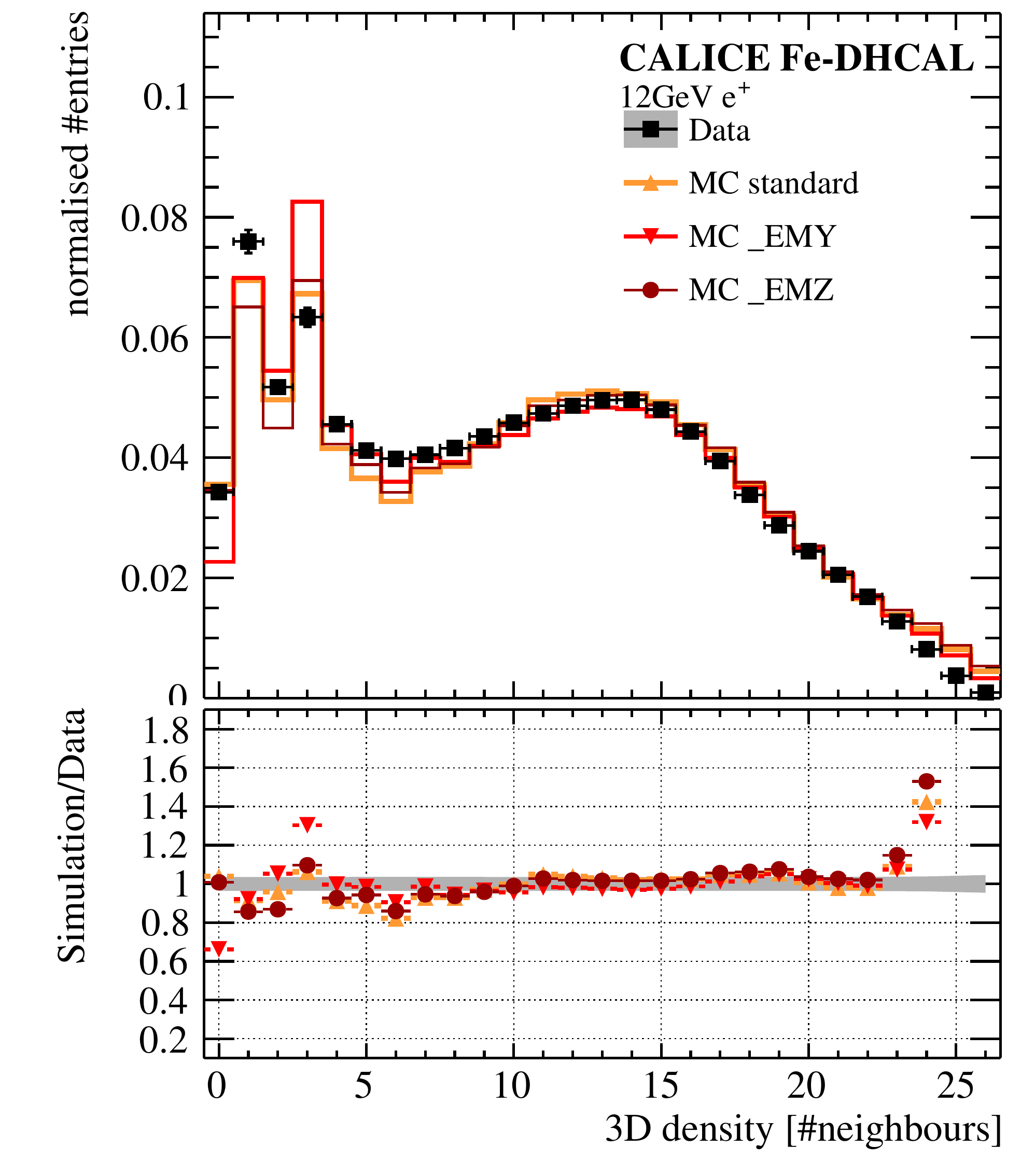}}
	\end{minipage}
\caption{The shower observables for 12\,GeV positrons; a) 2D hit density and b) 3D hit density. The data are represented as black squares and the grey error band corresponds to the systematic and statistical uncertainties added in quadrature. The bottom plots show the ratio between the simulations and data. 
}
\label{fig:digi12GeVeDensity}
\end{center}
\end{figure}

The 2D hit density is determined for each hit by counting the number of hits in the same layer and in an array of $3\times3$ pads surrounding a given hit, see Fig.~\ref{fig:digi12GeVeDensity}a. The 3D hit density additionally includes hits in the same x-y positions within $\pm1$ layer, see Fig.~\ref{fig:digi12GeVeDensity}b. The hit density distributions are in general well described by all three simulations. However, some differences remain between the simulations and in comparison with the data, notably at low densities.

\begin{figure}
\begin{center}
	\begin{minipage}{.5\textwidth}
	\subfloat [Longitudinal profile] {\label{fig:digi12GeVeLong} \includegraphics [width=1\textwidth]{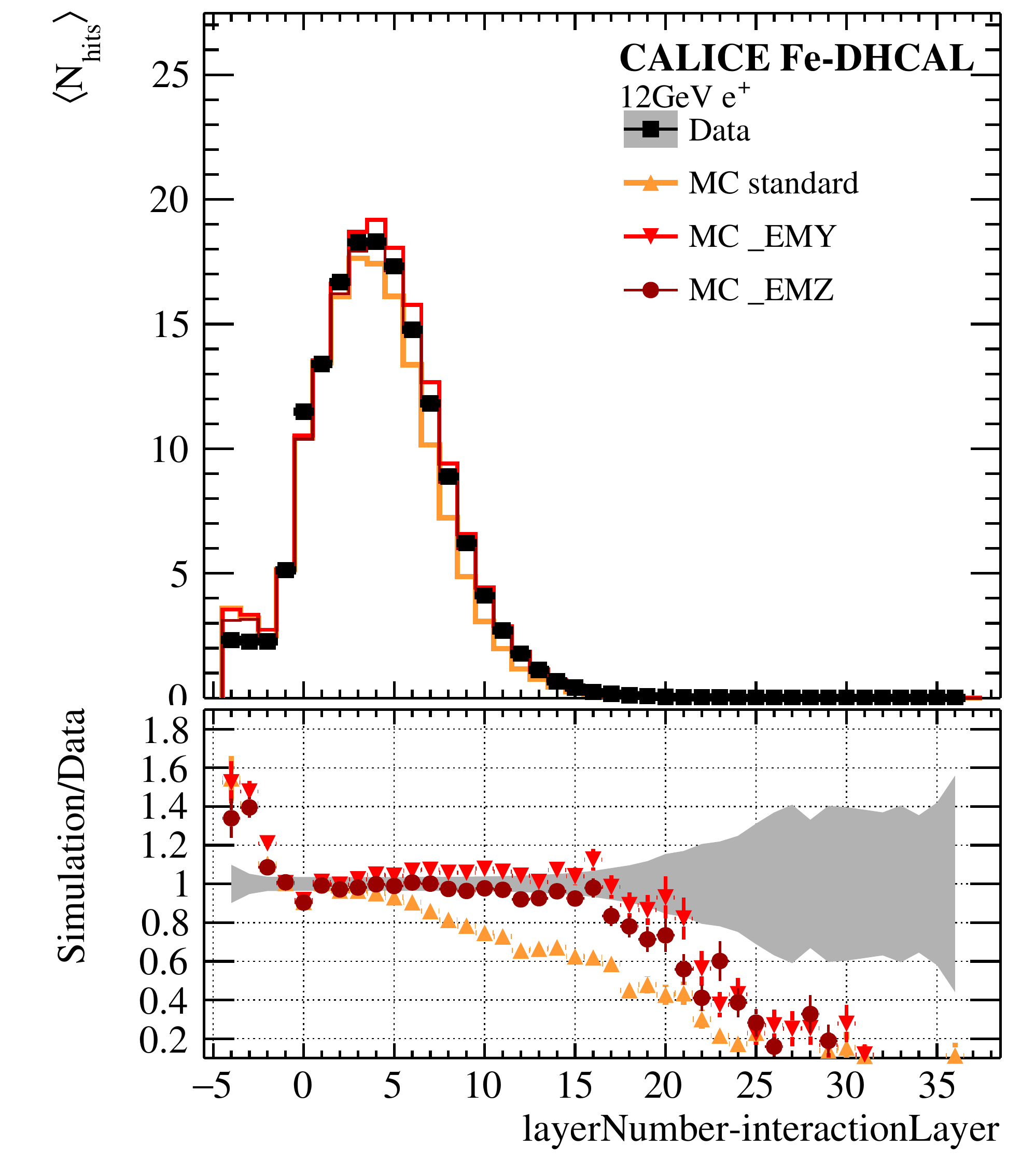}} 
	 \\
	\subfloat [Radial shower shape] {\label{fig:digi12GeVeRad} \includegraphics [width=1\textwidth]{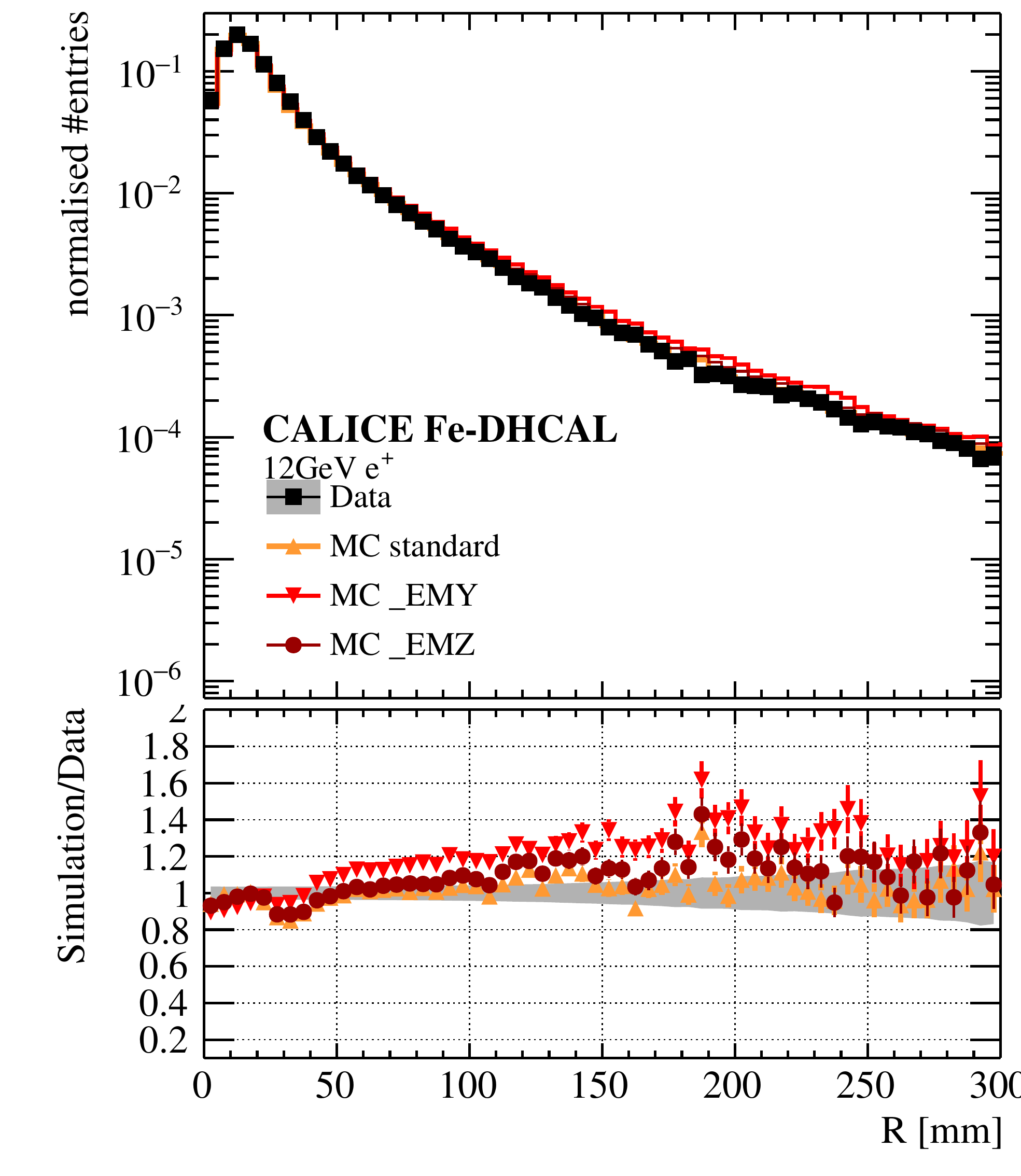}}
	\end{minipage}
\caption{The shower observables for 12\,GeV positrons; a) the longitudinal profile and b) the radial shower shape. The data are represented as black squares and the grey error band corresponds to the systematic and statistical uncertainties added in quadrature. The bottom plots show the ratio between the simulations and data. 
}
\label{fig:digi12GeVeProfiles}
\end{center}
\end{figure}

The longitudinal profile is defined as the average number of hits per layer with respect to the shower start and is shown in Fig.~\ref{fig:digi12GeVeLong}. The simulation with the standard EM physics list produces too few hits in the layers $>5$. Otherwise the profile is well reproduced. The disagreement between all simulations and the data in the tails for layers $>20$ can be explained by the noise level of 0.1\,hit per event in the data, which is not included in the simulation. 

The radial shower shape is defined as the distribution of the distance R of each hit $n$ from the shower axis
\begin{equation}
	R_{i,n}=\sqrt{ \left(x_{i,n}-cog_{x}\left(i\right)\right)^2 + \left(y_{i,n}-cog_{y}\left(i\right)\right)^2},
\end{equation} 
with an estimated shower axis obtained with a linear fit of the centre of gravity in x and y, $cog_{x,y}$, per layer $i$ to 
\begin{equation}
	cog_{x,y}\left(i\right)=a_{x,y}+b_{x,y}\cdot i.
\end{equation}
The radial shower shape is shown in Fig.~\ref{fig:digi12GeVeRad}. In general, the radial shower shapes show a good agreement between data and simulations, particularly at small radii. However, all simulations show a tendency to overestimate the number of hits in the outer parts of the shower. This behaviour is observed over the full energy range.

\subsection{Conclusion on the comparison of \textsc{Geant4} EM physics lists}
The digitiser of the RPC response, described in Sec.~\ref{sec:digi}, requires the tuning of several parameters in comparisons with positron and muon data to achieve a satisfying description of the hit multiplicities and EM shower profiles. Specialised EM models of \textsc{Geant4} were tested in order to reproduce the local hit distributions. After individual tuning, it is found that the simulation of the DHCAL requires the use of the EMZ physics list to obtain a good agreement with the testbeam data.

\section{Pion shower analysis}
\label{sec:pi}
The analysis of the positrons revealed a large variation of the simulation results using different \textsc{Geant4} EM physics lists. Hadron showers feature large fluctuations, which require sophisticated models to describe in detail. In the following, the $\pi^+$ showers are studied and compared to simulations, using the hadronic physics lists FTFP\_BERT and QGSP\_BERT, which have proven successful in other contexts~\cite{ValidationGeant4}. The two hadronic physics lists are tested for all three different EM physics list options, whereas the text will focus in the following on the results obtained with the EMZ model.

\subsection{Response and energy reconstruction}
\begin{figure}
\begin{center}
\begin{minipage}{.6\textwidth}
  		\subfloat [$N_{hits}$] {\label{fig:DHCALpiNhits} \includegraphics [width=1 \textwidth] {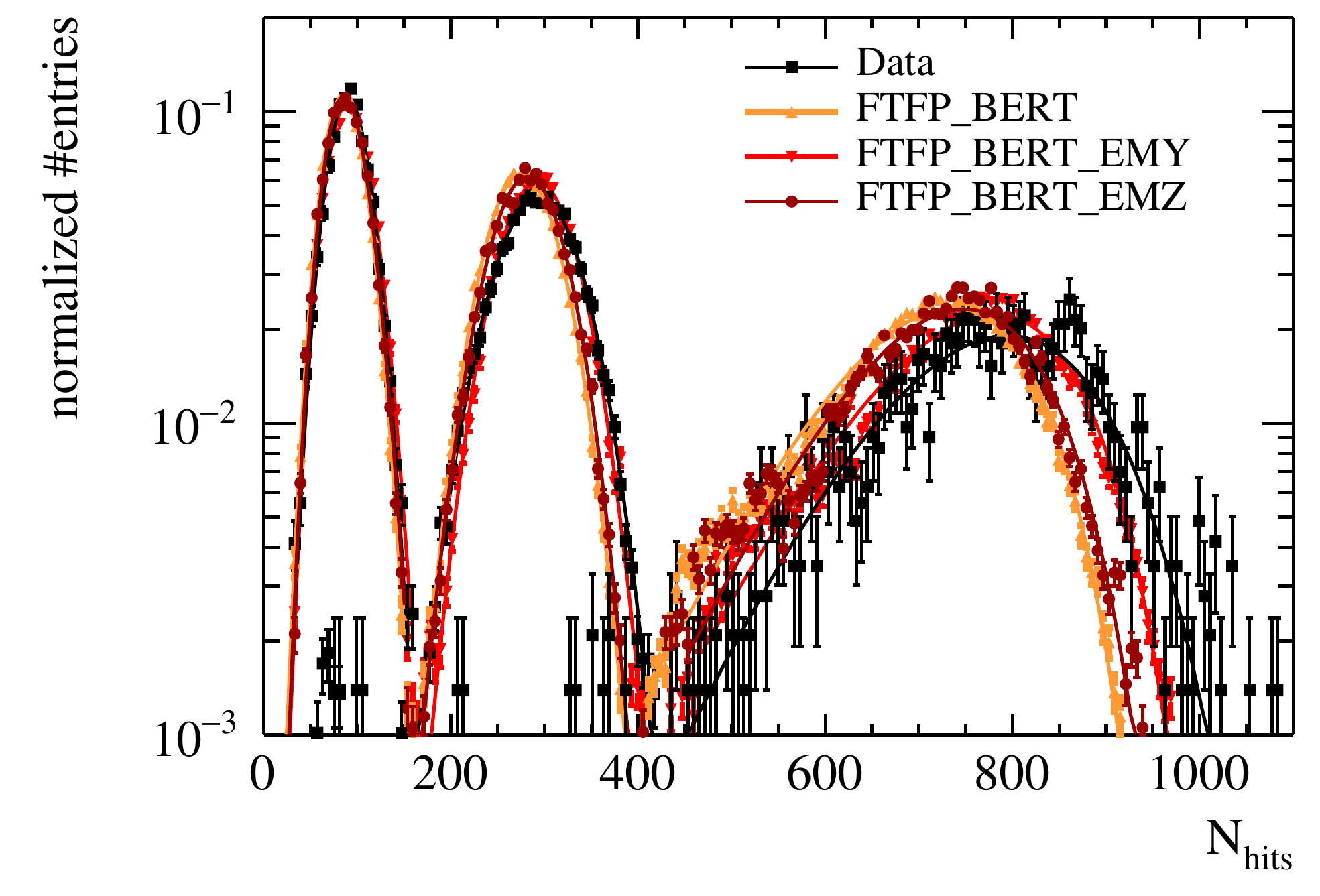}}
	\hfill
  		\subfloat [$\left<N_{hits}\right>$] {\label{fig:DHCALpiResponse} \includegraphics [width=1 \textwidth] {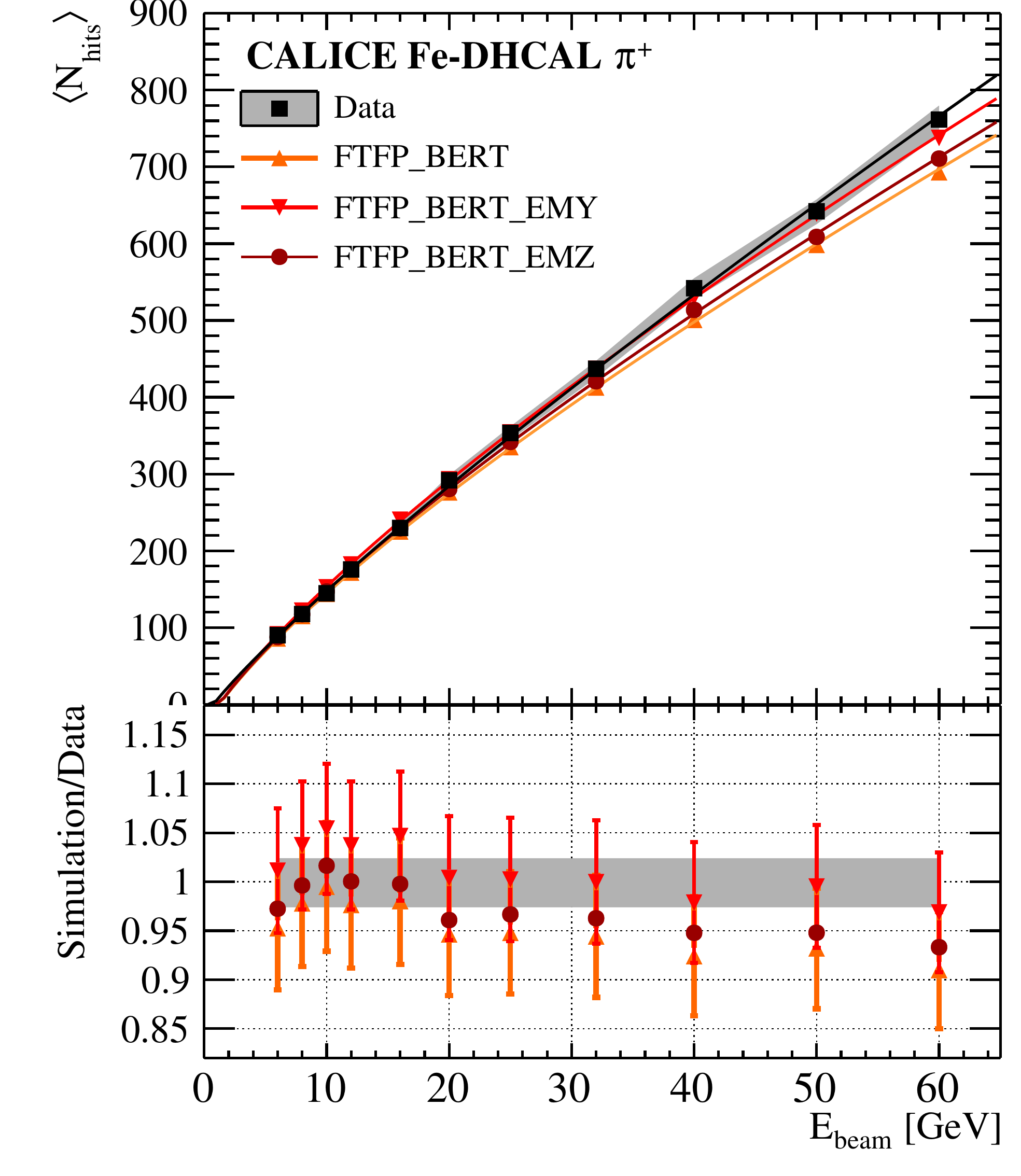}}
   	\end{minipage}
\caption{a) The response distributions in number of hits for 6, 20, and 60\,GeV $\pi^{+}$. The lines represent the Novosibirsk fits used for the determination of the mean response, shown in b) before the correction for the non-linearity. The curves show the power law fit function. The plot on the bottom shows the ratio between simulation and data.
The grey bands indicate the statistical and systematic uncertainties of the data. The statistical errors are smaller than the size of the markers. }
\label{fig:DHCALpi}
\end{center}
\end{figure}

\begin{figure}
\begin{center}
\begin{minipage}{.6\textwidth}
  		\subfloat [$N_{hits}$] {\label{fig:DHCALpiQNhits} \includegraphics [width=1 \textwidth] {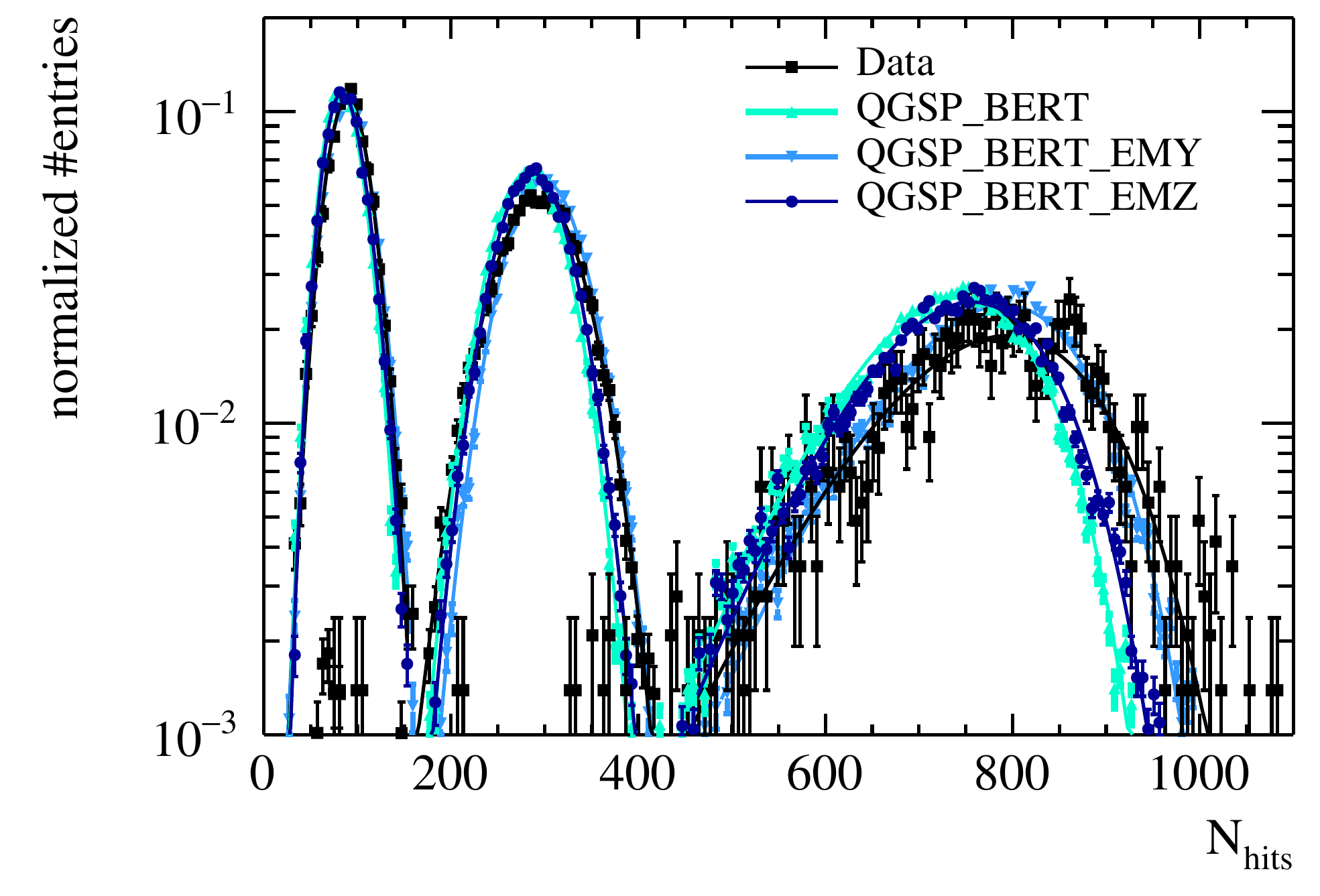}}
	\hfill
 		\subfloat [$\left<N_{hits}\right>$] {\label{fig:DHCALpiQResponse} \includegraphics [width=1 \textwidth] {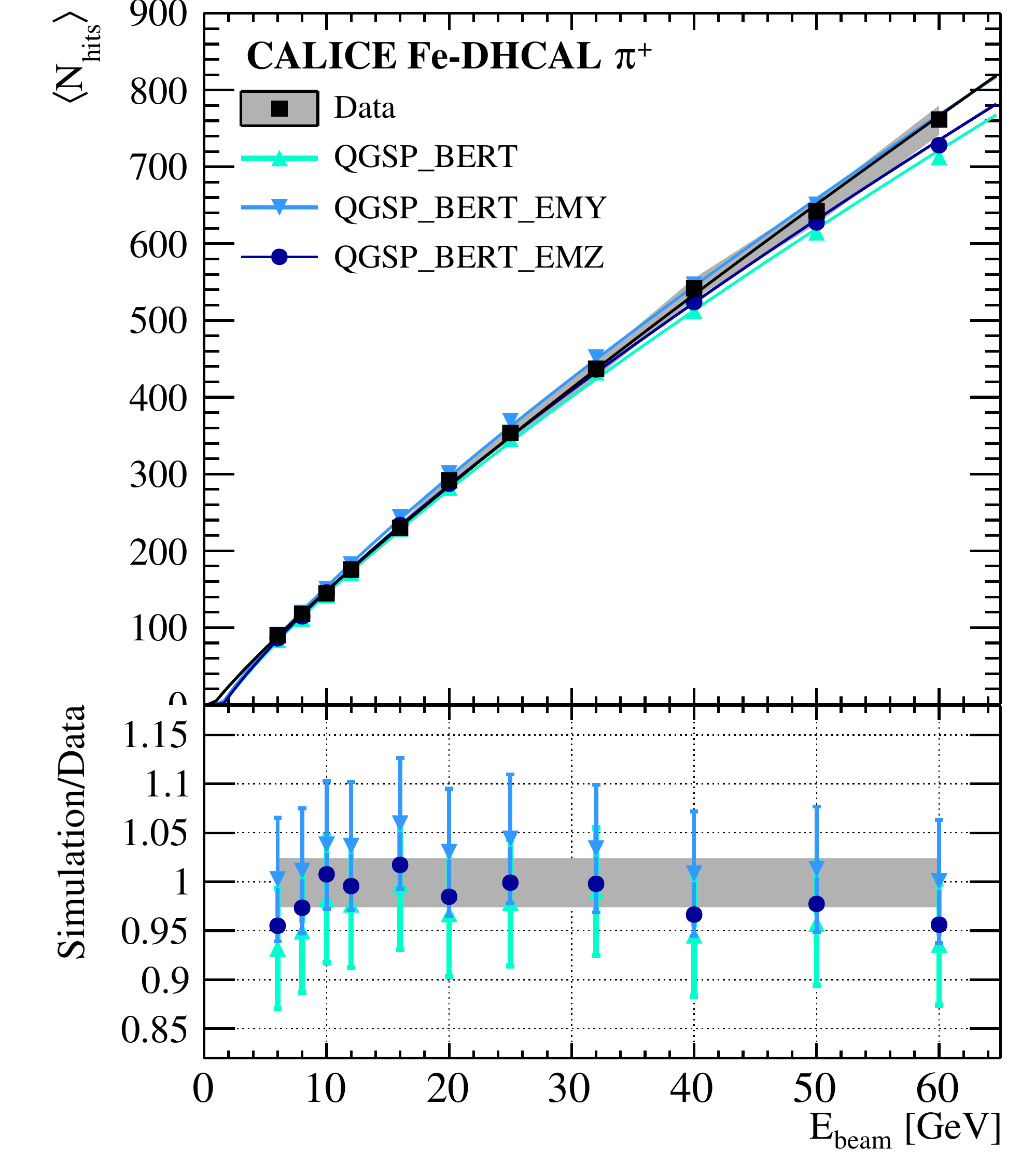}}
	\end{minipage}
\caption{a) The response distributions in number of hits for 6, 20, and 60\,GeV $\pi^{+}$. The lines represent the Novosibirsk fits used for the determination of the mean response, shown in b) before the correction for non-linearity. The curves show the power law fit function. The plot on the bottom shows the ratio between simulation and data. 
The grey bands indicate the statistical and systematic uncertainties of the data. The statistical errors are smaller than the size of the markers. }
\label{fig:DHCALpiQ}
\end{center}
\end{figure}

The distributions of the total number of hits for 6, 20 and 60\,GeV are shown for the FTFP\_BERT (QGSP\_BERT) simulations and the data in Fig.~\ref{fig:DHCALpiNhits} (\ref{fig:DHCALpiQNhits}). In both the simulations and the data, a tail towards smaller number of hits is seen for beam energies larger than 20\,GeV. This effect is most likely due to saturation effects. To include these tails in the estimation of the mean response, Novosibirsk fits are applied and are shown as curves in the figures. 

The mean number of hits $\left<N_{hits}\right>$ is shown as a function of beam energy in Fig.~\ref{fig:DHCALpiResponse} (\ref{fig:DHCALpiQResponse}) for the FTFP\_BERT (QGSP\_BERT) simulations compared to the data. Both hadronic physics lists exhibit a stronger saturation than seen in the data. The QGSP\_BERT\_EMZ simulation shows a good agreement with data for beam energies larger than 20\,GeV, as illustrated by the ratio of the simulation with the data in the bottom plot of Fig.~\ref{fig:DHCALpiQResponse}, while the FTFP\_BERT\_EMZ simulation describes the data within the errors for the lower energies, see Fig.~\ref{fig:DHCALpiResponse}. 

\begin{figure}
\begin{center}
\begin{minipage}{.6\textwidth}
  		\subfloat [$E_{rec}$] {\label{fig:DHCALpiErec} \includegraphics [width=1 \textwidth] {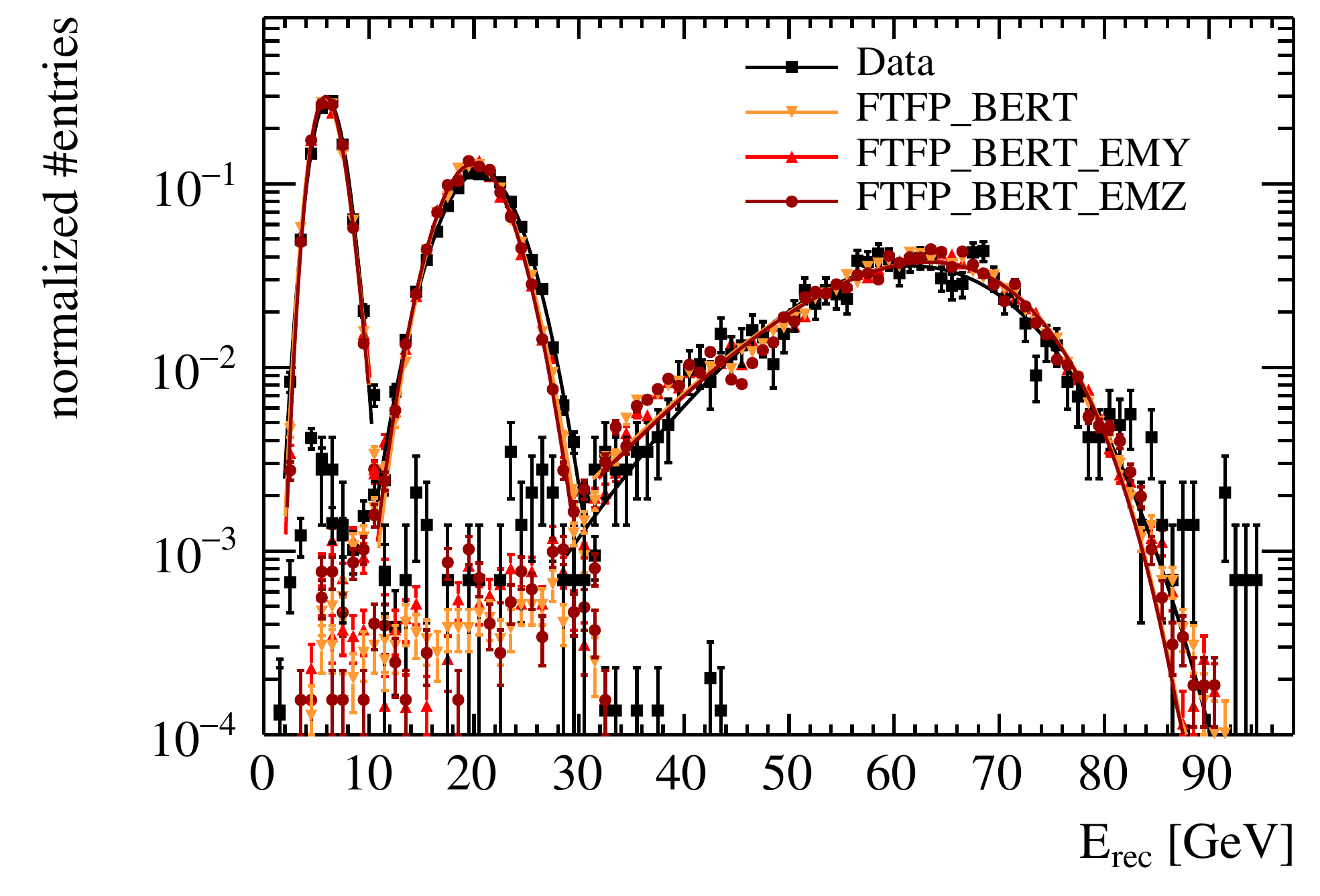}}
	\hfill
  		\subfloat [$\left<E_{rec}\right>$] {\label{fig:DHCALpiLin} \includegraphics [width=1 \textwidth] {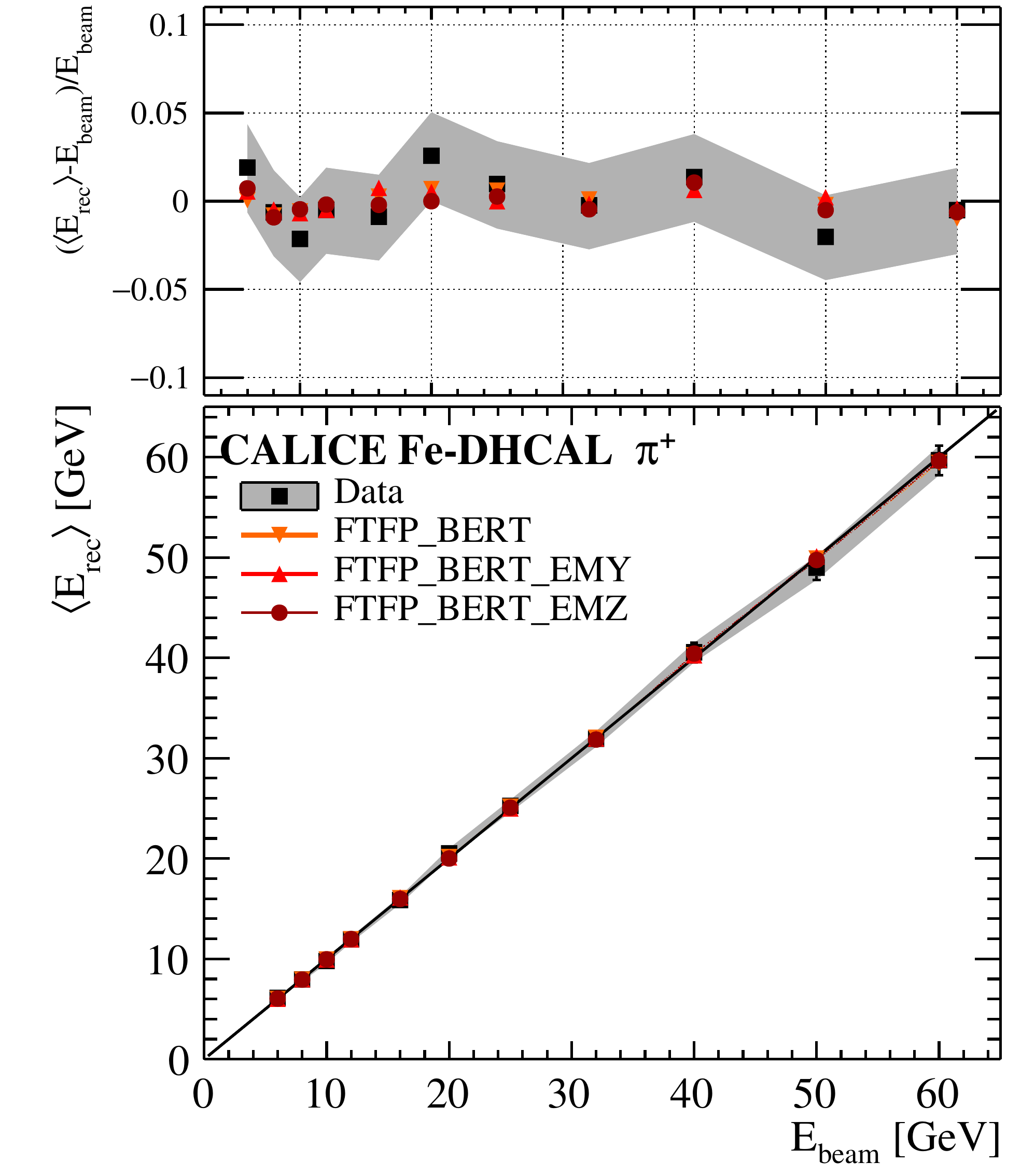}}
	\end{minipage}
\caption{a) The reconstructed energy distributions for 6, 20, and 60\,GeV $\pi^{+}$. The lines represent the Novosibirsk fits used for the determination of the linearity, shown in b) after the correction for the non-linearity. The plot on the top shows the residuals to the beam energy.
The grey bands indicate the statistical and systematic uncertainties of the data. The statistical errors are smaller than the size of the markers. }
\label{fig:DHCALpi}
\end{center}
\end{figure}

\begin{figure}
\begin{center}
\begin{minipage}{.6\textwidth}
  		\subfloat [$E_{rec}$] {\label{fig:DHCALpiQErec} \includegraphics [width=1 \textwidth] {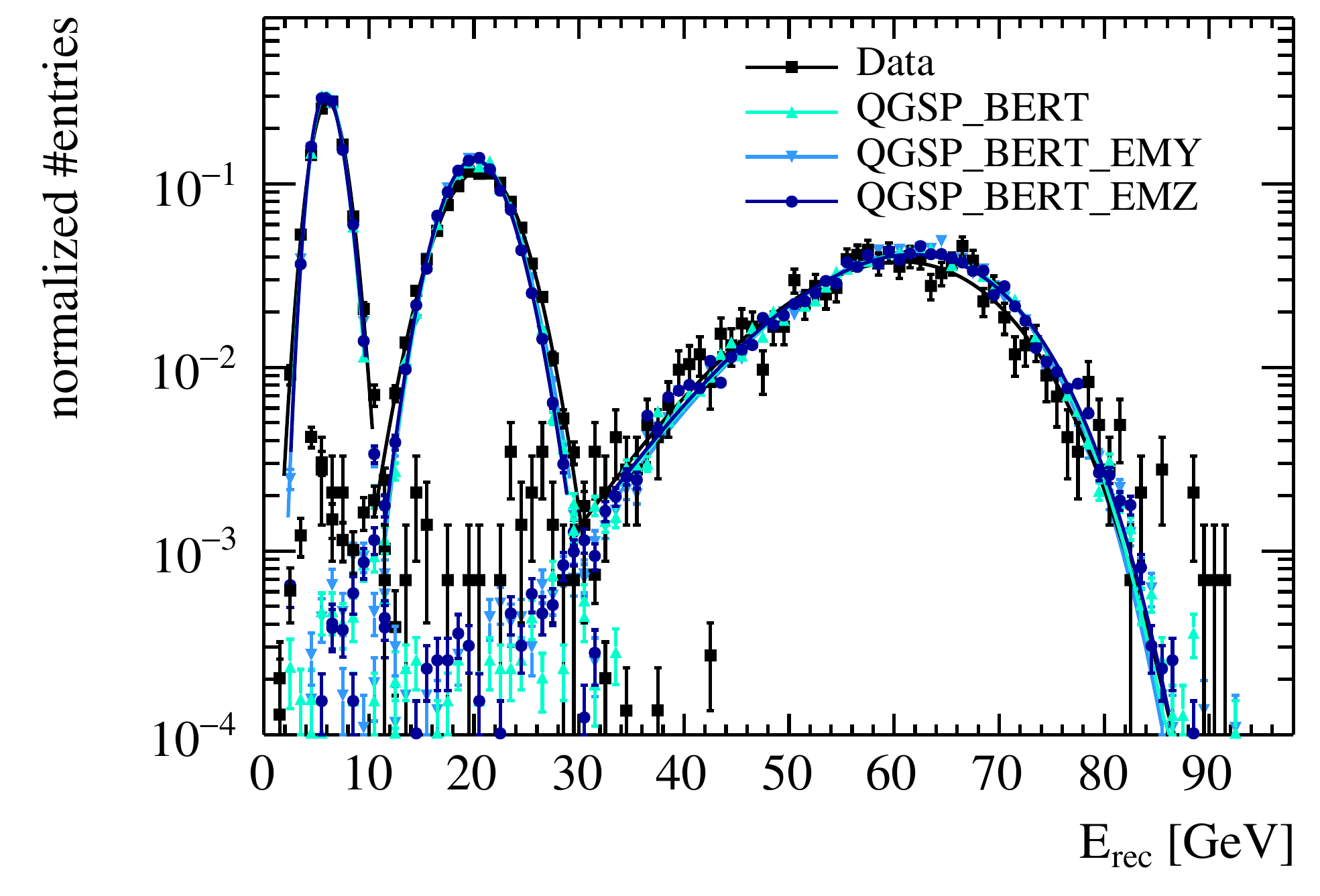}}
	\hfill
  		\subfloat [$\left<E_{rec}\right>$] {\label{fig:DHCALpiQLin} \includegraphics [width=1 \textwidth] {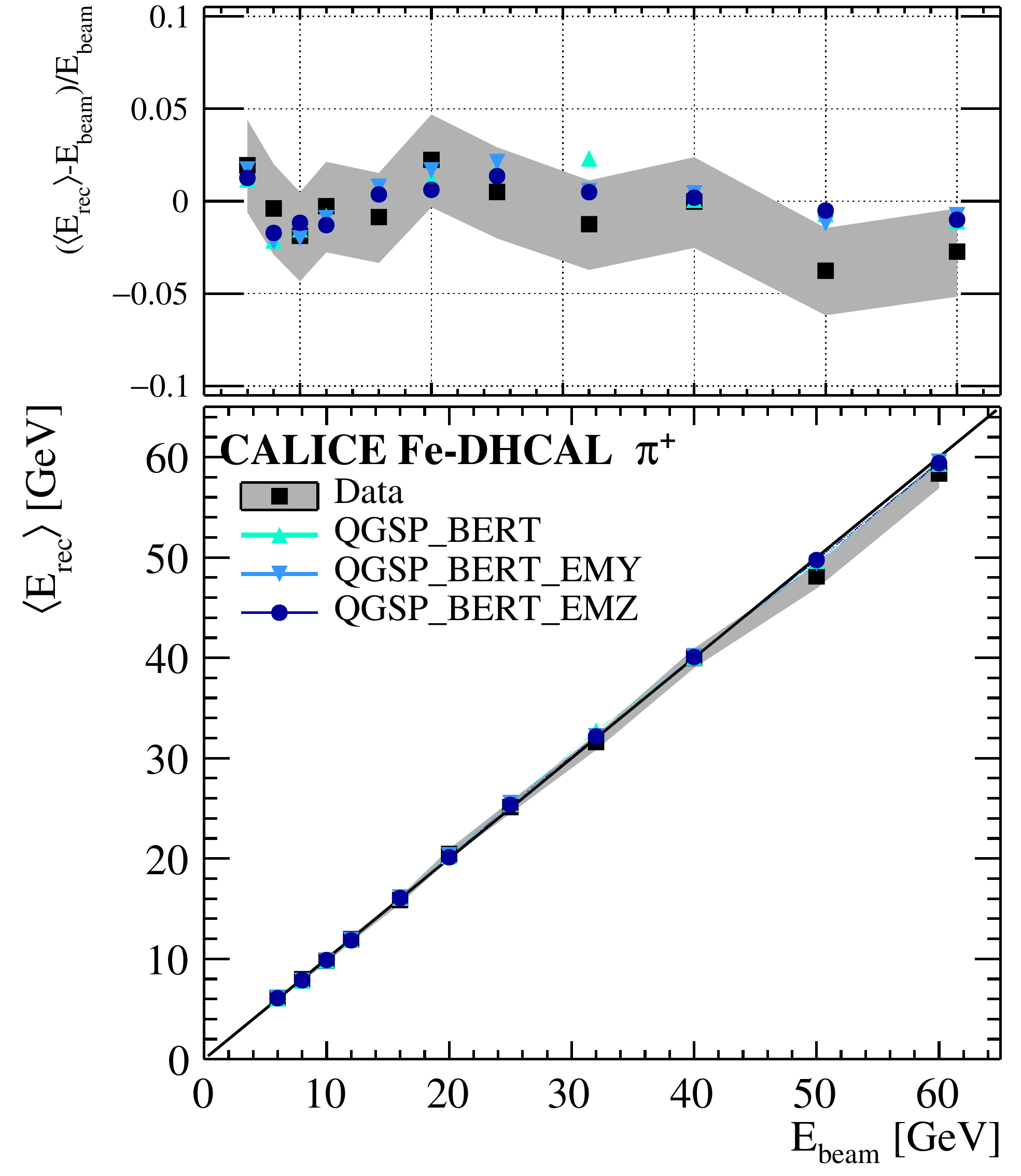}}
	\end{minipage}
\caption{a) The reconstructed energy distributions for 6, 20, and 60\,GeV $\pi^{+}$. The lines represent the Novosibirsk fits used for the determination of the linearity, shown in b) after the correction for non-linearity. The plot on the top shows the residuals to the beam energy.
The grey bands indicate the statistical and systematic uncertainties of the data. The statistical errors are smaller than the size of the markers. }
\label{fig:DHCALpiQ}
\end{center}
\end{figure}

To compare the energy resolution of the data and the simulations, a satisfactory linearity in the reconstructed energies is required. This is achieved as for the positrons by fitting a power law function $\left<N_{hits}\right>=a\cdot E_{beam}^{b}-c$ to the mean response, inverting the function and setting $E_{rec}=E_{beam}$. The inverted function, and the parameters of this fit are used to reconstruct the energy of each event. The reconstruction parameters are summarised in Table~\ref{tab:digiRecoParametersPi}, where the stronger saturation in the simulations is expressed by smaller $b$ parameters. The resulting reconstructed energy distributions are shown in Figs.~\ref{fig:DHCALpiErec} and~\ref{fig:DHCALpiQErec}.

The mean reconstructed energies as a function of the beam energy are shown in Figs.~\ref{fig:DHCALpiLin} and~\ref{fig:DHCALpiQLin}. The residuals to the beam energy show non-linearities smaller than 2\,\%. 


\begin{table}
\caption{The energy reconstruction parameters for $\pi^{+}$ events, extracted from the power law fit to the mean response in Fig.~\ref{fig:DHCALpiResponse} and~\ref{fig:DHCALpiQResponse}.}
\begin{center}
\begin{footnotesize}
\begin{tabular}{p{2.5cm}|c|c|c}
	 & $a$ [GeV$^{-b}$] & $b$ & $c$ [\#] \\
	\hline
	\hline
	Data 			& $21.1\pm3.0$ & $0.89\pm0.03$ & $15.0\pm9.8$ \\
	\hline
	FTFP\_BERT 		& $31.2\pm12.1$ &  $0.79\pm0.09$ & $36.9\pm31.8$ \\ 
	FTFP\_BERT\_EMY & $28.6\pm11.6$ & $0.79\pm0.09$ & $32.6\pm30.9$ \\
	FTFP\_BERT\_EMZ & $30.1\pm0.5$ & $0.785\pm0.004$ & $36.1\pm1.3$ \\
	\hline
	QGSP\_BERT  & 		$30.7\pm12.0$ & $0.79\pm0.09$ & $42.8\pm31.3$ \\
	QGSP\_BERT\_EMY & 	$30.9\pm11.5$ & $0.80\pm0.09$ & $40.6\pm30.9$ \\
	QGSP\_BERT\_EMY & 	$31.8\pm0.5$ & $0.782\pm0.004$ & $44.7\pm1.3$ \\
	\hline
	\hline
\end{tabular}
\end{footnotesize}
\end{center}
\label{tab:digiRecoParametersPi}
\end{table}%

\clearpage
\subsection{Energy resolution}\label{sec:DHCALpiResolution}
The energy resolution of the Fe-DHCAL for pions is shown in Fig.~\ref{DHCALpiResolution}. 
The resolution observed in data is showing a typical $1/\sqrt{E}$ behaviour for beam energies below 30\,GeV, which reaches a minimum of approximately 14\,\%. For beam energies above 30\,GeV the energy resolution degrades with increasing beam energy up to $\thicksim18$\,\% for 60\,GeV. Note that at the ILC, the relevant range of single particle energies lie mostly below 20\,GeV, where the energy resolution is dominated by the stochastic term~\cite{Thomson200925}.\\
The black curve in Fig.~\ref{DHCALpiResolution} shows the fit to the data up to the energy of 32\,GeV using Eq.~\ref{eq:EMresolution}.
The fit results in a stochastic term of $\left(51.5\pm1.5\right)\%\,\sqrt{E}$ and a constant term of $\left(10.6\pm0.5\right)\%$. 

The degradation of the resolution for $E_{beam}>30$\,GeV is due to the saturation in the response due to the digital readout combined with the cell size of $1\,\times\,1$\,cm$^2$. The efffect of leakage, longitudinal or lateral, is small, as shown in the longitudinal  and radial shower shapes (Figs.~\ref{fig:digiPiLongProfiles},~\ref{fig:digiPiQLongProfiles},~\ref{fig:digiPiRadProfiles},~\ref{fig:digiPiQRadProfiles}).

\begin{figure}
\label{DHCALpiResolution}
\begin{center}
	\begin{minipage}{.5\textwidth}
  		\subfloat [FTFP\_BERT] {\label{fig:resolutionFTFP} \includegraphics [width=1\textwidth] {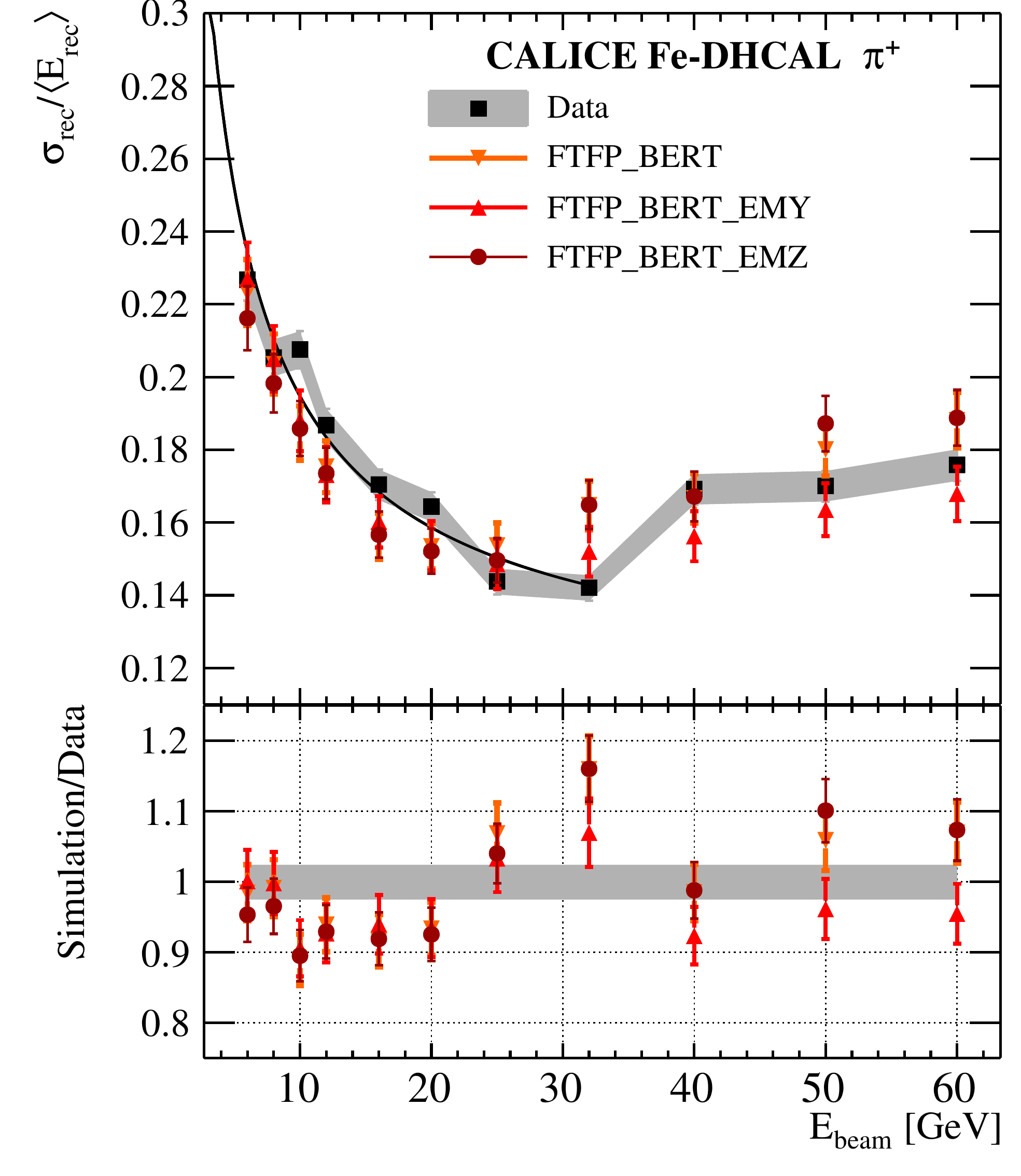}}
	\vspace*{0.5cm}
  		\subfloat [QGSP\_BERT] {\label{fig:resolutionQGSP} \includegraphics [width=1 \textwidth] {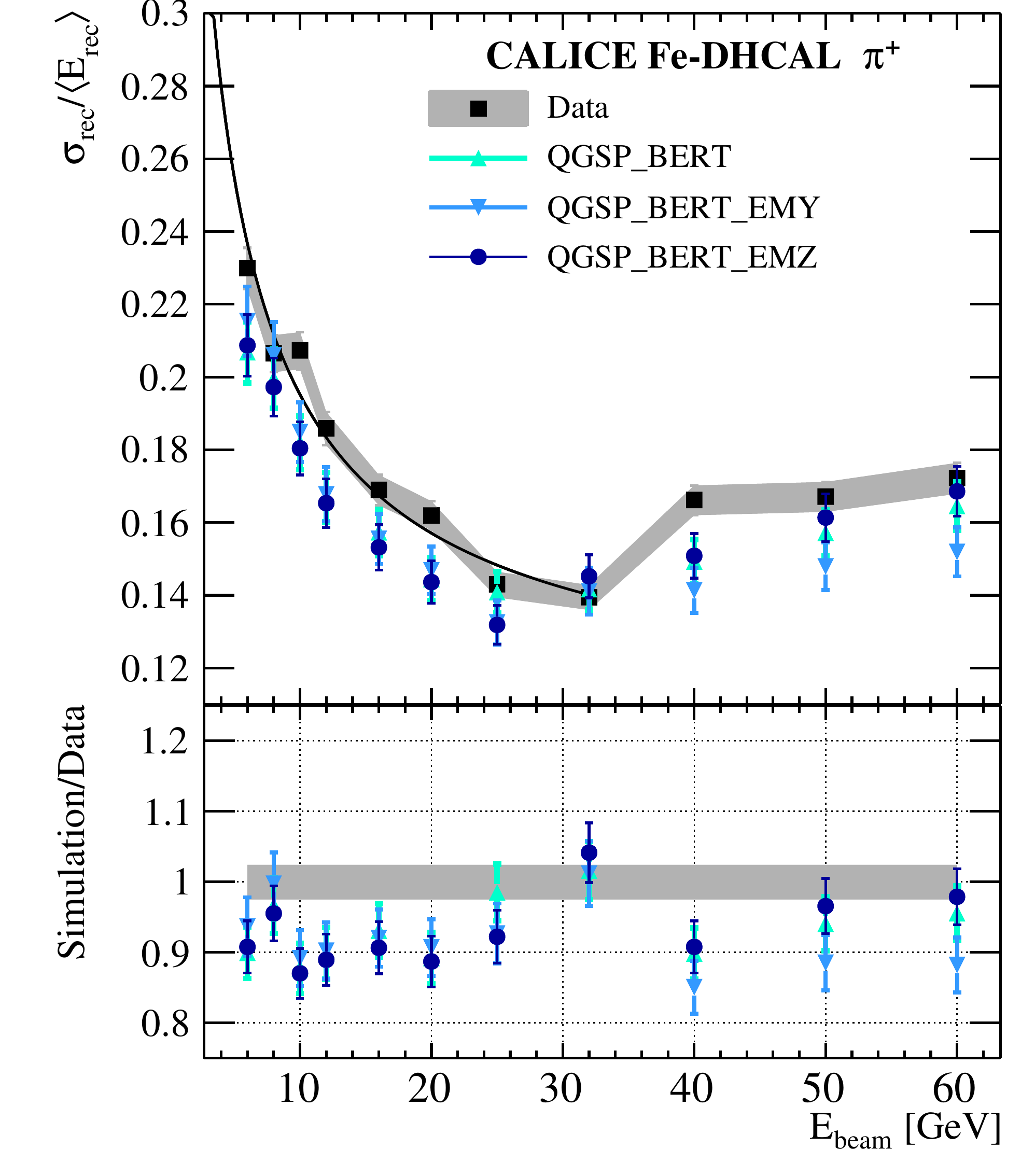}}
	\end{minipage}
\caption{The $\pi^{+}$ energy resolution of the Fe-DHCAL for beam energies from 6 to 60\,GeV. The bottom plots show the ratio of the simulations and data. The data is shown as black squares and the black curve represents the fit to Eq.~\ref{eq:EMresolution}. The error bands show the systematic and statical uncertainty added in quadrature. The statistical errors are smaller than the size of the markers. }
\label{DHCALpiResolution}
\end{center}
\end{figure}

The comparison of the simulated resolutions reveals a strong dependence on the EM and the hadronic physics lists. However, all simulations achieve an agreement with the data within 15\,\%. While the QGSP physics list shows the tendency to underestimate the pion resolution, originating from the overestimate of the total number of hits, the FTFP physics list shows stronger variation with the different EM physics lists. This could originate from a larger EM fraction of the hadronic showers described with the Fritiof String model~\cite{Geant4_PartonString}.  
The best agreement in the energy resolution between the data and MC is observed for the simulation using the QGSP\_BERT\_EMZ physics list, with a mean remaining difference of less than 5\,\%, see the bottom plot in Fig.~\ref{fig:resolutionQGSP}. 
Note that by applying a weighting scheme dependent on the hit density, the saturation of the response can be corrected and the energy resolution can thus be improved~\cite{SCdhcal}. However, further dedicated studies are necessary to determine to which extent this is possible.
\clearpage

\subsection{Pion shower shapes}
\begin{figure}
\begin{center}
	\begin{minipage}{.5\textwidth}
	\subfloat [2D hit density]  {\label{fig:digi40GeVpiD}\includegraphics [width=1\textwidth]{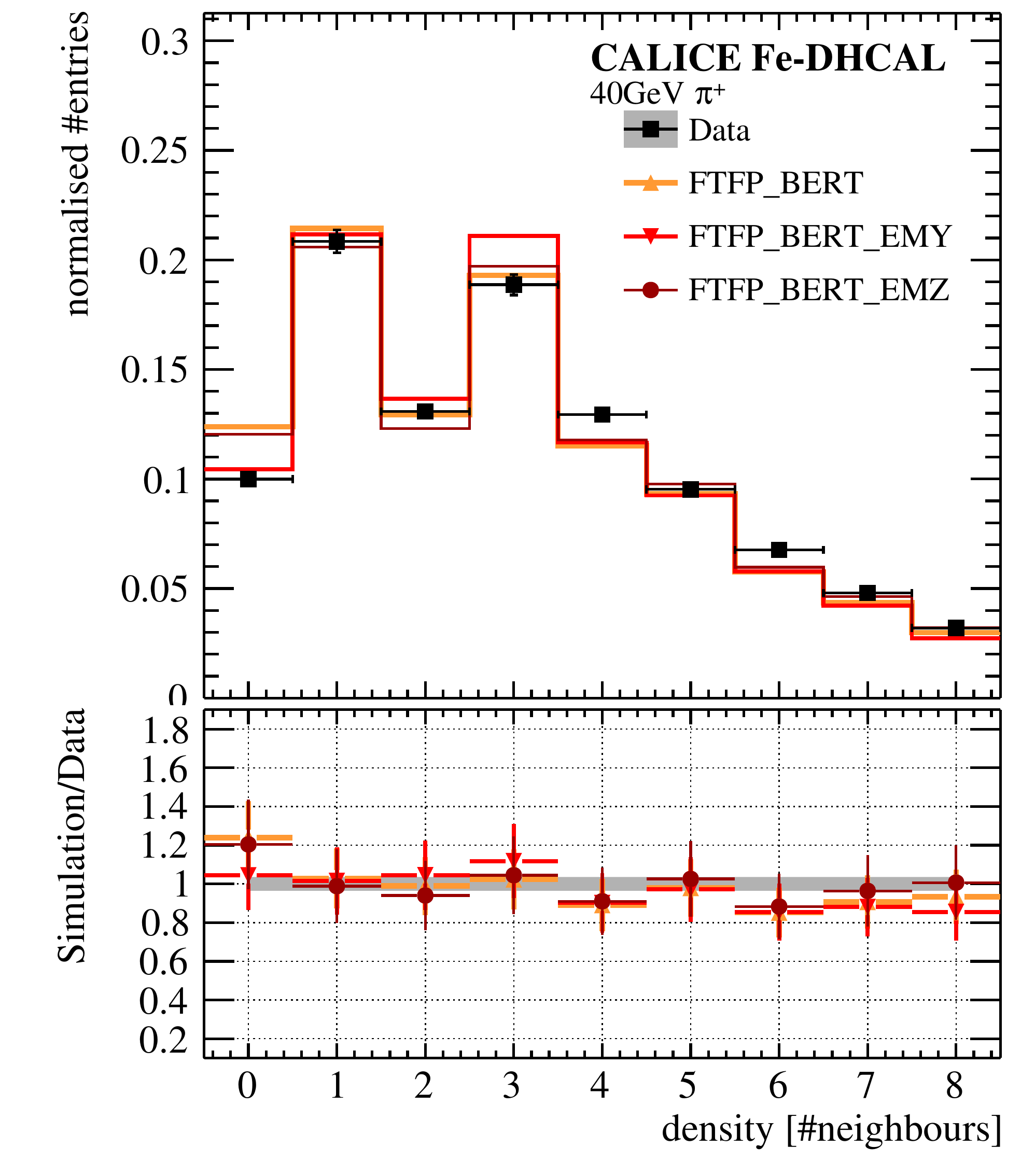}} 
	\\
	\subfloat [3D hit density]  {\label{fig:digi40GeVpi3DD}\includegraphics [width=1\textwidth]{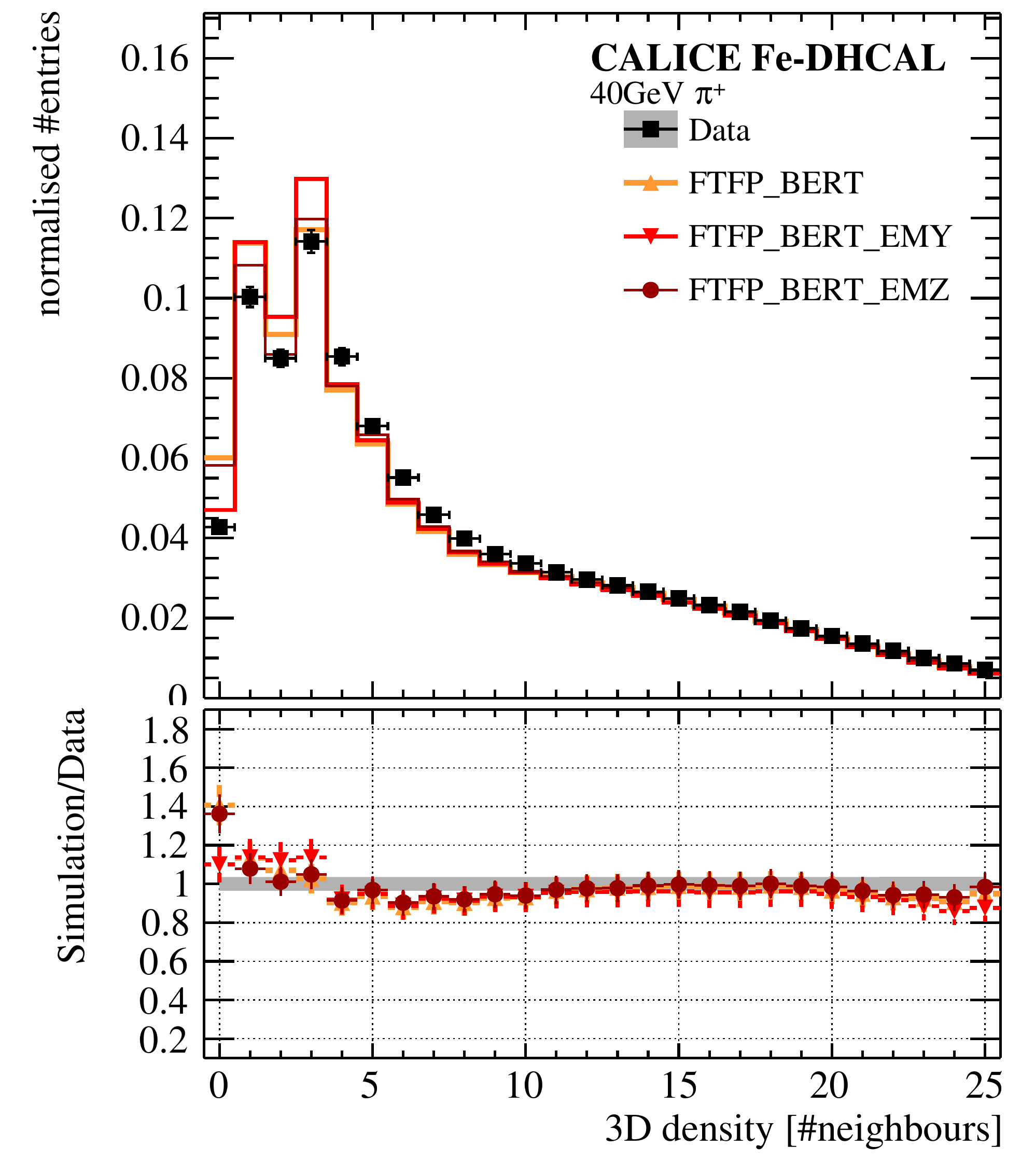}}
	\end{minipage}
\caption{The shower observables for 40\,GeV $\pi^{+}$ events; a) 2D hit density and b) 3D hit density. The data is represented as black squares and the grey error band corresponds to the systematic and statistical uncertainty added in quadrature. The ratios in the bottom plots show also the systematic uncertainty on the simulations.}
\label{fig:digiPiDensity}
\end{center}
\end{figure}

\begin{figure}
\begin{center}
	\begin{minipage}{.5\textwidth}
	\subfloat [2D hit density]  {\label{fig:digi40GeVpiDQ}\includegraphics [width=1\textwidth]{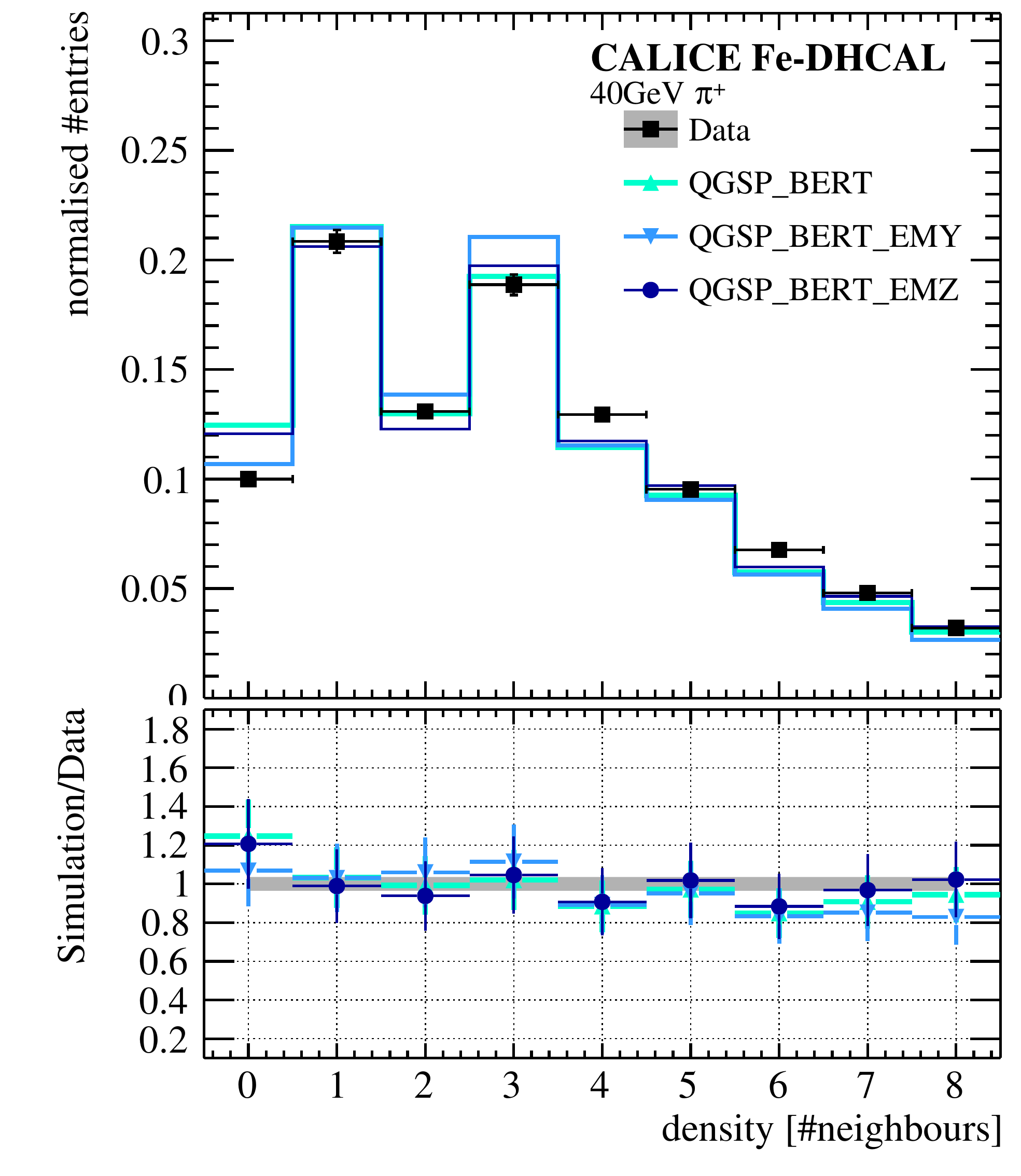}} 
	\\
	\subfloat [3D hit density]  {\label{fig:digi40GeVpi3DDQ}\includegraphics [width=1\textwidth]{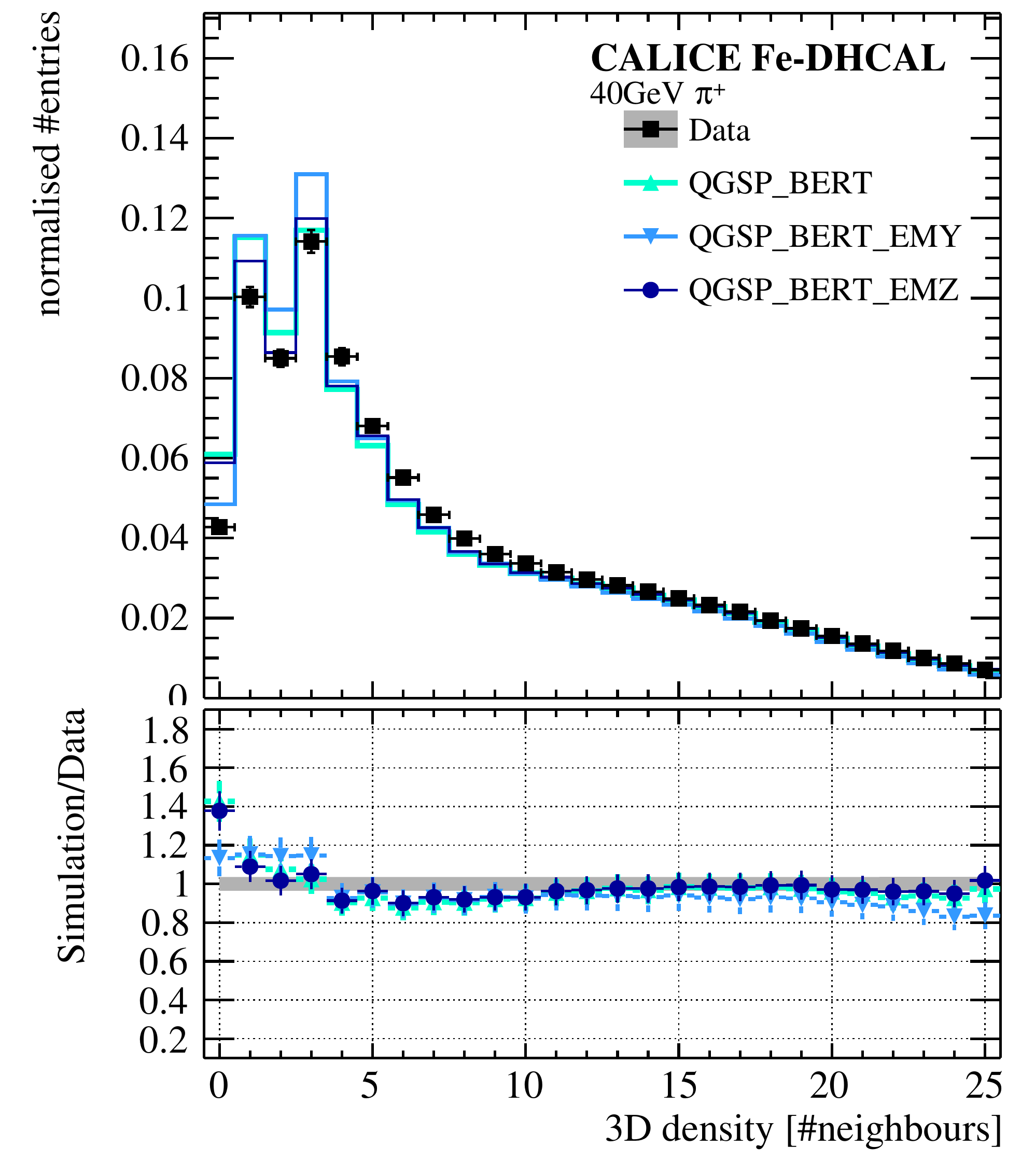}}
	\end{minipage}
\caption{The shower observables for 40\,GeV $\pi^{+}$ events; a) 2D hit density and b) 3D hit density. The data is represented as black squares and the grey error band corresponds to the systematic and statistical uncertainty added in quadrature. The ratios in the bottom plots show also the systematic uncertainty on the simulations.}
\label{fig:digiPiQDensity}
\end{center}
\end{figure}

\begin{figure}
\begin{center}
	\begin{minipage}{.5\textwidth}
	\hspace*{-2.5cm}
		\subfloat {\label{fig:digi6GeVpiLong} \includegraphics [width=1\textwidth]{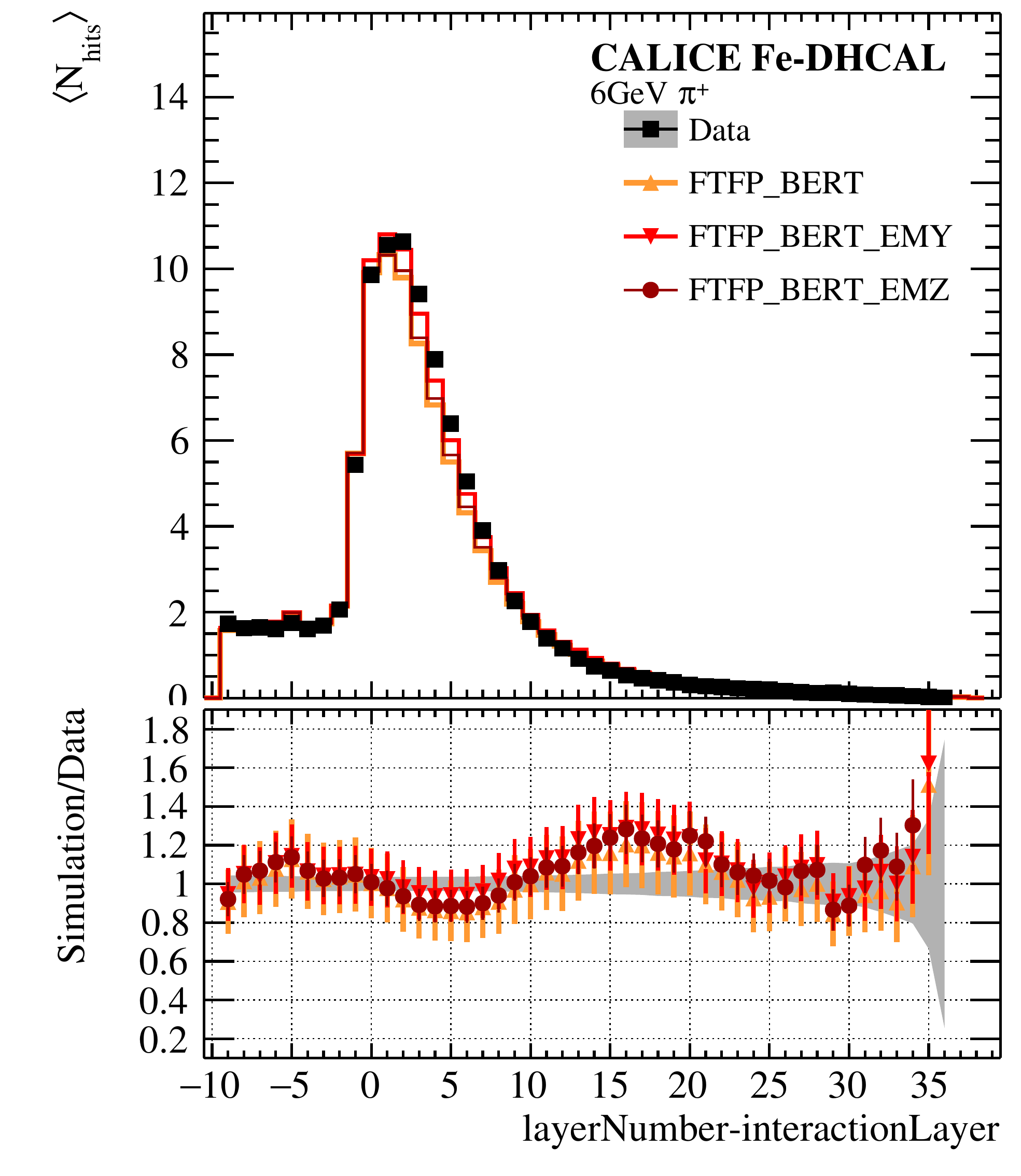}} 	
		\subfloat {\label{fig:digi10GeVpiLong}\includegraphics [width=1\textwidth]{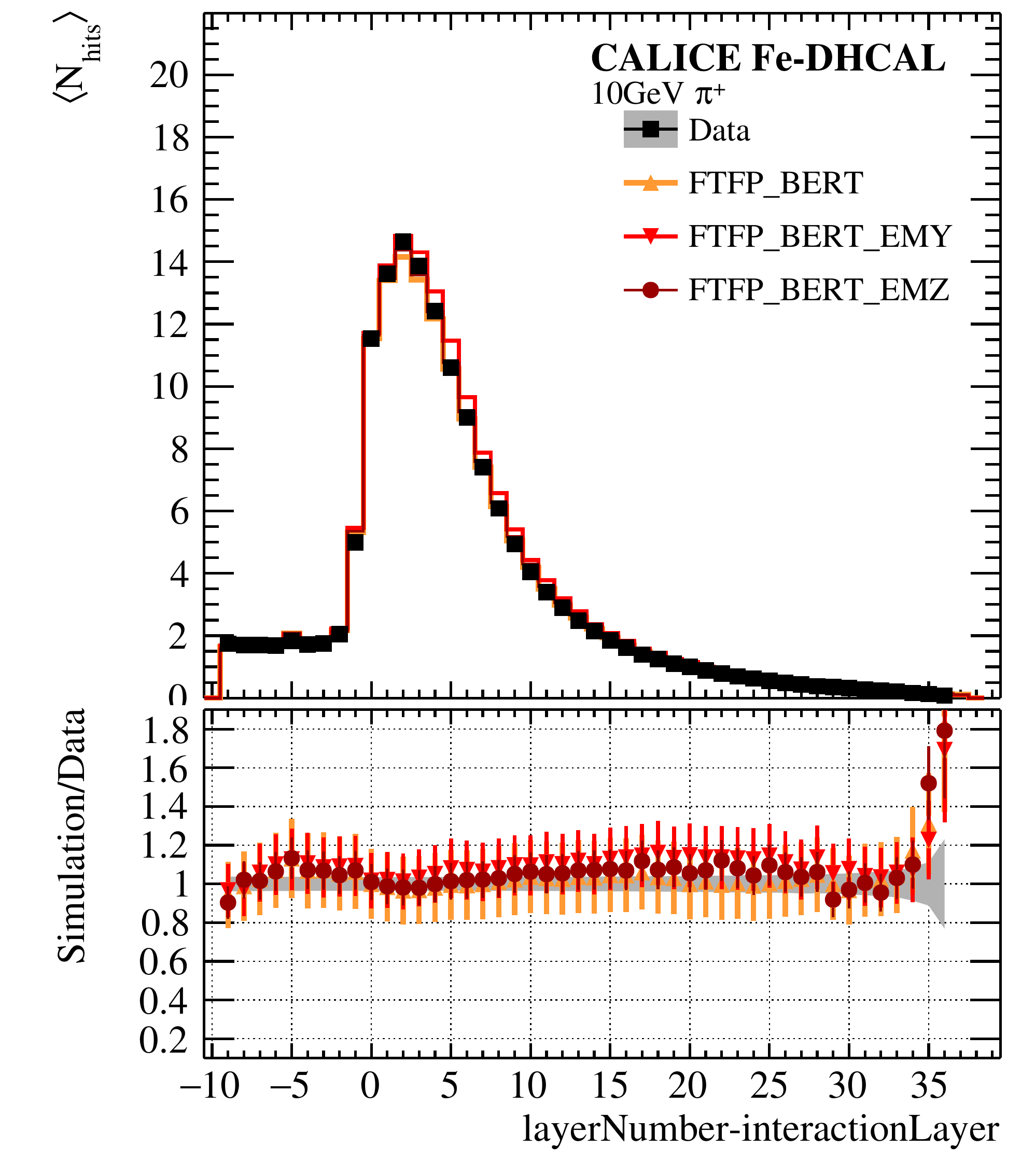}}
	 \end{minipage}	
	 \\
	\begin{minipage}{.5\textwidth}
	\hspace*{-2.5cm}
		\subfloat {\label{fig:digi20GeVpiLong}\includegraphics [width=1\textwidth]{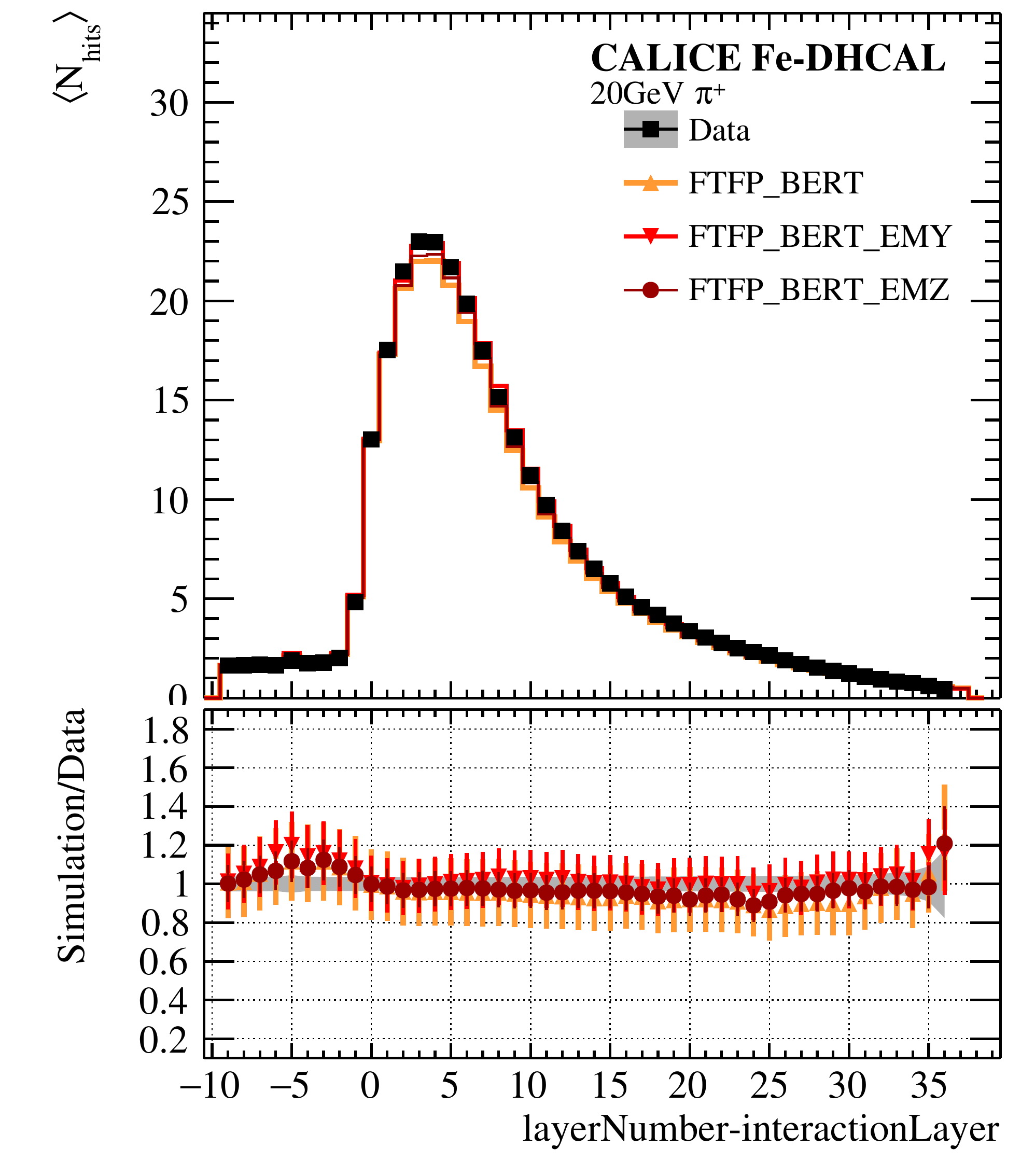}}		\subfloat {\label{fig:digi40GeVpiLong}\includegraphics [width=1\textwidth]{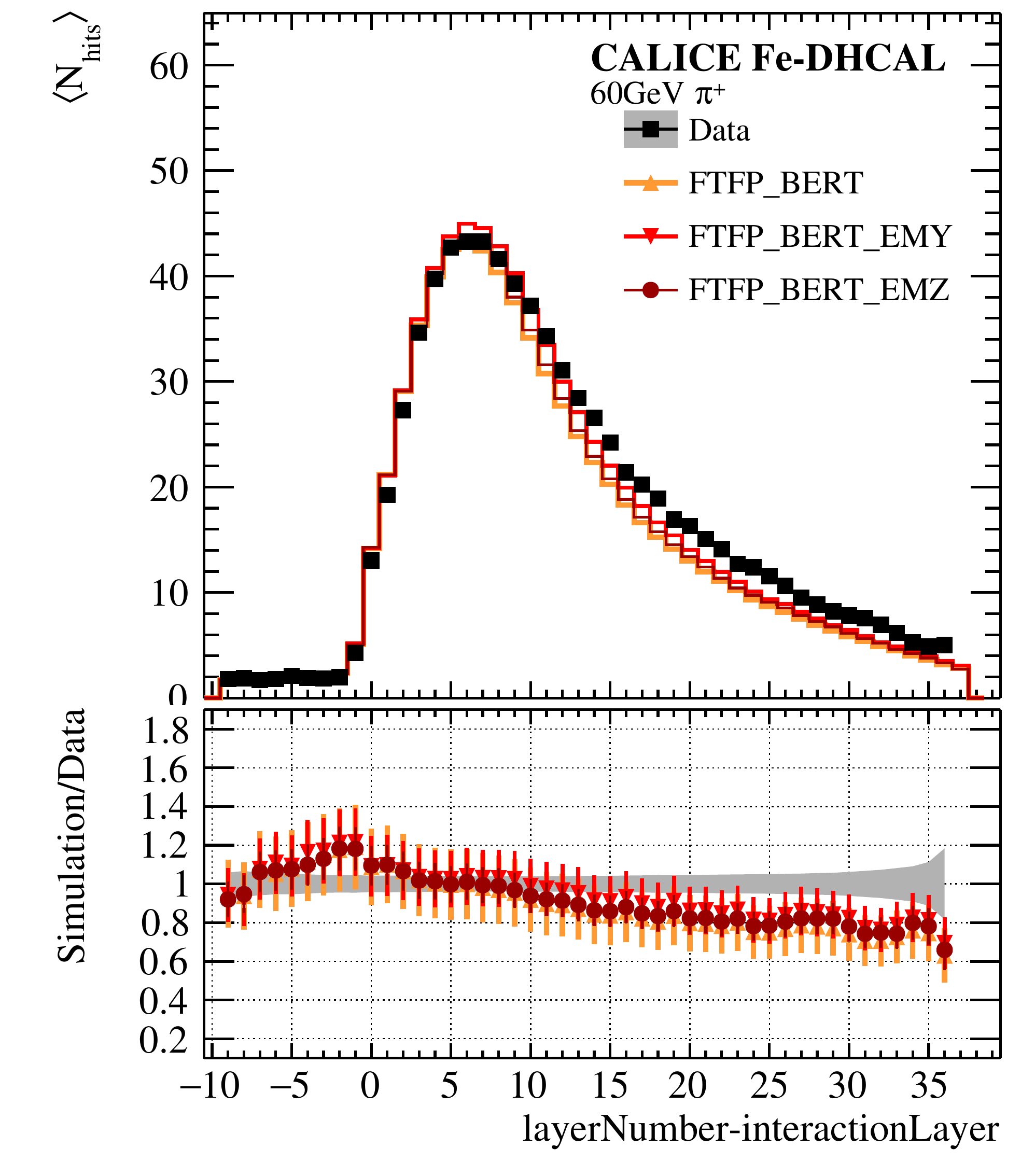}}
	\end{minipage}
\caption{The longitudinal profiles of 6, 10, 20 and 60\,GeV $\pi^{+}$ events. The data is represented as black squares and the grey error band corresponds to the systematic and statistical uncertainty added in quadrature. The ratios in the bottom plots show also the systematic uncertainty on the simulations.}
\label{fig:digiPiLongProfiles}
\end{center}
\end{figure}

\begin{figure}
\begin{center}
	\begin{minipage}{.5\textwidth}
	\hspace*{-2.5cm}
		\subfloat {\label{fig:digi6GeVpiLongQ} \includegraphics [width=1\textwidth]{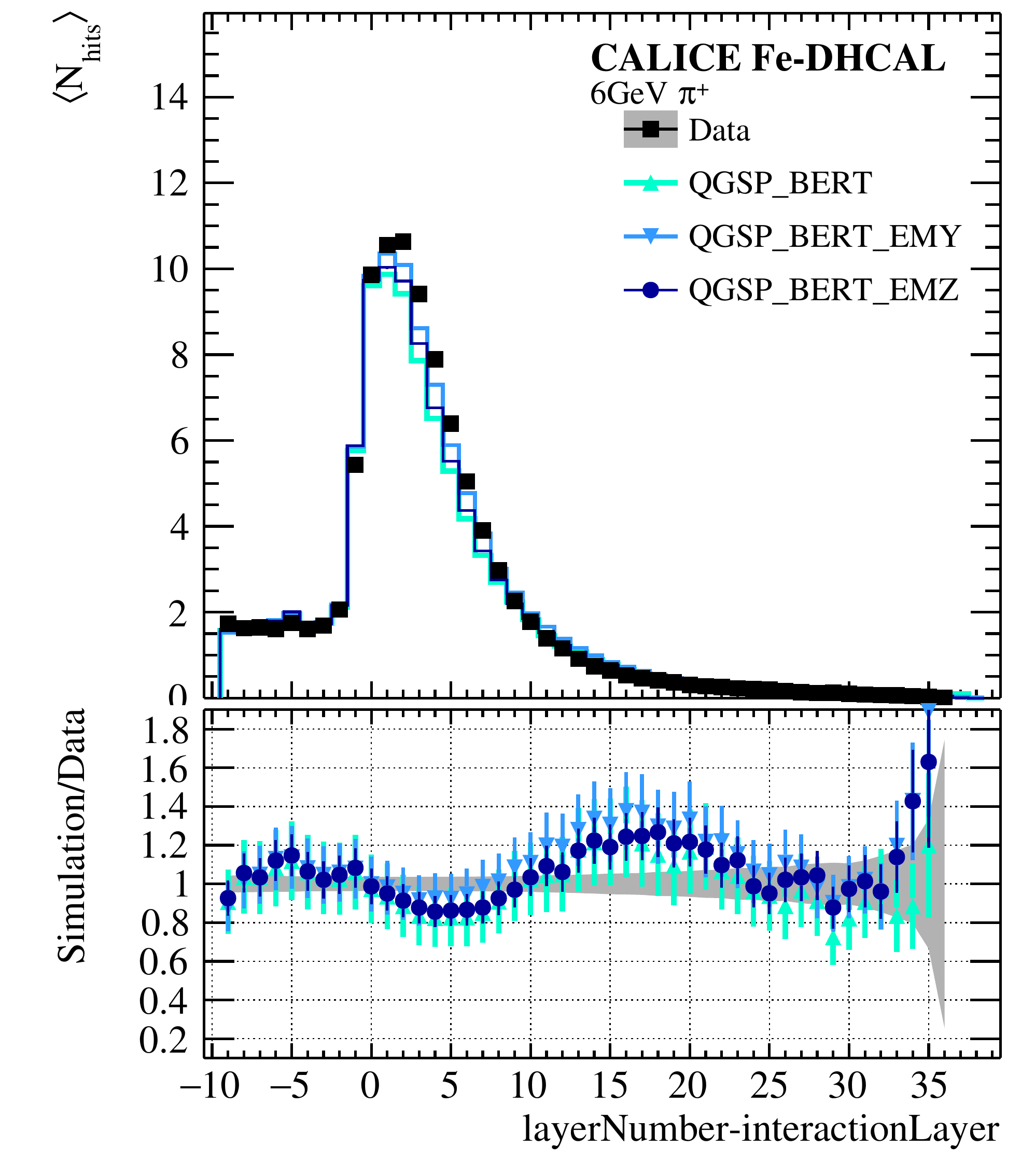}} 	
		\subfloat {\label{fig:digi10GeVpiLongQ}\includegraphics [width=1\textwidth]{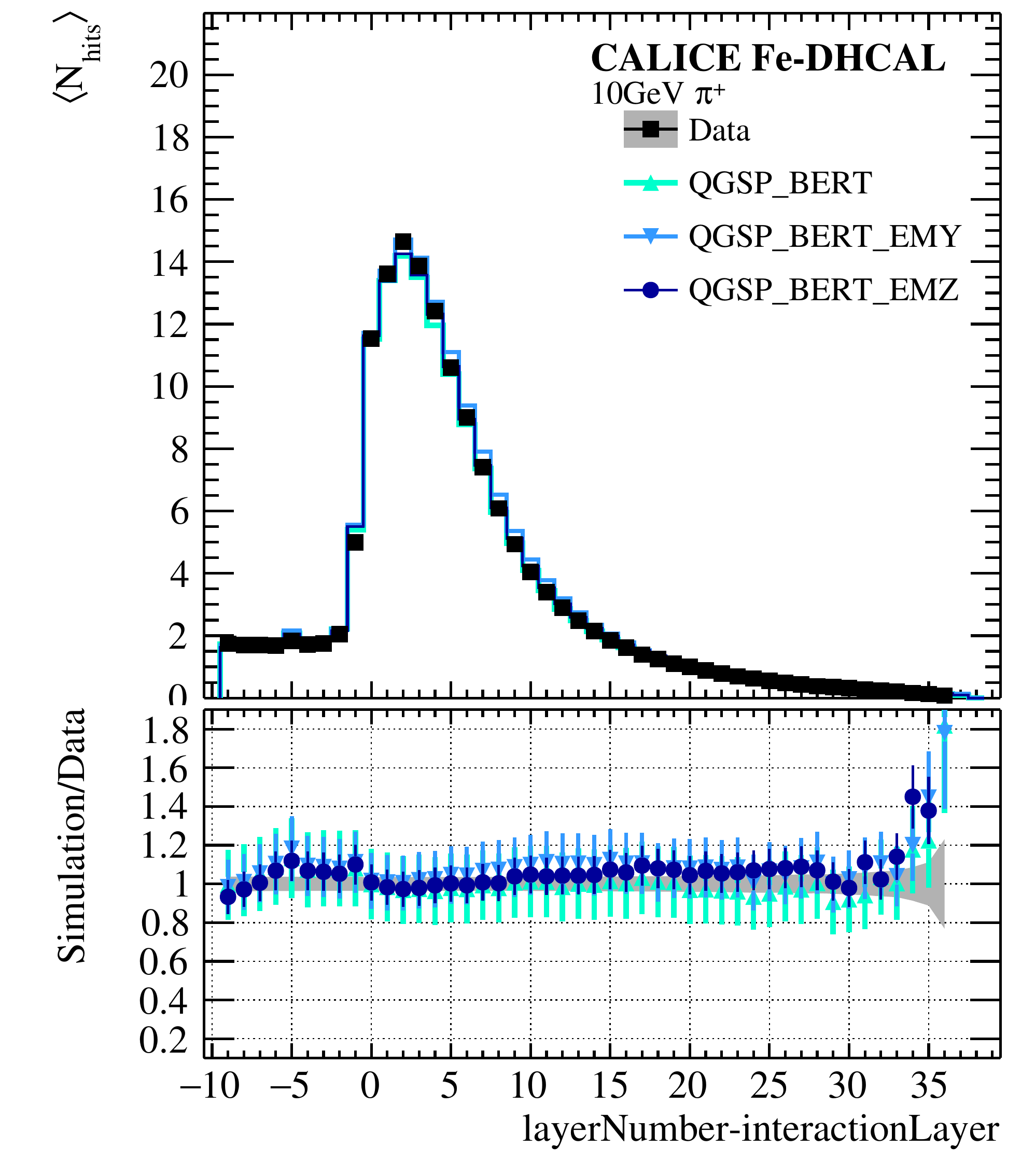}}
	 \end{minipage}	
	 \\
	\begin{minipage}{.5\textwidth}
	\hspace*{-2.5cm}
		\subfloat {\label{fig:digi20GeVpiLongQ}\includegraphics [width=1\textwidth]{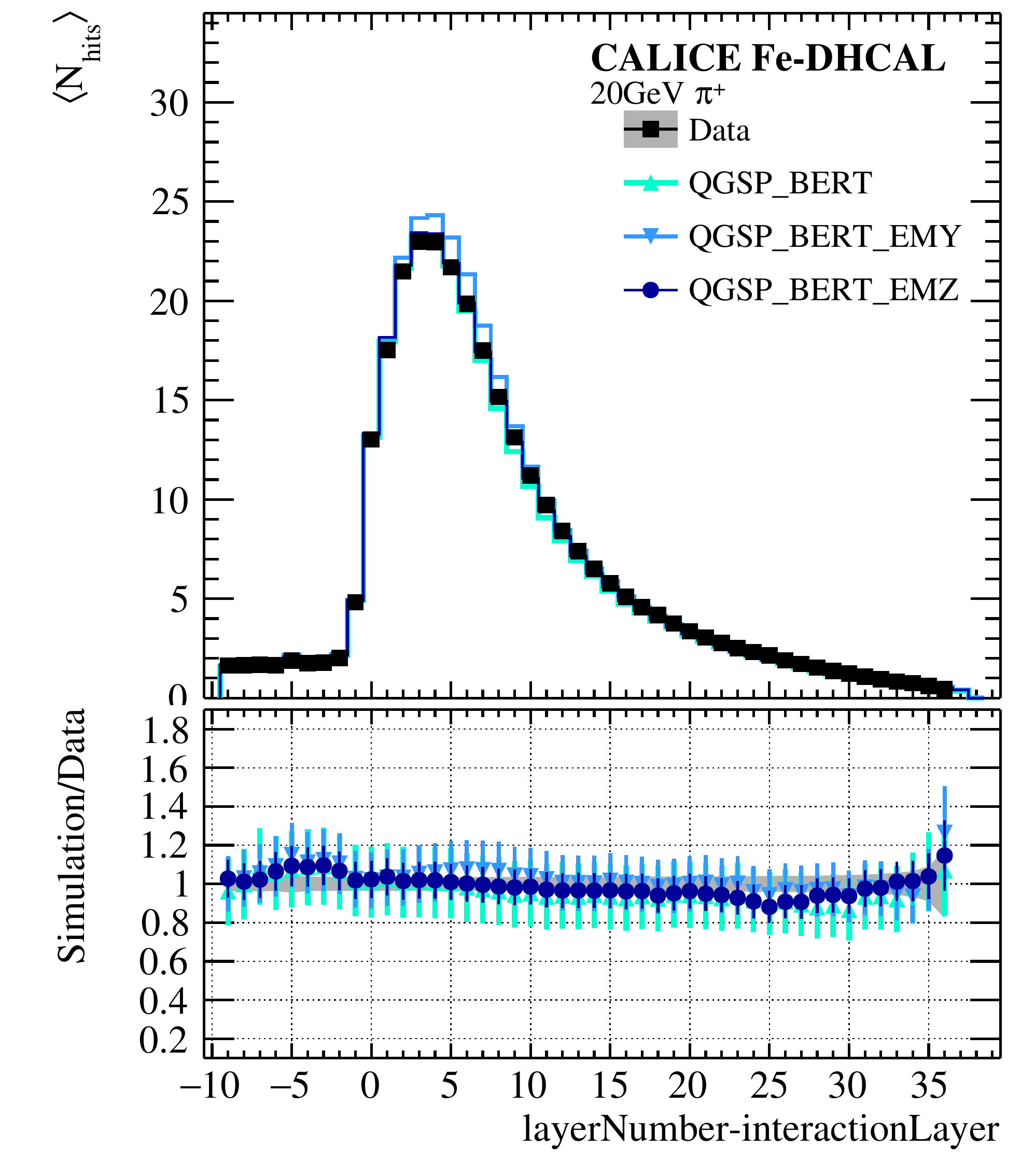}}
		\subfloat {\label{fig:digi40GeVpiLongQ}\includegraphics [width=1\textwidth]{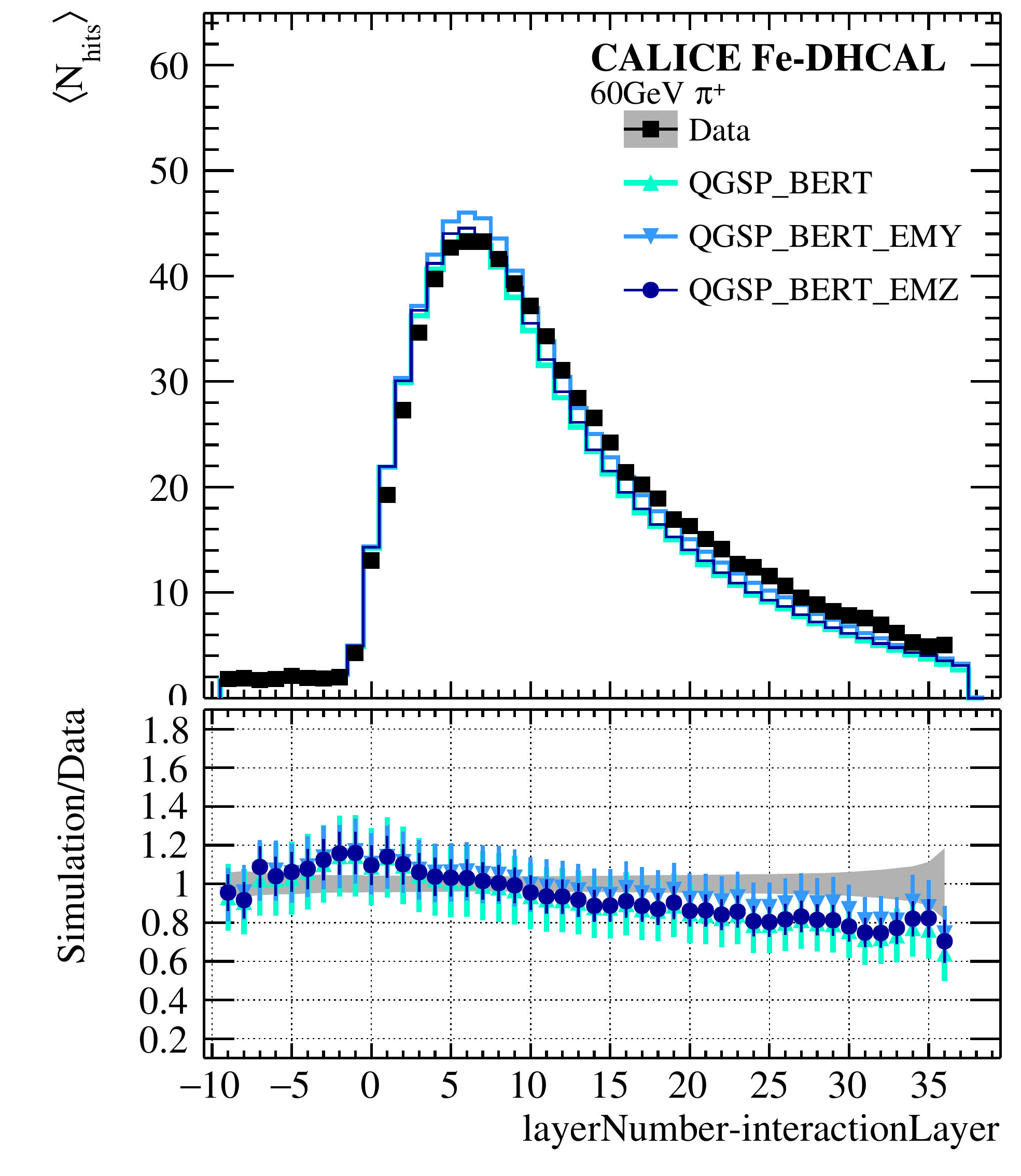}}
	\end{minipage}
\caption{The longitudinal profiles of 6, 10, 20, and 60\,GeV $\pi^{+}$ events. The data is represented as black squares and the grey error band corresponds to the systematic and statistical uncertainty added in quadrature. The ratios in the bottom plots show also the systematic uncertainty on the simulations.}
\label{fig:digiPiQLongProfiles}
\end{center}
\end{figure}

The hit densities and shower shapes of the pion showers are studied for all energies. The comparison of the data and simulation in the shower observables 2D density and 3D density is shown as an example for 40\,GeV $\pi^{+}$ events and compared to the FTFP\_BERT physics list in Fig.~\ref{fig:digiPiDensity} and QGSP\_BERT physics list in Fig.~\ref{fig:digiPiQDensity}. 
The simulations are repeated using the standard, \_EMY and \_EMZ EM physics lists, which show similar effects independent of the hadronic physics lists. In general, data and MC are in good agreement. No significant difference between the hadronic physics lists could be observed. While the largest impact of the different EM sub-shower descriptions is observed in the very low density bins, mostly originating from single electron tracks, and very high density bins, corresponding to the EM shower center. The shower center is best described by the EMZ physics list.
The longitudinal profiles are shown for 6, 10, 20 and 60\,GeV in Figs.~\ref{fig:digiPiLongProfiles} and~\ref{fig:digiPiQLongProfiles}, and are compared to the FTFP\_BERT and QGSP\_BERT physics lists, respectively. 

The longitudinal profiles are well described by the simulations within the relatively large errors of the simulation. However, all simulations show a depletion of hits in the 10 to 35th layer from the shower start for energies above 20\,GeV. For the lowest energies, $<10\,GeV$, a trend to an excess of number of hits in the tails is observed. For the beam energies from 10 to 20\,GeV, the longitudinal profiles show good agreement for all studied physics lists. The longitudinal shower shapes with energies above 20\,GeV are best described by the QGSP\_BERT physics list.

\begin{figure}
\begin{center}
	\begin{minipage}{.5\textwidth}
	\hspace*{-2.5cm}
		\subfloat {\label{fig:digi6GeVpiLong} \includegraphics [width=1\textwidth]{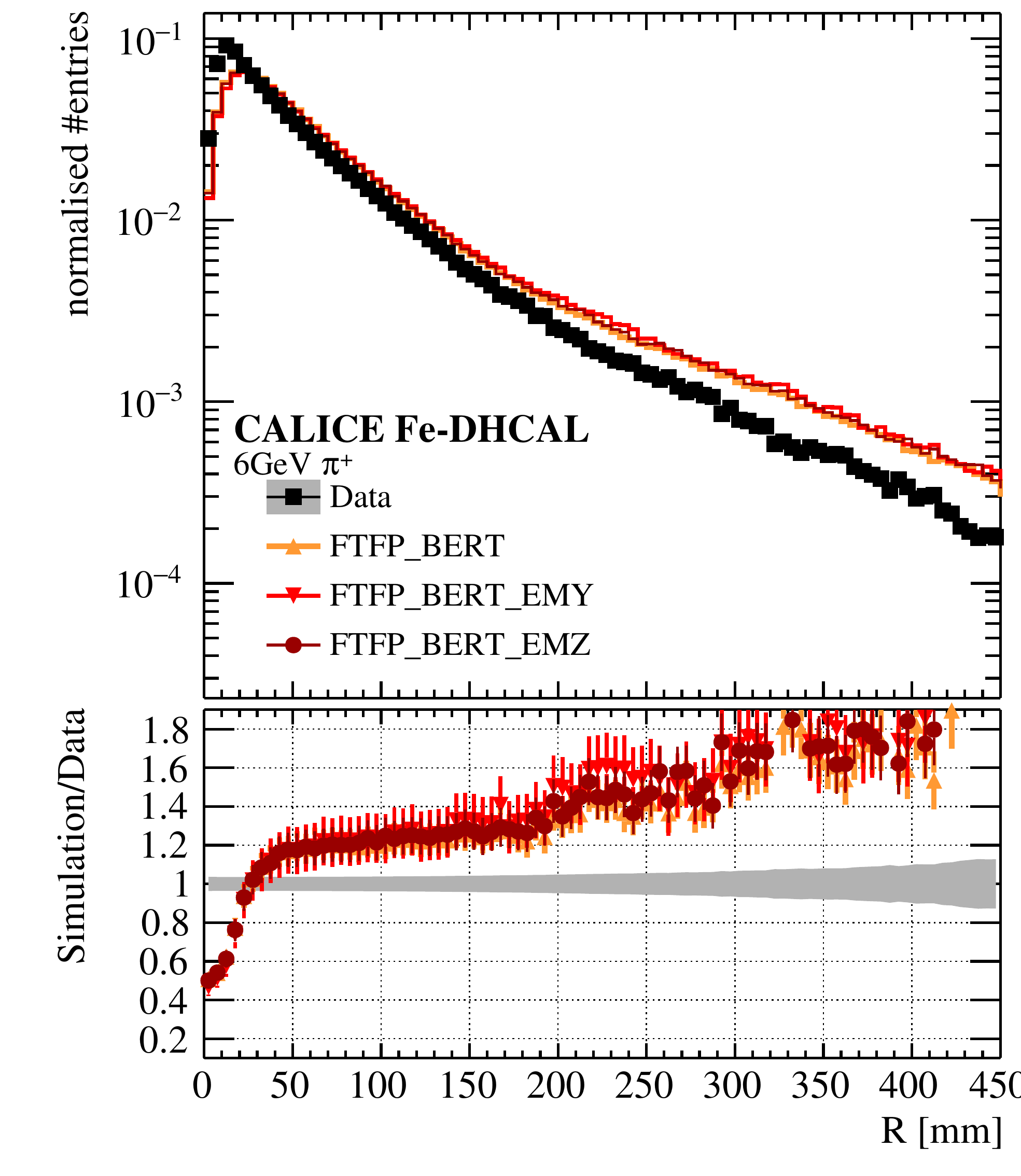}} 
		\subfloat {\label{fig:digi10GeVpiLong} \includegraphics [width=1\textwidth]{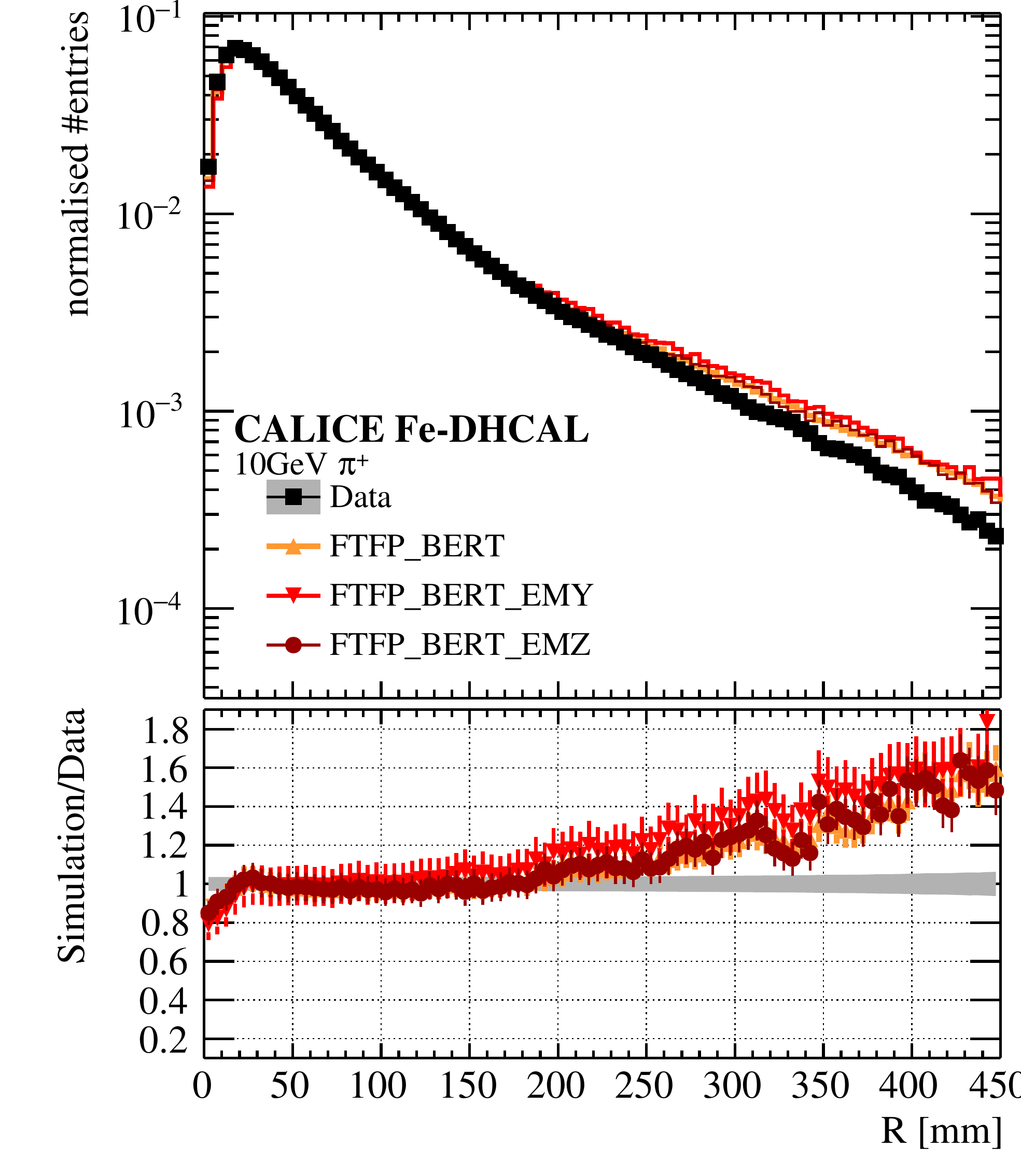}} 
	\end{minipage}
	 \\
	\begin{minipage}{.5\textwidth}
	\hspace*{-2.5cm}
		\subfloat {\label{fig:digi20GeVpiRad}\includegraphics [width=1\textwidth]{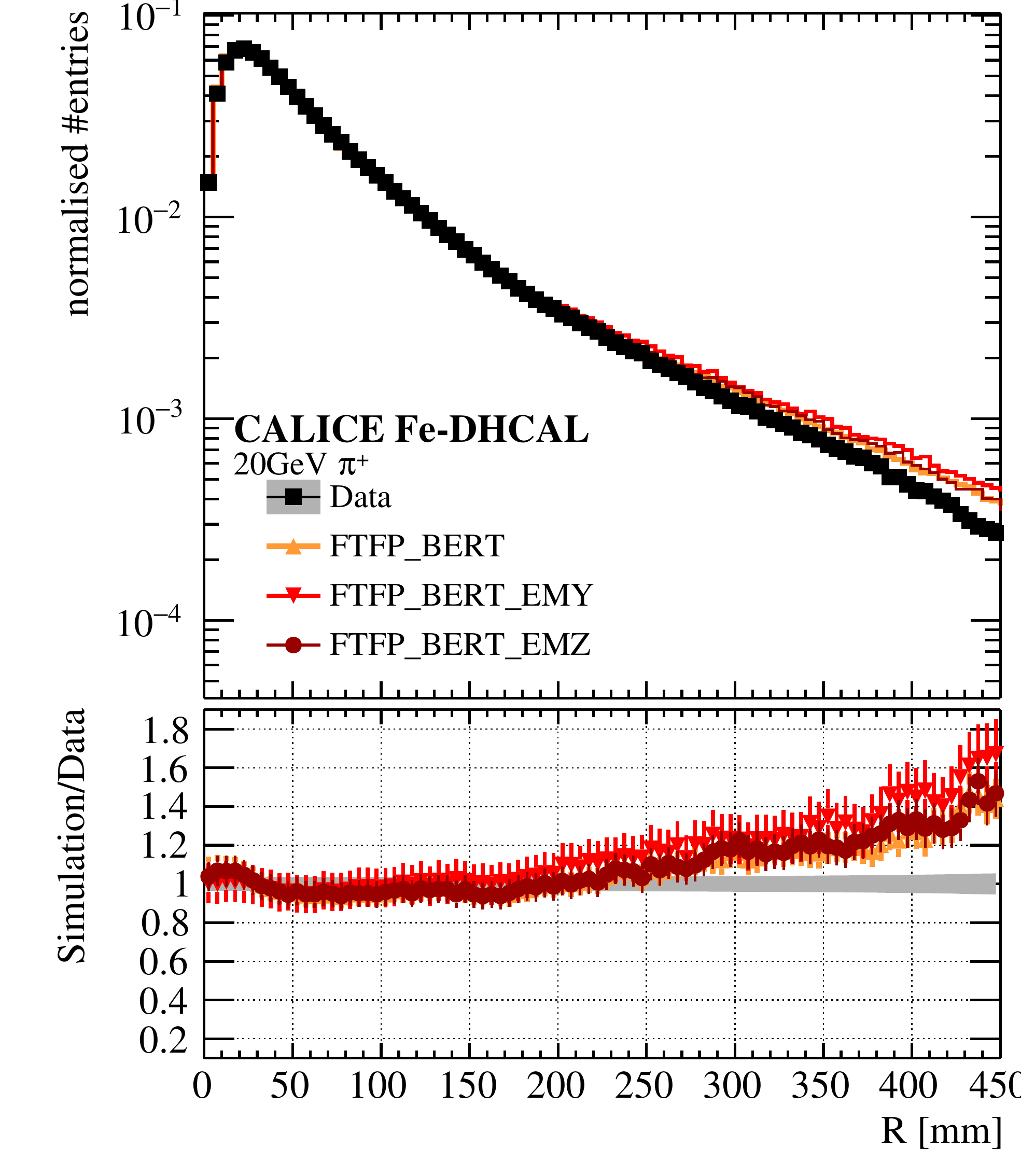}}
		\subfloat {\label{fig:digi40GeVpiRad}\includegraphics [width=1\textwidth]{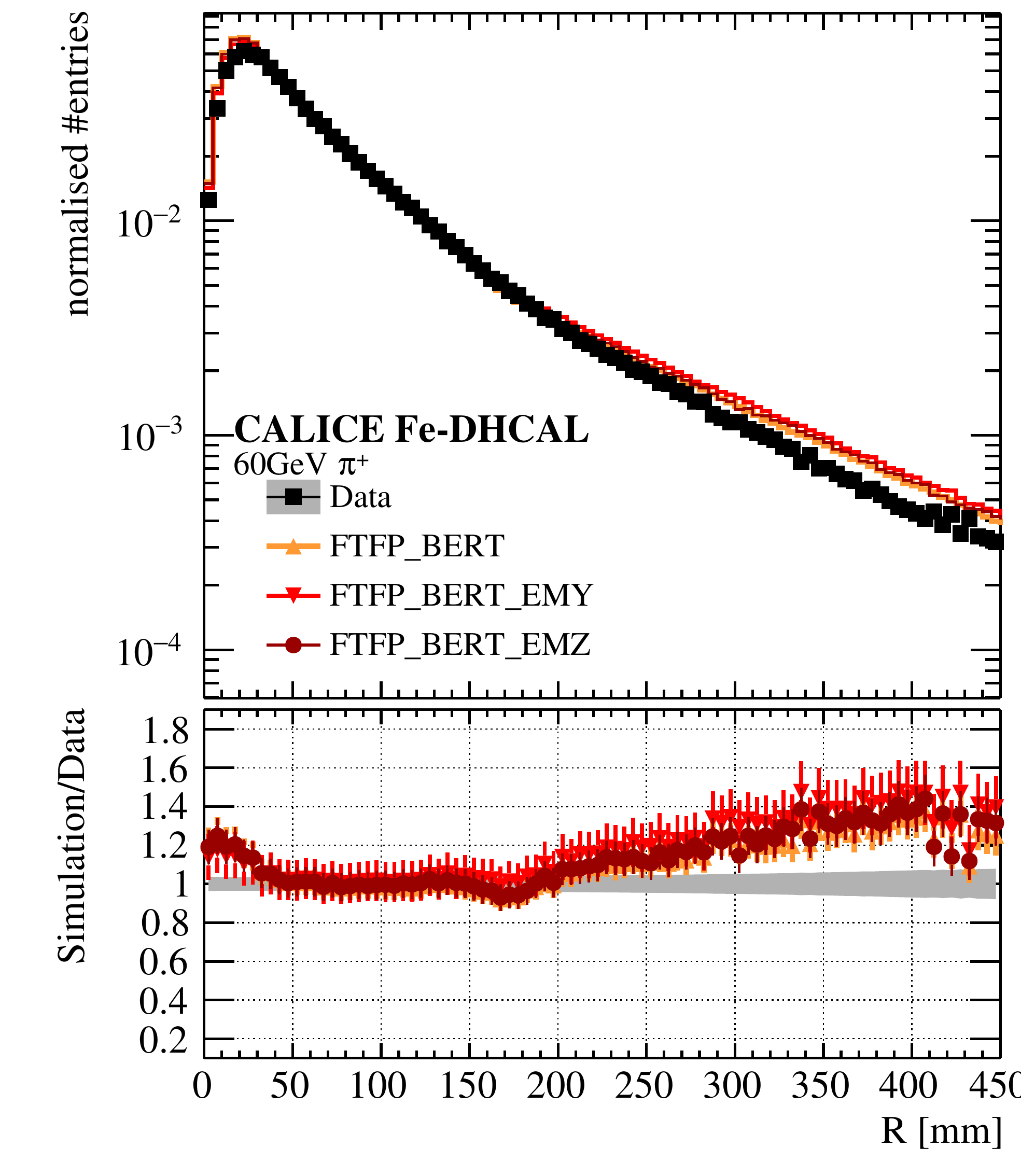}}
	\end{minipage}
\caption{The radial shower shapes for 6, 10, 20, and 60\,GeV $\pi^{+}$ events. The data is represented as black squares and the grey error band corresponds to the systematic and statistical uncertainty added in quadrature. The ratios in the bottom plots show also the systematic uncertainty on the simulations.}
\label{fig:digiPiRadProfiles}
\end{center}
\end{figure}

\begin{figure}
\begin{center}
	\begin{minipage}{.5\textwidth}
	\hspace*{-2.5cm}
		\subfloat {\label{fig:digi6GeVpiRadQ} \includegraphics [width=1\textwidth]{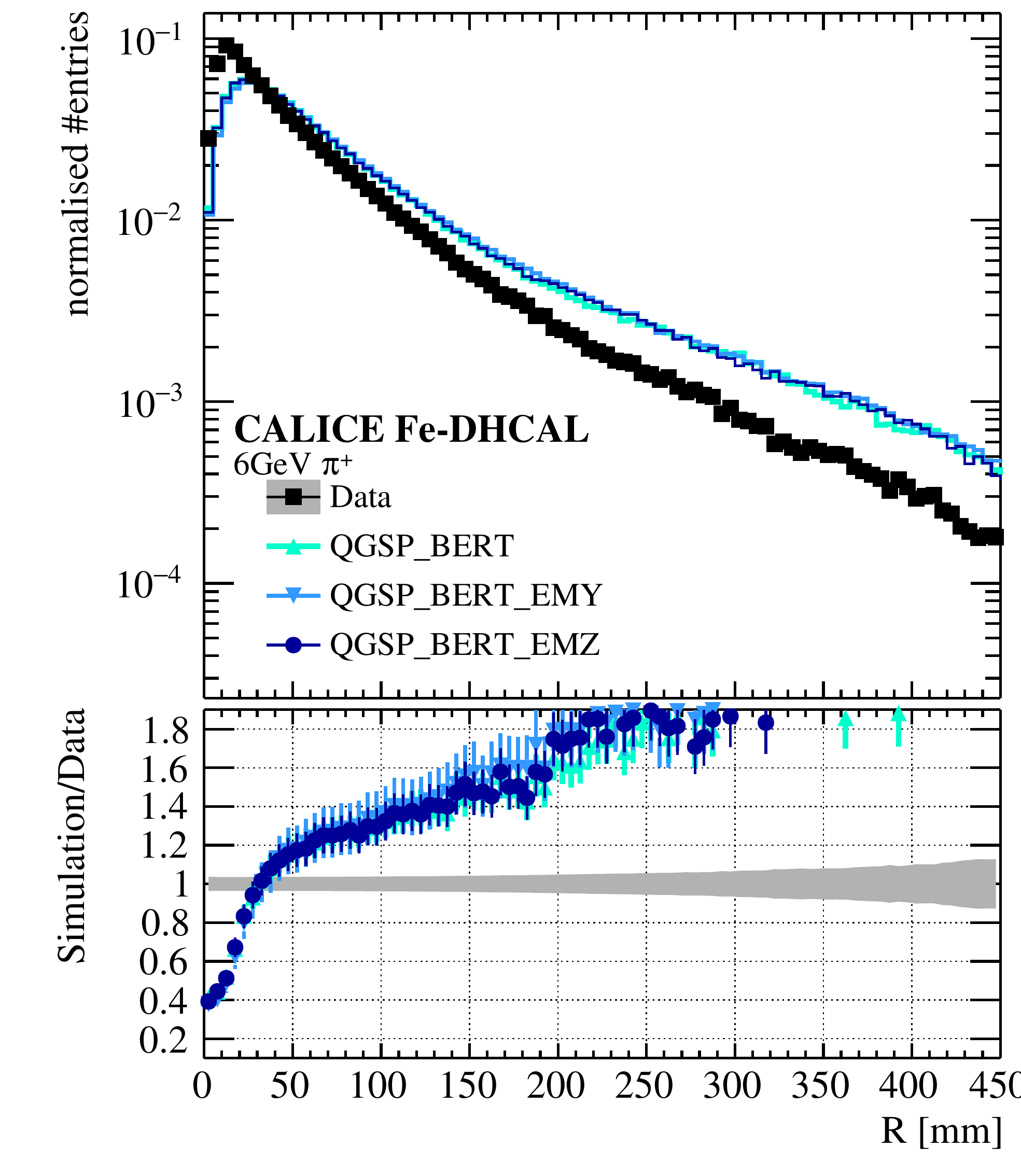}} 	
		\subfloat {\label{fig:digi10GeVpiRadQ}\includegraphics [width=1\textwidth]{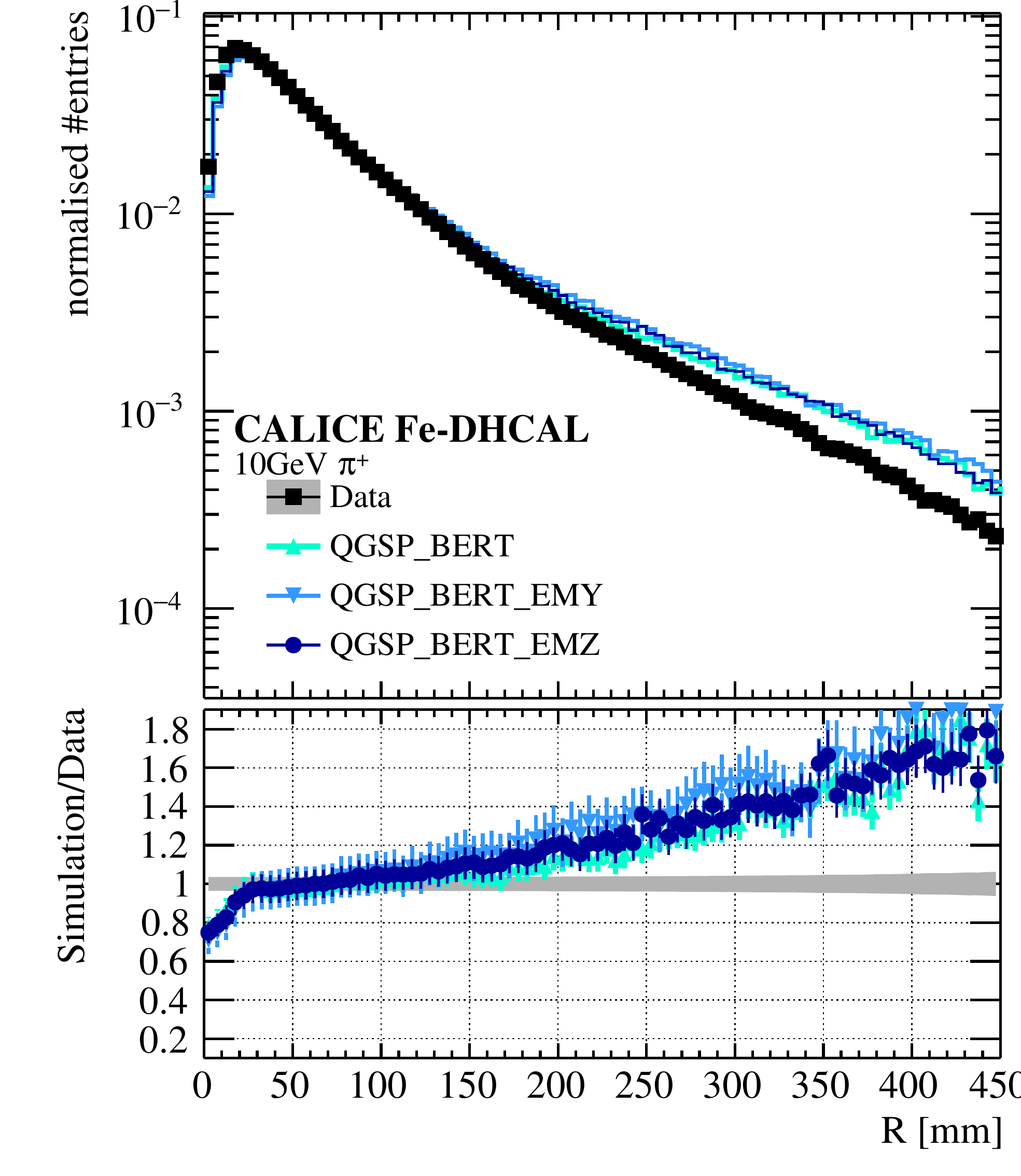}}
	 \end{minipage}	
	 \\
	\begin{minipage}{.5\textwidth}
	\hspace*{-2.5cm}
		\subfloat {\label{fig:digi20GeVpiRadQ}\includegraphics [width=1\textwidth]{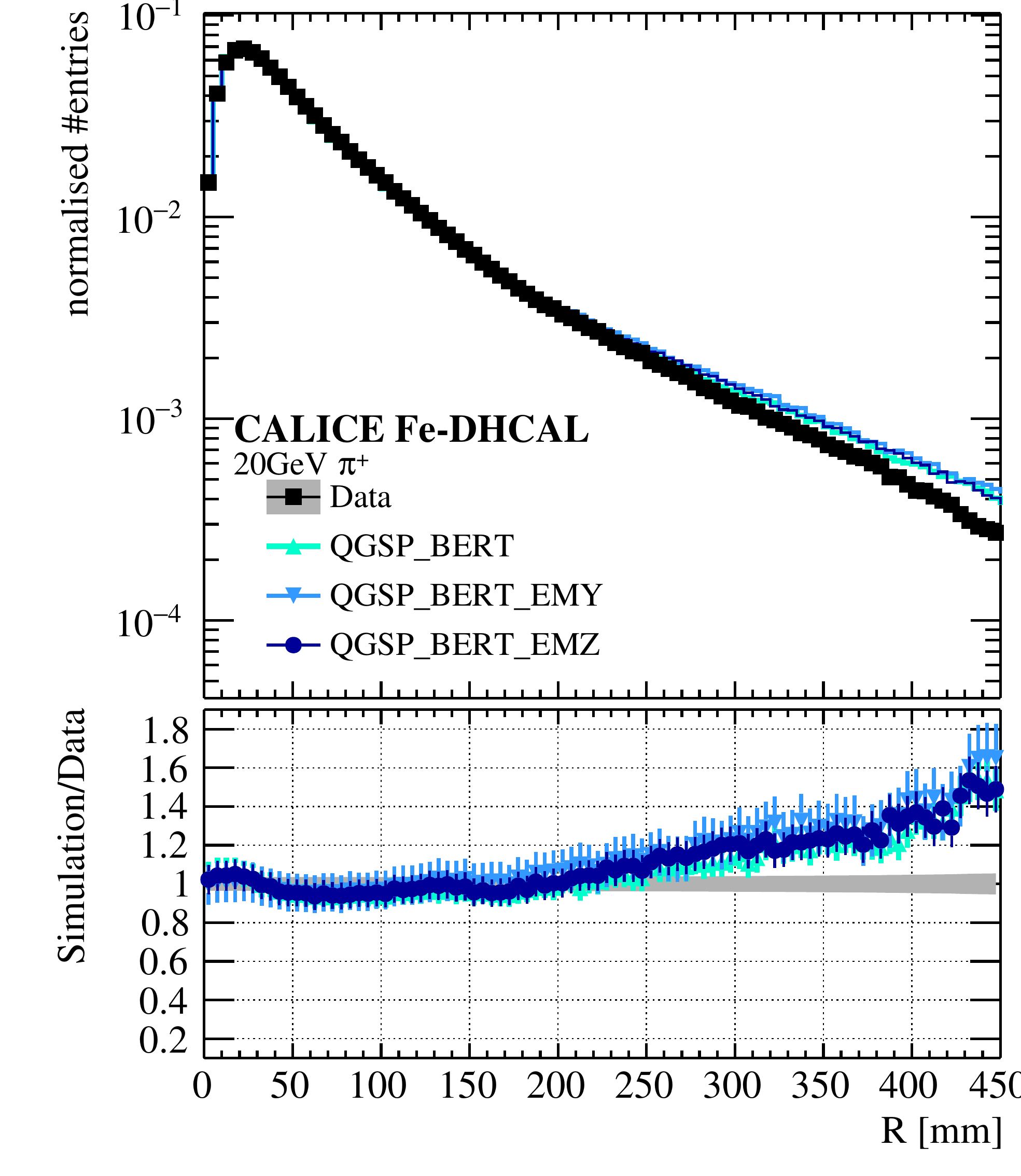}}
		\subfloat {\label{fig:digi40GeVpiRadQ}\includegraphics [width=1\textwidth]{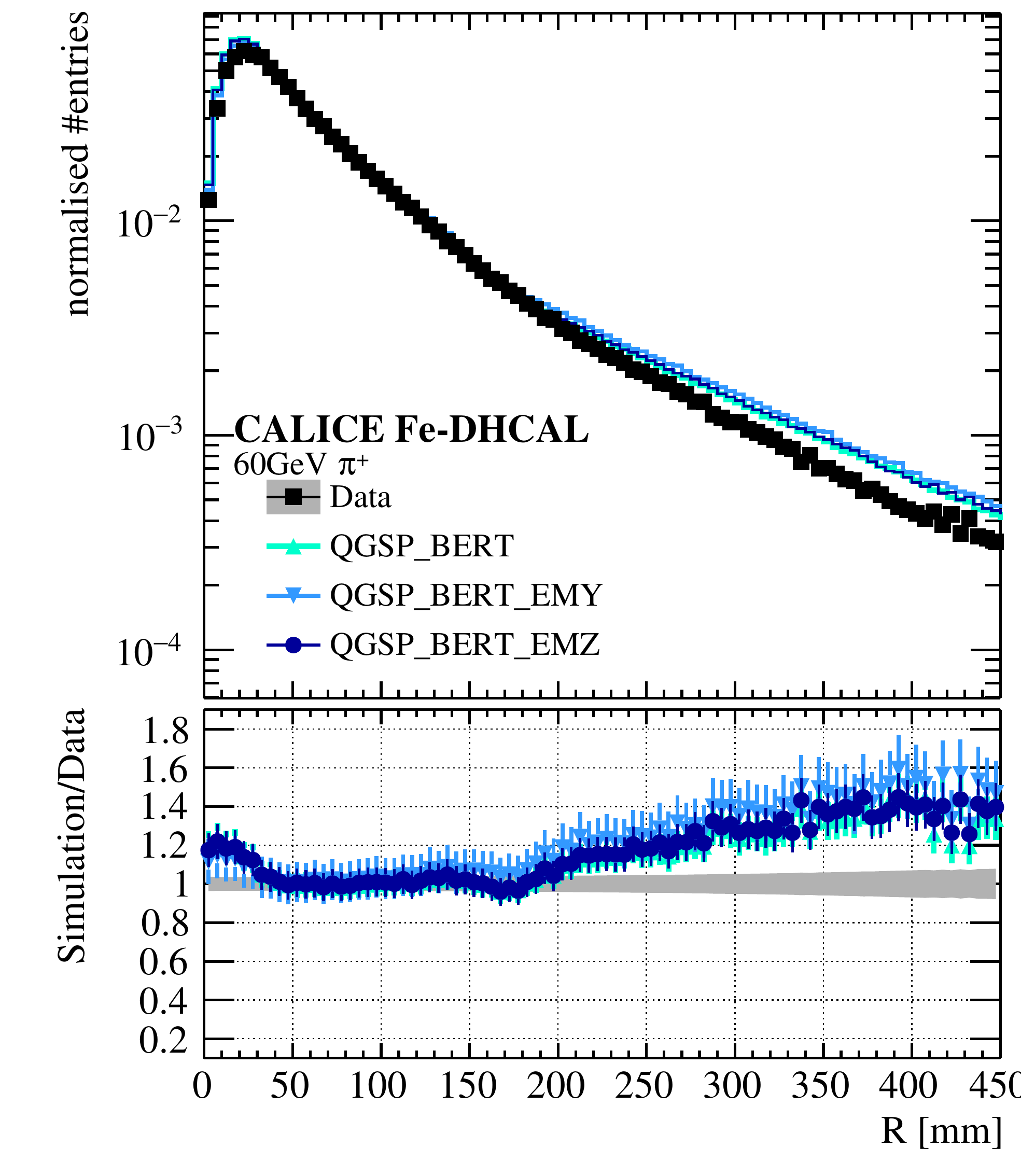}}
	\end{minipage}
\caption{The radial shower shapes for 6, 10, 20, and 60\,GeV $\pi^{+}$ events. The data is represented as black squares and the grey error band corresponds to the systematic and statistical uncertainty added in quadrature. The ratios in the bottom plots show also the systematic uncertainty on the simulations.}
\label{fig:digiPiQRadProfiles}
\end{center}
\end{figure}

The radial shower shapes of the simulated pion showers, shown in Figs.~\ref{fig:digiPiRadProfiles} and~\ref{fig:digiPiQRadProfiles}, are in good agreement with the data for particle shower energies above 10\,GeV. However, in general the simulated showers tend to exhibit a slightly broader shower core and a larger radial dispersion than the measured showers. 
The somewhat higher density of the simulated shower core is consistent with the saturation observed for the simulated high energy pions ($>32$\,GeV).

\clearpage


\clearpage

\section{$e/\pi$ ratio of the DHCAL}
\label{sec:eOverPi}
\begin{figure}
\begin{center}
	\includegraphics[width=.7\textwidth]{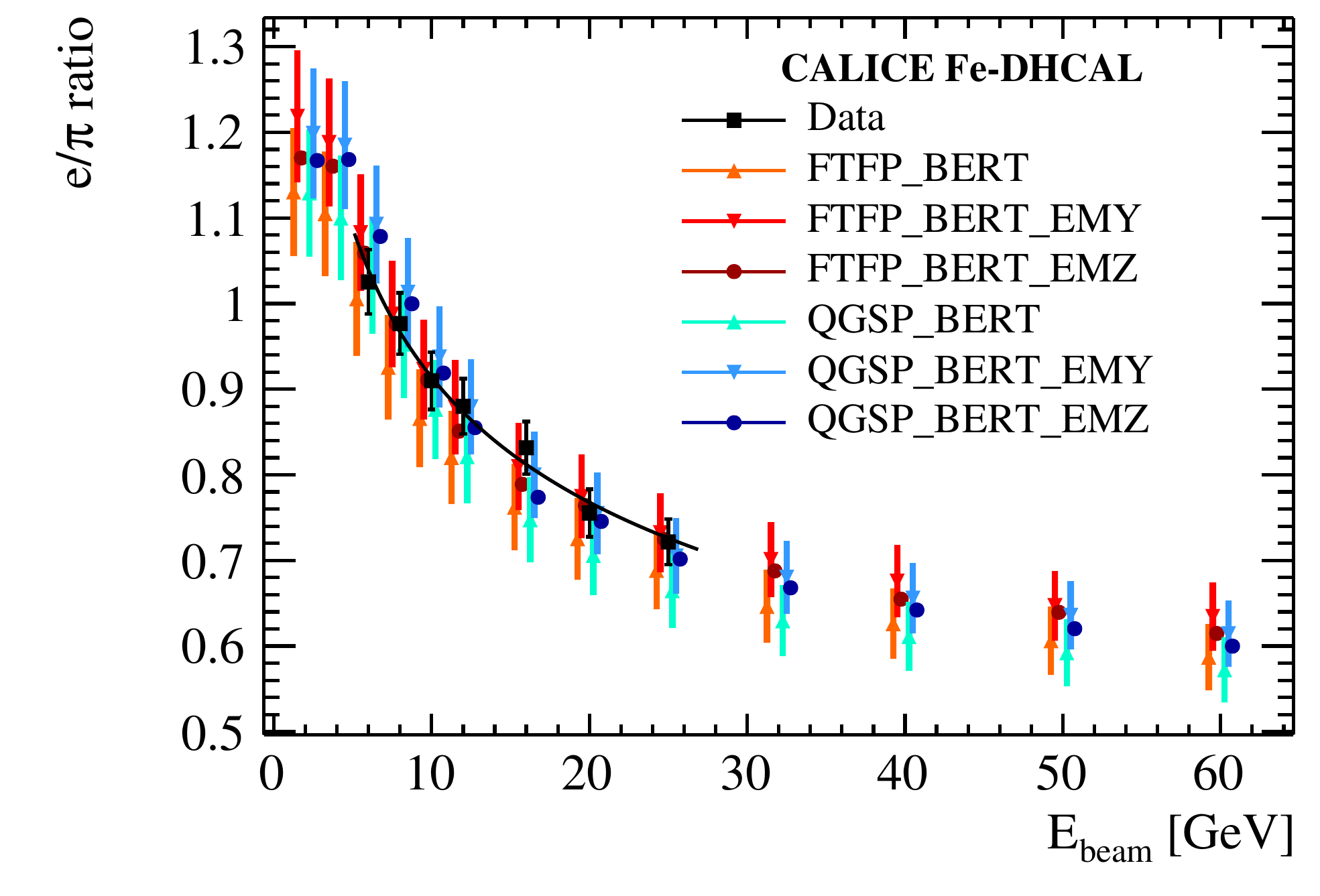}
\caption{The $e/\pi$ ratio determined from the mean response to positrons and pions in the Fe-DHCAL. The black curve shows the fit to data with Equation~\ref{eq:pol}. The markers of the simulations are shifted in $E_{beam}$ for clarity. The error bars represent the systematic and statistical uncertainties added in quadrature.}
\label{fig:ePi}
\end{center}
\end{figure}

The $e/\pi$ ratio of the Fe-DHCAL is determined from the mean response to positrons and pions, before non-linearity correction. The results are shown in Fig.~\ref{fig:ePi} for the data and the simulations. The $e/\pi$ ratio of the Fe-DHCAL is energy dependent and varies from 1.03 to 0.74 between 6 and 25\,GeV. The $e/\pi$ ratio of a sampling calorimeter is usually larger than 1 due to the higher response to electrons. The Fe-DHCAL shows a different behaviour because of the digital readout and the hence resulting saturation in the response to dense electromagnetic showers. However, this ratio is close to unity around 8\,GeV, which is near the average energy of neutral hadrons expected at the ILC~\cite{Magill:2002ss}. 
All simulations agree within the errors with the data. 
The $e/\pi$ ratio can be parameterised as~\cite{wigmans}:
\begin{equation}\label{eq:pol}
\frac{e}{\pi}=\frac{e/h}{1-\left[1-\left(\frac{E_{beam}}{E_{0}}\right)^{k-1}\right]\cdot\left(1-e/h\right)},
\end{equation}
with $e/h$ the ratio between the response to electromagnetic and non-electromagnetic shower components, $E_{0}$ the energy threshold for $\pi^0$ production and the factor $k$, that is related to the multiplicity of $\pi^0$s~\cite{wigmans}. The fit to the data is shown as a black curve in Fig.~\ref{fig:ePi} resulting in the following parameter values: $e/h=0.61\pm0.02$, $E_{0}=\left(1.1\pm0.8\right)$\,GeV and $k=0.74\pm0.03$. The values of $E_{0}$ and $k$ are in agreement with the values in the literature of $E_{0}=0.8\,GeV$ for iron and $k\sim0.75-0.85$~\cite{wigmans}. 

The increasing non-compensation of the Fe-DHCAL with higher beam energies degrades the energy resolution for pion (hadron) showers and motivates the development of software compensation algorithms. These algorithms can correct for the lower EM response by weighting hits belonging to EM sub-showers and hits in the hadronic shower parts differently~\cite{SCPaper, SCdhcal}. 

\section{Conclusions}
The Fe-DHCAL was operated in a mixed particle beam at Fermilab. During the data taking, the changing environmental conditions affected the gain of the chambers. A calibration method based on through going tracks was successfully used to equalise the response of each RPC in the stack. The performance of the method was only limited by the sparse statistics of the tracks in the bottom and top RPCs of each layer. 

The imaging capabilities of the DHCAL are successfully used in the event selection to separate muon, positron and pion events, using their shower topologies, without biasing the data samples.

The simulation of the Fe-DHCAL testbeam setup is based on \textsc{Geant4}, which allows the test of a variety of physics lists. The simulation of the RPC response (digitisation) is done, assuming all RPCs are operated in the same conditions. The Fe-DHCAL response to muons and positrons from calibrated data samples is used as reference. The tuning of the digitisation parameters required some assumptions on the physical range of the parameter values. 

The comparison between the data and the simulations reveals a strong dependence of the response and energy resolution on the EM physics lists for the positron and the pion showers. The simulation of the positrons using the \_EMZ physics list shows overall the best agreement with data. The best agreement for pion showers between data and simulation is achieved using the QGSP\_BERT\_EMZ physics list. 

The electromagnetic and hadronic longitudinal shower shapes are well described by all simulations within their uncertainties.

The Fe-DHCAL achieves an energy resolutions for single positrons of $\left(34.6\pm0.9\right)\%\,\sqrt{E}$ in the stochastic term and $\left(12.5\pm0.3\right)\%$ in the constant term within an energy range of 2 to 25\,GeV. For single pions, the Fe-DHCAL reaches a resolution with a stochastic term of $\left(51.5\pm1.5\right)\%\,\sqrt{E}$ and a constant term of $\left(10.6\pm0.5\right)\%$ for pion energies up to 32\,GeV. For higher energies, the resolution degrades due to saturation effects. 

The presented analysis of the Fe-DHCAL testbeam data and simulation presents the first detailed study of \textsc{Geant4} Monte Carlo models with recorded data of a prototype calorimeter based on RPCs. It has been shown that a better description of the EM shower core of hadron showers is needed to precisely reproduce the data. 

\section*{Acknowledgements}

We gratefully acknowledge the DESY, CERN and FNAL managements for their support and hospitality, and their accelerator staff for the reliable and efficient beam operation. We would also like to thank the Fermilab test beam crew, in particular Aria Soha, Erik Ramberg, Doug Jensen, Rick Coleman and Chuck Brown, for providing us with excellent beam and all the necessary support.
This work was supported 
by the 'Quarks and Leptons' Programme of CNRS/IN2P3 France;
by the FWO, Belgium; 
by the Natural Sciences and Engineering Research Council of Canada;
by the Ministry of Education, Youth and Sports of the Czech Republic;
by the Alexander von Humboldt Stiftung (AvH), Germany;
by the Bundesministerium f\"ur Bildung und Forschung (BMBF), Germany; 
by the Deutsche Forschungsgemeinschaft (DFG), Germany; 
by the Helmholtz-Gemeinschaft (HGF), Germany; 
by the I-CORE Program of the Planning and Budgeting Committee, Israel;
by the Nella and Leon Benoziyo Center for High Energy Physics, Israel;
by the Israeli Science Foundation, Israel;
by the National Research Foundation of Korea;
by the Korea-EU cooperation programme of National Research Foundation of Korea, Grant Agreement 2014K1A3A7A03075053; 
by the Science and Technology Facilities Council, UK;
by the Nuclear Physics, Particle Physics, Astrophysics and Cosmology Initiative, a Laboratory Directed Research and Development program at the Pacific Northwest National Laboratory, USA.
This publication is based upon work supported by the U.S. Department of Energy, Office of Science, Office of High Energy Physics, under contract number DE-AC02-06CH11357.

\bibliographystyle{elsarticle-num} 
\bibliography{Literature}





\end{document}

%% file: authors.tex

\title{Analysis of Testbeam Data of the Highly Granular RPC-Steel CALICE Digital Hadron Calorimeter and Validation of \textsc{Geant4} Monte Carlo Models} 

\address[add1]{Laboratoire d'Annecy-le-Vieux de Physique des Particules, Universit\'{e} de Savoie, CNRS/IN2P3, 9 Chemin de Bellevue BP110, F-74941 Annecy-le-Vieux CEDEX, France}
\address[add2]{Argonne National Laboratory, 9700 S.\ Cass Avenue, Argonne, IL 60439-4815, USA}
\address[add11]{CERN, 1211 Gen\`{e}ve 23, Switzerland}
\address[add15]{NICADD, Northern Illinois University, Department of Physics, DeKalb, IL 60115, USA}
\address[add18]{DESY,  Notkestr. 85, D-22607 Hamburg, Germany}
\address[add19]{Univ. Hamburg, Physics Department, Institut f\"ur Experimentalphysik, Luruper Chaussee 149, 22761 Hamburg, Germany} 
\address[add21]{University of Iowa, Dept. of Physics and Astronomy, 203 Van Allen Hall, Iowa City, IA 52242-1479, USA}
\address[add22]{University of Kansas, Department of Physics and Astronomy, Malott Hall, 1251 Wescoe Hall Drive, Lawrence, KS 66045-7582, USA}
\address[add23]{Department of Physics and Research Center for Advanced Particle Physics, Kyushu University, 744 Motooka, Nishi-ku, Fukuoka 819-0395, Japan}

\author[]{{\bf The CALICE Collaboration}}
\author[add1]{M.\,Chefdeville}

\author[add2]{J.\,Repond}
\author[add2]{J. Schlereth}
\author[add2]{J.R. Smith\fnref{fn14}} 
\author[add2]{D. Trojand\fnref{fn15}} 
\author[add2]{L.\,Xia } 
\author[add2]{Q. Zhang\fnref{fn16}}

%
%
%
%
%
%
%

\author[add11]{J.\,Apostolakis}
\author[add11]{C.\,Grefe\fnref{fn1}}
\author[add11]{V.\,Ivantchenko}
\author[add11]{G.\,Folger}
\author[add11]{A.\,Ribon}
\author[add11]{V.\,Uzhinskiy}

%
%

\author[add15]{G. C.\,Blazey}
\author[add15]{A.\,Dyshkant}
\author[add15]{K.\,Francis}
\author[add15]{V.\,Zutshi}

%

\author[add18]{O.\,Bach}
\author[add18]{V.\,Bocharnikov}
\author[add18]{E.\,Brianne}
\author[add18]{K.\,Gadow}
\author[add18]{P.\,G\"{o}ttlicher}
\author[add18]{O.\,Hartbrich\fnref{fn2}}
\author[add18]{D.\,Heuchel}
\author[add18]{F.\,Krivan}
\author[add18]{K.\,Kr\"{u}ger}
\author[add18]{J.\,Kvasnicka\fnref{fn3}}
\author[add18]{S.\,Lu}
\author[add18]{C.\,Neub\"{u}ser\corref{cor0}\fnref{fn0,fn6}}
\ead{coralie.neubuser@cern.ch}
\author[add18]{O.\,Pinto}
\author[add18]{A.\,Provenza}
\author[add18]{M.\,Reinecke}
\author[add18]{F.\,Sefkow}
\author[add18]{S.\,Schuwalow}
\author[add18]{Y.\,Sudo}
\author[add18]{H.L.\,Tran}

\author[add19]{P.\,Buhmann}
\author[add19]{E.\,Garutti}
\author[add19]{D.\,Lomidze}
\author[add19]{S.\,Martens}
\author[add19]{M.\,Matysek}
 
 
\author[add21]{B.\,Bilki\fnref{fn4}}
\author[add21]{D.\,Northacker}
\author[add21]{Y.\,Onel }

\author[add22]{B.\,van Doren}
\author[add22]{G.W.\,Wilson}
 
\author[add23]{K.\,Kawagoe}
\author[add23]{Y.\,Miura}
\author[add23]{R.\,Mori}
\author[add23]{I.\,Sekiya}
\author[add23]{T.\,Suehara}
\author[add23]{T.\,Yoshioka}


\author[add28]{D.\,Belver}
\author[add28]{E.\,Calvo Alamillo}
\author[add28]{M.C.\,Fouz}
\author[add28]{H.\,Garc\'\i{}a Cabrera}
\author[add28]{J.\,Mar\'\i{}n}
\author[add28]{J.\,Navarrete}
\author[add28]{J.\,Puerta Pelayo}
\author[add28]{A.\,Verdugo}
\address[add28]{CIEMAT, Centro de Investigaciones Energ\'eticas, Medioambientales y Tecnol\'ogicas, Madrid, Spain}


\author[add30]{F.\,Corriveau}
\author[add30]{B.\,Freund\fnref{fn6}}
\address[add30]{Department of Physics, McGill University, Ernest Rutherford Physics Bldg., 3600 University Ave., Montr\'{e}al, Qu\'{e}bec, Canada H3A 2T8}

\author[add31]{M.\,Chadeeva\fnref{fn7}}
\author[add31]{M.\,Danilov\fnref{fn7}}
\address[add31]{P.\,N.\, Lebedev Physical Institute, Russian Academy of Sciences, 117924 GSP-1 Moscow, B-333, Russia}



\author[add34]{M.\,Gabriel}
\author[add34]{L.\, Emberger}
\author[add34]{C.\,Graf}
\author[add34]{Y.\,Israeli}
\author[add34]{F.\,Simon}
\author[add34]{M.\,Szalay}
\author[add34]{H.\,Windel}
\address[add34]{Max-Planck-Institut f\"ur Physik, F\"ohringer Ring 6, D-80805 Munich, Germany}


\author[add36]{S.\,Bilokin\fnref{fn9}}
\author[add36]{J.\,Bonis}
\author[add36]{A.\,Irles}
\author[add36]{R.\,P\"oschl}
\author[add36]{A.\,Thiebault}
\author[add36]{F.\,Richard}
\author[add36]{D.\,Zerwas}
\address[add36]{Laboratoire de l'Acc\'elerateur Lin\'eaire, CNRS/IN2P3 et Universit\'e de Paris-Sud XI, Centre Scientifique d'Orsay B\^atiment 200, BP 34, F-91898 Orsay CEDEX, France}

\author[add39]{J.\,Cvach}
\author[add39]{M.\,Janata}
\author[add39]{M.\,Kovalcuk}
\author[add39]{I.\,Polak}
\author[add39]{J.\,Smolik}
\author[add39]{V.\,Vrba}
\author[add39]{J.\,Zalesak}
\author[add39]{J.\,Zuklin}
\address[add39]{Institute of Physics, The Czech Academy of Sciences, Na Slovance 2, CZ-18221 Prague 8, Czech Republic}


%
\author[add42]{T.\,Takeshita}
\address[add42]{Shinshu Univ.\,, Dept. of Physics, 3-1-1, Asahi, Matsumoto-shi, Nagano 390-8621, Japan}
\author[add43]{A.\,Elkhalii}
\author[add43]{M.\,G\"otze}
\author[add43]{C.\,Zeitnitz}
\address[add43]{Bergische Universit\"{a}t Wuppertal Fakult\"at 4 / Physik, Gaussstrasse 20, D-42097 Wuppertal, Germany}

\author[add44]{S.\,Chang}
\author[add44]{A.\,Khan\fnref{fn12}}
\author[add44]{D.H.\,Kim}
\author[add44]{D.J.\,Kong}
\author[add44]{Y.D.\,Oh}
\address[add44]{Department of Physics, Kyungpook National University, Daegu, 702-701, Republic of Korea}

%
%
%

\cortext[cor0]{corresponding author}
\fntext[fn0]{now at CERN, 1211 Gen\`{e}ve 23, Switzerland}
\fntext[fn1]{now at Bonn University, Bonn, Germany}
\fntext[fn2]{now at University of Hawaii at Manoa, High Energy Physics Group, 2505 Correa Road, HI, Honolulu 96822, USA}
\fntext[fn3]{also at Institute of Physics, The Czech Academy of Sciences}
\fntext[fn4]{also at Beykent University, Istanbul, Turkey}
\fntext[fn6]{also at Argonne National Laboratory}
\fntext[fn7]{also at MEPhI}
\fntext[fn9]{now at IPHC Strasbourg, 23 rue du loess - BP28, 67037 Strasbourg cedex 2}
\fntext[fn12]{now at Islamia College University Peshawar}
\fntext[fn14]{now at University of Maryland}
\fntext[fn15]{now at University of Windsor}
\fntext[fn16]{now at Xi'an Jiaotong University}

%% file: PaperFeDHCAL_20190121.bbl
\begin{thebibliography}{10}
\expandafter\ifx\csname url\endcsname\relax
  \def\url#1{\texttt{#1}}\fi
\expandafter\ifx\csname urlprefix\endcsname\relax\def\urlprefix{URL }\fi
\expandafter\ifx\csname href\endcsname\relax
  \def\href#1#2{#2} \def\path#1{#1}\fi

\bibitem{ILC}
\href{https://www.linearcollider.org}{{{International Linear Collider (ILC)
  collaboration}}}.
\newline\urlprefix\url{https://www.linearcollider.org}

\bibitem{CLIC}
\href{http://clic-study.web.cern.ch}{{{Compact Linear Collider (CLIC)
  collaboration}}}.
\newline\urlprefix\url{http://clic-study.web.cern.ch}

\bibitem{Thomson200925}
M.~Thomson, {{Particle flow calorimetry and the PandoraPFA algorithm}}, NIM A
  611~(1) (2009) 25 -- 40.
\newblock \href {https://doi.org/http://dx.doi.org/10.1016/j.nima.2009.09.009}
  {\path{doi:http://dx.doi.org/10.1016/j.nima.2009.09.009}}.

\bibitem{CALICE}
\href{https://twiki.cern.ch/twiki/bin/view/CALICE/WebHome}{{{CALICE
  collaboration}}}.
\newline\urlprefix\url{https://twiki.cern.ch/twiki/bin/view/CALICE/WebHome}

\bibitem{Adams}
{{C. Adams et al.}}, {{Design, construction and commissioning of the Digital
  Hadron Calorimeter---DHCAL}}, JINST 11~(07) (2016) P07007.

\bibitem{RPCs}
{{G. Drake et al.}}, {{Resistive Plate Chambers for hadron calorimetry: Tests
  with analog readout}}, NIM A 578~(1) (2007) 88 -- 97.
\newblock \href {https://doi.org/https://doi.org/10.1016/j.nima.2007.04.160}
  {\path{doi:https://doi.org/10.1016/j.nima.2007.04.160}}.

\bibitem{Fermilab}
\href{http://ftbf.fnal.gov/beam-delivery-path/}{{{Fermilab Test Beam
  Facility}}}.
\newline\urlprefix\url{http://ftbf.fnal.gov/beam-delivery-path/}

\bibitem{TCMT}
{{C. Adloff et al. (CALICE collaboration)}},
  \href{http://stacks.iop.org/1748-0221/7/i=04/a=P04015}{{{Construction and
  performance of a silicon photomultiplier/extruded scintillator tail-catcher
  and muon-tracker}}}, JINST 7~(04) (2012) P04015.
\newline\urlprefix\url{http://stacks.iop.org/1748-0221/7/i=04/a=P04015}

\bibitem{DHCALmuons_pro}
{{J. Repond for the CALICE collaboration}}, {{Calibration of the DHCAL with
  Muons}}, XI workshop on Resistive Plate Chambers and Related Detectors
  (RPC2012) PoS (2012)~(RPC2012) (2012) 076.

\bibitem{EnvDepRPCs}
{{Q. Zhang et al. }}, {{Environmental Dependence of the Performance of
  Resistive Plate Chambers}}, JINST 5~(P02007).

\bibitem{CalibrationDhcalPaper}
{{B. Bilki et al.}},
  \href{http://stacks.iop.org/1748-0221/3/i=05/a=P05001}{{{Calibration of a
  digital hadron calorimeter with muons}}}, JINST 3~(05) (2008) P05001.
\newline\urlprefix\url{http://stacks.iop.org/1748-0221/3/i=05/a=P05001}

\bibitem{tracksAHCAL}
{{C. Adloff et al. (CALICE collaboration)}},
  \href{http://stacks.iop.org/1748-0221/8/i=09/a=P09001}{Track segments in
  hadronic showers in a highly granular scintillator-steel hadron calorimeter},
  JINST 8~(09) (2013) P09001.
\newline\urlprefix\url{http://stacks.iop.org/1748-0221/8/i=09/a=P09001}

\bibitem{PhDthesis}
{{C. Neub{\"u}ser}}, {{The Comparison of Two Highly Ganular Hadronic
  Calorimeter Concepts}}, Ph.D. thesis, Universit{\"a}t Hamburg (2016).

\bibitem{DHCALinteractionLayer}
C.~G. Reichelt, {{Particle ID Studies in a Highly Granular Hadron
  Calorimeter}}, Cern summer student program, CERN - European Organization for
  Nuclear Research, Switzerland (September 2013).

\bibitem{GEANT4}
{{S. Agostinelli et al.}}, {{Geant4---a simulation toolkit}}, Nuclear
  Instruments and Methods in Physics Research Section A: Accelerators,
  Spectrometers, Detectors and Associated Equipment 506~(3) (2003) 250 -- 303.
\newblock \href {https://doi.org/https://doi.org/10.1016/S0168-9002(03)01368-8}
  {\path{doi:https://doi.org/10.1016/S0168-9002(03)01368-8}}.

\bibitem{Geant4_emLists}
\href{https://geant4.web.cern.ch}{{{\textsc{Geant4} electromagnetic physics
  lists}}}.
\newline\urlprefix\url{https://geant4.web.cern.ch}

\bibitem{Geant4_PartonString}
{{G. Folger, J.P. Wellisch}}, {{String parton models in GEANT4}}, Proceedings,
  13th International Conference on Computing in High-Enery and Nuclear Physics
  (CHEP 2003): La Jolla, California CHEP~(MOMT007).

\bibitem{Geant4_reccomendation}
{{\url{http://geant4.cern.ch/support/proc_mod_catalog/physics_lists/useCases.shtml}}}.

\bibitem{ValidationGeant4}
{{C. Adloff et al. (CALICE collaboration)}}, {{Validation of \textsc{Geant4}
  Monte Carlo models with a highly granular scintillator-steel hadron
  calorimeter}}, JINST 8~(07) (2013) P07005.

\bibitem{SDHCALdigitizer}
{{Z. Deng et al. (CALICE collaboration)}},
  \href{http://stacks.iop.org/1748-0221/11/i=06/a=P06014}{{{Resistive Plate
  Chamber digitization in a hadronic shower environment}}}, JINST 11~(06)
  (2016) P06014.
\newline\urlprefix\url{http://stacks.iop.org/1748-0221/11/i=06/a=P06014}

\bibitem{RPCSimulation}
{{M. Abbrescia et al.}}, {{The simulation of resistive plate chambers in
  avalanch mode: charge spectra and efficiency}}, NIM A~(431) (1999) 413--427.

\bibitem{Burak}
{{B. Bilki (University of Iowa)}}, {{Private communication}} (2014).

\bibitem{Geant4_private}
{{A. Dotti (\textsc{Geant4} working group @ SLAC)}}, {{Private communication}}
  (2014).

\bibitem{CAN-31}
L.~Xia, {{CALICE DHCAL Noise Analysis}}, {{CALICE Analysis Note CAN-31}} (March
  2011).

\bibitem{Min-DHCAL}
{{B. Freund et al. (CALICE collaboration)}},
  \href{http://stacks.iop.org/1748-0221/11/i=05/a=P05008}{{{DHCAL with minimal
  absorber: measurements with positrons}}}, JINST 11~(05) (2016) P05008.
\newline\urlprefix\url{http://stacks.iop.org/1748-0221/11/i=05/a=P05008}

\bibitem{NovoFit}
{{H. Ikeda et al.}},
  \href{http://www.sciencedirect.com/science/article/pii/S0168900299009924}{{{A
  detailed test of the CsI(Tl) calorimeter for BELLE with photon beams of
  energy between 20MeV and 5.4GeV}}}, Nuclear Instruments and Methods in
  Physics Research Section A: Accelerators, Spectrometers, Detectors and
  Associated Equipment 441~(3) (2000) 401 -- 426.
\newblock \href {https://doi.org/https://doi.org/10.1016/S0168-9002(99)00992-4}
  {\path{doi:https://doi.org/10.1016/S0168-9002(99)00992-4}}.
\newline\urlprefix\url{http://www.sciencedirect.com/science/article/pii/S0168900299009924}

\bibitem{CAN-49a}
C.~Neub{\"u}ser, {{Comparison of Energy Reconstruction Schemes and Different
  Granularities in the CALICE AHCAL}}, {{CALICE Analysis Note CAN-49a}} (May
  2016).

\bibitem{SCPaper}
{{C. Adloff et al. (CALICE collaboration)}}, {{Hadronic energy resolution of a
  highly granular scintillator-steel hadron calorimeter using software
  compensation techniques}}, JINST 7~(P09017).

\bibitem{SCdhcal}
C.~Neub{\"u}ser.
\newblock
  \href{\url{https://agenda.linearcollider.org/event/6557/contributions/31752/attachments/26182/40131/20150420_CaliceMeetingKEK_SCforDHCAL.pdf}}{{{Fe-DHCAL:
  Software Compensation}}} [online] (\\ Asian Linear Collider Workshop
  (ALCW2015) \\
  \url{https://agenda.linearcollider.org/event/6557/contributions/31752/attachments/26182/40131/20150420_CaliceMeetingKEK_SCforDHCAL.pdf}).
\newblock \\ April 19-24th 2015.

\bibitem{Magill:2002ss}
{{S. Magill et al.}}, {E-flow optimization of the hadron calorimeter for future
  detectors}, in: {Calorimetry in particle physics. Proceedings, 10th
  International Conference, CALOR 2002, Pasadena, USA, March 25-29, 2002},
  2002, pp. 806--813.

\bibitem{wigmans}
R.~Wigmans, {{Calorimetry}}, Oxford Science Publications, 2000.

\end{thebibliography}
